\documentclass[acmsmall]{acmart}

\AtBeginDocument{%
  \providecommand\BibTeX{{%
    \normalfont B\kern-0.5em{\scshape i\kern-0.25em b}\kern-0.8em\TeX}}}

\setcopyright{acmlicensed}
\acmJournal{TKDD}
\acmYear{2021} \acmVolume{1} \acmNumber{1} \acmArticle{1} \acmMonth{1} \acmPrice{15.00}\acmDOI{10.1145/3494561}

\usepackage{graphicx}
\usepackage{amsmath}
\usepackage{amsfonts}
\usepackage{epsfig}
\usepackage{subfigure}
\usepackage{url}
\usepackage{balance}
\usepackage{algorithmic}
\usepackage{algorithm}
\usepackage{multirow}
\usepackage{xspace}
\usepackage{color}
\usepackage{bbding}
\usepackage[tableposition=top]{caption}

\newtheorem{theor}{Theorem}

\newtheorem{lem}{Lemma}

\newtheorem{exam}{Example}
\newtheorem{problem}{Problem}

\newcommand{\bigO}{\mathcal{O}}

\newcommand{\spara}[1]{\smallskip\noindent{\bf #1}}

\newcommand{\squishlist}{
	\begin{list}{$\bullet$}
		{  \setlength{\itemsep}{0pt}
			\setlength{\parsep}{3pt}
			\setlength{\topsep}{3pt}
			\setlength{\partopsep}{0pt}
			\setlength{\leftmargin}{2em}
			\setlength{\labelwidth}{1.5em}
			\setlength{\labelsep}{0.5em}
	} }
	\newcommand{\squishlisttight}{
		\begin{list}{$\bullet$}
			{ \setlength{\itemsep}{0pt}
				\setlength{\parsep}{0pt}
				\setlength{\topsep}{0pt}
				\setlength{\partopsep}{0pt}
				\setlength{\leftmargin}{2em}
				\setlength{\labelwidth}{1.5em}
				\setlength{\labelsep}{0.5em}
		} }
		
		\newcommand{\squishdesc}{
			\begin{list}{}
				{  \setlength{\itemsep}{0pt}
					\setlength{\parsep}{3pt}
					\setlength{\topsep}{3pt}
					\setlength{\partopsep}{0pt}
					\setlength{\leftmargin}{1em}
					\setlength{\labelwidth}{1.5em}
					\setlength{\labelsep}{0.5em}
			} }
			
			\newcommand{\squishend}{
			\end{list}
		} 

\begin{document}
%\title{Graph Summary: From Single to Multi-Relations}
\title{Multi-relation Graph Summarization}
\author{Xiangyu Ke}
\email{xiangyu001@ntu.edu.sg}
\affiliation{%
  \institution{Nanyang Technological University}
  \country{Singapore}
}

\author{Arijit Khan}
\email{arijit.khan@ntu.edu.sg}
\affiliation{%
  \institution{Nanyang Technological University}
  \country{Singapore}
}

\author{Francesco Bonchi}
\email{francesco.bonchi@isi.it}
\affiliation{%
  \institution{ISI Foundation, Italy}
  \country{and Eurecat, Spain}
}

\begin{abstract}
Graph summarization is beneficial in a wide range of applications, such as visualization, interactive and exploratory analysis, approximate query processing, reducing the on-disk storage footprint, and graph processing in modern hardware. However, the bulk of the literature on graph summarization surprisingly overlooks the possibility of having edges of different types.
In this paper, we study the novel problem of producing summaries of multi-relation networks, i.e., graphs where multiple edges of different types may exist between any pair of nodes. Multi-relation graphs are an expressive model of real-world activities, in which a relation can be a topic in social networks,
an interaction type in genetic networks, or a snapshot in temporal graphs.

The first approach that we consider for multi-relation graph summarization is a two-step method based on summarizing each relation in isolation, and then aggregating the resulting summaries in some clever way to produce a final unique summary. In doing this, as a side contribution, we provide the first polynomial-time approximation algorithm based on the {\sf $k$-Median} clustering for the classic problem of lossless  single-relation graph summarization.

Then, we demonstrate the shortcomings of these two-step methods, and propose holistic approaches, both approximate and heuristic algorithms, to compute a summary directly for
multi-relation graphs. In particular, we prove that the approximation bound of {\sf $k$-Median} clustering for the single relation solution can be maintained in a multi-relation graph with proper aggregation operation over adjacency matrices corresponding to its multiple relations. Experimental results and case studies (on co-authorship networks and brain networks) validate the effectiveness and efficiency of the proposed algorithms.
\end{abstract}




\maketitle \sloppy

\section{Introduction}
\label{sec:intr}
Fueled by the unprecedented growth rate of knowledge graphs, social networks, and Internet-of-Things\footnote{The Knowledge and Action Graph of Microsoft has 21 billion facts, 18 billion action links, and over five billion relationships between more than one billion people, places, and things \cite{JKLYE15}. Facebook has 800 million active users \cite{KE18}. Graph-of-Things (GoT), which is a live knowledge graph system for Internet-of-Things, has been adding millions of records per hour, and roughly about 10 billion RDF triples per month \cite{PhuocQQNH16}.}, the problems of storing, managing, and mining very large graph data have received an enormous deal of attention in the data mining research community in recent years. At the current rate of data volume increase, in fact, it is becoming highly impractical to store, manage, process, and visualize such big graphs. Graph summarization alleviates these issues by producing a concise graph representation (i.e., summary) that still can be meaningfully explored and queried.  Graph summarization has shown to be beneficial in a wide range of applications, such as visualization, interactive and exploratory analysis, approximate query processing, reducing the on-disk storage footprint, graph embedding, and graph processing in modern hardware \cite{LiuSDK18,KBB17,KoutraVB18,BH18,YYW21}.

Surprisingly, little attention has been paid to the problem of summarizing multi-relation graphs.
In real-world, entities are often correlated in multiple ways, either explicitly or implicitly. For instance,
{\sf BioGRID} (\url{thebiogrid.org})
describes seven different types of genetic interactions between genes in Homo Sapiens.
%: direct interaction, physical association,
%suppressive genetic interaction defined by inequality, association, co-localization, additive genetic interaction defined by inequality,
%and synthetic genetic interaction defined by inequality.
{\sf STRING} (\url{string-db.org})  models protein-to-protein
interactions with six types of correlations statistically learned from existing protein databases, revealing that most
protein interactions are associated with at least two types of correlations. Other applications where multiple
relations exist between entities include social and financial networks \cite{GBG17}, urban and transportation systems \cite{CGZRPPB12},
ecology research \cite{SASGA17}, and recommender systems \cite{MaoLZZ17,JGHJL20,XHXDZYPB21}.

When multiple relations exist between entities, data is modeled as
{\em multi-relation networks} (also known as {\em multi-layer}, {\em multiplex}, or {\em multi-dimensional networks}) \cite{dickison_magnani_rossi_2016}.
This graph model has been attracting increasing research interest in graph
analytics, such as in shortest path finding \cite{ZhangO19}, core decomposition and densest subgraph discovery \cite{GBG17,GalimbertiBGL20},
node clustering \cite{BGHS12},   frequent subgraphs mining \cite{YanZH05}, and in social networks analysis \cite{CosciaRPCG13}, just to mention a few.

In this paper, we
% first revisit the traditional single-relation graph summarization problem, and provide a polynomial-time approximation algorithm based on the {\sf $k$-Median} clustering. Then, we introduce and
 study, for the first time, the problem of multi-relation graph summarization.
Before discussing the contributions of our work and how they collocate in the state of the art, we need to provide some background notions and formally define the problems.

%
%If understanding and querying
%single-relation graphs are difficult, would they not be even harder over multi-relation networks? Hence, there is a
%critical need to summarize multi-relation graphs for interactive query processing and visualization, as well as
%to discover complex, hidden, correlated inter-dependencies across different entities in the dataset.
%It is, however, a challenging problem. We first demonstrate that an intuitive approach
%to aggregate the individual graph summaries across relations
%would be inadequate for an effective summarization of multi-relation graphs. We, therefore, design holistic algorithms
%to compute more compact and efficient summary for multi-relation graphs, with theoretically bounded size.

\subsection{Background and related work}
\label{subsec:background}

\begin{figure}[tb!]
	\centering
	\includegraphics[scale=0.75,angle=90]{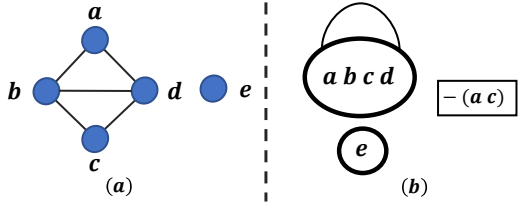}
	\caption{\small A simple graph and its summary.}
	\label{fig:example_s}
\end{figure}

We consider graph summarization obtained by aggregating nodes into \emph{supernodes}.
In particular, we adopt \emph{lossless} summarization as introduced by Navlakha et al. \cite{NRS08}, for simple undirected single-relation graphs. Given a graph $G = (V,E)$, a lossless summary  $\mathcal{S}=\langle G_S, \mathcal{C}_{S} \rangle$ consists of a \emph{summary graph} $G_S=(V_S,E_S)$ and a set of \emph{edge corrections} $\mathcal{C}_{S}=\langle \mathcal{C}_{S}^+, \mathcal{C}_{S}^-\rangle$, where:
\squishlist
\item $V_S = \{S_1, \ldots, S_k\}$ is the set of \emph{supernodes} inducing a partition of $V$, i.e., $\bigcup_{i = 1}^k S_i = V$ and  $\bigcap_{i = 1}^k S_i = \emptyset$;
\item $E_S \subseteq V_S \times V_S$ is a set of \emph{superedges} between supernodes (possibly including self-loops);
\item $\mathcal{C}_{S}^+ \subseteq E$ is the set of edges to be inserted to reconstruct $G$, while $\mathcal{C}_{S}^- \subseteq (V \times V) \setminus E$ is the set of edges to
be deleted.
\squishend
The summarization is lossless because given the summary $\mathcal{S}$, we can reconstruct the original graph \emph{exactly},
by (1) ``exploding'' each superedge  $(U,W) \in E_S$, i.e., creating an edge $(u,w)$ for each pair of nodes $u,w \in U \times W$, (2) adding each edge in $\mathcal{C}_{S}^+$, and (3) removing the edges in $\mathcal{C}_{S}^-$. More formally:
$E = \{(u,w) \mid u\in U, w\in W, (U,W) \in E_S\} \cup  \mathcal{C}_{S}^+ \setminus \mathcal{C}_{S}^- .$
An example of lossless summary is provided in Figure~\ref{fig:example_s}.

The problem studied by Navlakha et al. \cite{NRS08} is to find the smallest possible summary of an input graph.

\begin{problem}[Lossless-Sum \cite{NRS08}]
\label{prob:single}
Given a simple, undirected, graph $G=(V,E)$, find its smallest lossless summary $\mathcal{S}=\langle G_S, \mathcal{C}_{S} \rangle$, i.e., the one that minimizes
$|E_S|+|\mathcal{C}_{S}|$.
\end{problem}

The cost of storing the mapping from nodes in $V$ to supernodes in $V_S$ is disregarded in the objective function \cite{NRS08}, since it remains constant across different summaries, i.e., $\bigO(|V|)$. %Navlakha et al. \cite{NRS08} shows that this cost is generally small compared to the storage cost of $|E_S|+|\mathcal{C}|$.
The problem can be seen through the lenses of the  {\em Minimum Description Length} (MDL) principle \cite{JR78}, which states that the
best theory to infer from a set of data is the one which minimizes the sum of the size of the theory and  the size of the data when encoded through the theory. Here, the data is the input graph $G$, the theory is the summary
graph represented by supernodes $V_S$ and superedges $E_S$, and the correction list $\mathcal{C}_{S}$ is the encoding
of the data with regards to the theory.
For instance, in the example in Figure ~\ref{fig:example_s} the cost of the summary is 2 (1 superedge + 1 correction), against a cost of 5 of the original graph (i.e., 5 edges).

As observed in \cite{NRS08}, a summary is entirely defined by the partitioning of nodes into supernodes.
In fact, once provided such partitioning, superedges can be deterministically decided by simply checking whether they induces advantages w.r.t. the summary cost or not. Finally, once decided the superedges, the corrections are straightforwardly identified. Therefore, Problem \ref{prob:single} is essentially a graph partitioning problem: given a single-relation graph $G$ with $m$ nodes and $n$ edges, there are $\sum_{k=1}^{n}{n\choose k}=\bigO(2^n)$ possibilities to partition these $n$ nodes into supernodes $V_S$. To solve it, Navlakha et al. \cite{NRS08} proposed a simple greedy agglomerative heuristic {\em without quality guarantee}. They also studied a lossy version of the problem. To the best of our knowledge, the computational complexity of Problem~\ref{prob:single} has not been addressed in the literature \cite{NRS08,SGKR19}.
We keep the problem of investigating the computational complexity of Problem~\ref{prob:single} open here, and consider it as an interesting future work.

A similar approach is followed by LeFevre and Terzi  \cite{LeFevreT10} who study summaries obtained by aggregating nodes into supernodes. However, %they do not consider having a correction table, instead
they keep on each superedge the information about how many original edges it represents (but not which ones): this is clearly a lossy summarization. Their objective is then to find the summary minimizing the loss for a given number $k$ of allowed supernodes (which implicitly controls the compression rate). The loss is represented by the \emph{reconstruction error}, i.e., the difference between the original graph and the probabilistic graph that one can reconstruct form the summary.
Similar to \cite{NRS08}, %LeFevre and Terzi
\cite{LeFevreT10}  propose a simple greedy agglomerative heuristic {\em with no quality guarantee}.
Later Riondato et al. \cite{RGB14,RGB17} %, while studying the same problem of LeFevre and Terzi  \cite{LeFevreT10}, exposed a connection between graph summarization and geometric clustering problems (e.g., $k$-median), thanks to which they achieve
propose {\em the first polynomial-time approximation algorithm} for the problem of \cite{LeFevreT10}. In this paper we show that, following a similar intuition, we can achieve {\em the first polynomial-time approximation algorithm} for the classic lossless summarization problem of \cite{NRS08}.

\subsection{Multi-relation graph summarization}
In this work, we extend the notion of lossless summarization over multi-relation graphs.
An undirected, multi-relation graph $G$ is a triplet $(V,E,R)$, where $V$ is a set of $n$ nodes,
$R$ is a set of $q$ relations, and $E \subseteq V \times V \times R$ is a set of $m$ undirected edges.
Therefore, in a multi-relation graph, each edge is a triplet: e.g., an edge between nodes $u$ and $v$ in relation $r\in R$ is represented by $(u,v,r)$. A summary $\mathcal{S}=\langle G_S, \mathcal{C}_{S} \rangle$ is defined as in the single-relation case, the only difference is that the correction edges in $\mathcal{C}_{S}$ are triplets and  also superedges are now triplets, i.e., $E_S \subseteq V_S \times V_S \times R$.

\begin{exam}
 Figure~\ref{fig:example_m} provides an example of a multi-relation graph and its summary.
The graph on the left-hand side is defined over 3 relations, contains 5 nodes and 16 edges. The summary on the right-hand side is obtained by grouping $\{a,c\}$ and $\{b,d\}$ as two supernodes and keeping $\{e\}$ as a supernode over all 3 relations. No correction is required here. The cost of such summary is thus 6 (given by 6 superedges + 0 correction), while the cost of the original graph was 16 (i.e., 16 edges).

\begin{figure}[tb!]
	\centering
	\includegraphics[scale=0.85,angle=90]{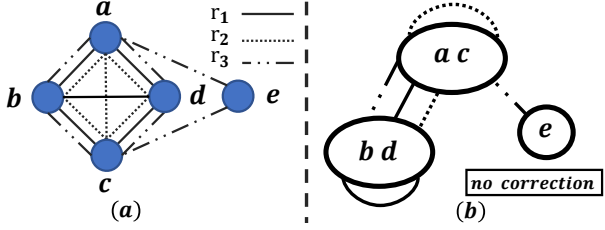}
	\vspace{-2mm}
	\caption{\small A multi-relation graph and its summary.}
	\label{fig:example_m}
\end{figure}
\begin{figure}[tb!]
	\centering
	{\includegraphics[scale=0.7,angle=90]{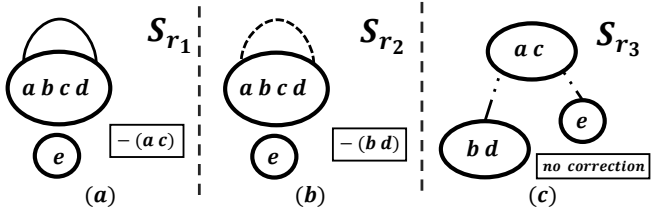}}
	\vspace{-2mm}
	\caption{\small The individual summary for each relation in Figure~\ref{fig:example_m}(a).}
	\label{fig:s1_graphs}
\end{figure}
\end{exam}

The problem we study in this article is formally defined as follows.

\begin{problem}[Lossless-Sum-Multi]
\label{prob:multi}
 Given a multi-relation graph $G=(V,E,R)$, find its smallest lossless summary $\mathcal{S}=\langle G_S, \mathcal{C}_{S} \rangle$, i.e., the one that minimizes
$|E_s|+|\mathcal{C}_{S}|$.
\end{problem}

In many applications one might be interested in a summary with a predefined number $k$ of supernodes.
Moreover, one might solve Problem \ref{prob:multi} by using an algorithm which takes in input the number of supernodes $k$, with a wrapper for selecting the optimal value of $k$. Therefore, in this article we also tackle the following problem.

\begin{problem}[$k$-Lossless-Sum-Multi]
\label{prob:k_multi}
Given a multi-relation graph $G=(V,E,R)$ and $k \in \mathbb{N}$, find the lossless summary $\mathcal{S}=\langle G_S, \mathcal{C}_{S} \rangle$, such that $G_S=(V_S,E_S,R)$  and $|V_S|=k$, that minimizes $|E_S|+|\mathcal{C}_{S}|$.
\end{problem}

\subsection{Why not keeping an individual summary for each relation?}
%
%As discussed, %earlier,
%graph summarization is essentially a type of graph partitioning problem. The additional complexity brought by dealing with multi-relation graph
%%resides
%is in the necessity of finding a node partitioning which is as good as possible, over multiple relations.
Figure~\ref{fig:s1_graphs} shows the optimal summary for each relation of the multi-relation graph in Figure~\ref{fig:example_m}(a). One could argue that storing these three individual summaries
would also serve as a lossless summarization for the given multi-relation graph. However, this is not a good option due to two reasons:
{\bf (1)} The optimal multi-relation summary in Figure~\ref{fig:example_m}(b) provides us more insights about the input network. For example,
by only looking at the individual summaries in Figure~\ref{fig:s1_graphs}, we cannot easily determine the fact that the nodes in set $\{a,c\}$
are fully connected with the nodes in set $\{b,d\}$ via all relations. However, the nodes within each of these sets interact with themselves in
different ways. Thus, it is better to characterize them as two strongly connected supernodes with different self-loops as in the
multi-relation summary in Figure~\ref{fig:example_m}(b). {\bf (2)} More storage is required for maintaining all individual summaries,
because each individual summary might require storing of a different node mapping (i.e., the mapping from nodes in $V$ to supernodes in $V_S$). This is demonstrated in our experiments (\S~\ref{sec:base}).

\subsection{Contributions and roadmap}
The considerations above also suggest a natural  two-step approach to multi-relation graph summarization: first  summarize the input graph one relation at a time, then aggregate the various summary graphs into a single summary. We follow this intuition and, as a first solution, we develop such a two-step approach. For both steps we compare different methods, producing several baseline two-step algorithms.

Among the algorithms we design for the first step, we also consider a $k$-median algorithm to produce single-relation summary, inspired by \cite{RGB14}. For this algorithm we prove approximation guarantees: this is, to the best of our knowledge, the \emph{first polynomial-time approximation algorithm for the classic lossless graph summarization problem} of \cite{NRS08}.

Finally we show, by means of an example, an effectiveness limitation suffered by the two-steps approach.
Therefore, following the intuition behind the example, we move on to design \emph{holistic} approaches, which are experimentally shown to be faster and more accurate than the two-step approaches. Moreover, our holistic {\sf $k$-Median$^+$} algorithm maintains the same approximation guarantee for multi-relation graph summarization (\S~\ref{sec:algo_multi}). Our final {\sf Hybrid} algorithm combines the {\sf Greedy} method \cite{NRS08} and the approximate
solution {\sf $k$-Median}, to provide the most compact summary in practice (\S\ref{sec:hybrid}).

Our main contribution is to initiate investigation into multi-relation graph summarization.
Besides, this paper achieves the following contributions:

\begin{itemize}

\item We revise the classic single-relation graph lossless summarization problem, and provide the first polynomial-time approximation algorithm (\S\ref{sec:bound_single}).

\item We design basic two-step algorithms, which first generate a lossless summary for each relation, then properly aggregate them to obtain one uniform summary. We also highlight the limits of this approach  (\S\ref{sec:algo_single}).

\item We propose holistic algorithms for more compact and efficient summary generation. Our holistic {\sf $k$-Median$^+$} algorithm maintains the same approximation guarantee for multi-relation graph summarization (\S\ref{sec:algo_multi}).

\item We combine the traditional {\sf Greedy} method \cite{NRS08} and the approximation soluton {\sf $k$-Median} as the final proposed algorithm {\sf Hybrid}, which empirically produces the most compact summary (\S\ref{sec:hybrid}).

\item  Our empirical evaluation on four real-world networks confirms that our holistic algorithms can produce more compact summaries and are faster than the two-step approaches (\S\ref{sec:exp}).

  \item  Real-world applications on visualization, classification, and query processing demonstrate the effectiveness
    and efficiency of our proposal (\S\ref{sec:app}).
\end{itemize}

Next section cover additional related literature. Section \ref{sec:conclusion} concludes the paper and discusses future work.

\section{Other related work}\label{sec:other_RW}
Graph summarization has been used for a wider range of problems related to static plain graphs \cite{BV04,CKLMPR09,BRSV11,BLN09,RG03,KKVF15,CS12,KF11},
static attributed graphs \cite{SBDY12,THP08,CYZHY08,ZhangBNCZ15}, dynamic and stream graphs \cite{SKZGF15,SBD14,TMBB16,Bandyopadhyay10,ZhaoAW11,CormodeM05,FeigenbaumKMSZ08,AhnGM12,AGM12,TCM16,KhanA17,Gou0Z019}, probabilistic and distributed graph summarization \cite{HassanlouST13,LTHLM14}.
We refer the reader to excellent surveys and tutorials \cite{LFCY010,TianP10,LYL13,LiuSDK18,KBB17,KoutraVB18,BH18}. Regardless this wide literature, \emph{no prior work has tackled graph summarization in the multi-relation graph setting}. Therefore, the rest of this section covers the literature about single-relation graph summarization.
In Section \ref{subsec:background},  we already reviewed the most important related papers which constitute the background for our work. Table \ref{tab:rw} collocates our contribution within the most important related work.

\begin{table} [t!]
\caption{Characterization of the most related papers.}
\centering
\begin{tabular} { c|c|c|c}
\hline
\textsf{Paper} & \textsf{Lossless} & \textsf{Multi-Relation} & {\sf Approx. Guarantees}\\
\hline
\cite{NRS08}  & \checkmark  & $\;$ & $\;$ \\
 \cite{LeFevreT10} & $\;$ & $\;$ & $\;$ \\
\cite{RGB14}  & $\;$ &  $\;$& \checkmark \\
\cite{BegAZK18} & $\;$ &  $\;$& \checkmark \\
\cite{SGKR19}  & \checkmark  & $\;$ & $\;$\\
this work &  \checkmark   & \checkmark & \checkmark \\
%{\sf $k$-Median$^+$ [ours]}  & \checkmark  & \checkmark   & \checkmark & \checkmark \\
%{\sf Greedy$^+$ [ours]}   & \checkmark   & \checkmark  & $\;$ & $\;$ \\
%{\sf Randomized$^+$ [ours]}    & \checkmark  & \checkmark & $\;$ & $\;$ \\
\hline
\end{tabular}
\label{tab:rw}
\end{table}

\spara{Aggregation-based graph summary.} Notable techniques under this category are pattern mining and community based
summarization \cite{BC08,KKVF14,RZ18}, supernode and edge-correction (thus lossless) \cite{NRS08,SGKR19}, supernode and reconstruction-error (thus lossy) \cite{LeFevreT10,RGB14,BegAZK18}. Supernode based aggregation methods \cite{NRS08,SGKR19,LeFevreT10,RGB14,BegAZK18} are
most similar to ours and are summarised in Table~\ref{tab:rw}.

Very recently, SSumM \cite{LJKLS20} proposes lossy graph summarization to minimize the reconstruction error, however the constraint is on size of the summary graph in bits (and not on the number of supernodes as in \cite{LeFevreT10,RGB14}). Notice that our focus in this work is lossless summarization over multi-relation graphs, which is different from \cite{LJKLS20}.  We experimentally demonstrate the summary cost in bits based on various storage formats in \S~\ref{sec:exp}.

\spara{Web graph and social networks compression.} Boldi and Vigna \cite{BV04} show that web graphs are compressible
down to almost two bits per edge. Chierichetti et al. \cite{CKLMPR09} use shingle ordering instead of lexicographical ordering of web pages, in order to tackle
social networks. Finding an order of nodes, which captures the ``regularity'' of the network,
is a challenging problem. Boldi et al. \cite{BRSV11} introduce a layered label propagation
algorithm for reordering very large graphs. Other interesting works include \cite{BLN09,RG03,KKVF15,CS12,KF11}.
These methods focus on reducing the number of bits needed to encode an edge, and none compute graph summaries.
%Hence, these approaches do not provide
%any insight into graph structure.

\spara{Attribute-based graph summary.} Nodes and edges of many real-world graphs
are annotated with attributes. Hence, there exist graph summarization works considering both topology
and semantics of the node and edge attributes \cite{KNL14}.
{\sf FUSE}~\cite{SBDY12} is a functional summarization technique for protein interaction networks,
and this helps comprehending high-level functional relationships in disease-related PPI networks such as Alzheimer's disease network.
{\sf SNAP} \cite{THP08} and {\sf OLAP} \cite{CYZHY08} allow interactive summarization at various resolutions over heterogeneous
networks. Topology and attribute-based summarization of a large collection of small graphs (e.g., chemical compounds) and its application in constructing data-driven visual graph query interfaces are discussed in~\cite{ZhangBNCZ15}. These methods are not directly comparable to ours,
since our summarization deals with the graph structure. For example, a superedge in \cite{THP08} exists between a pair of supernodes if any node in a supernode has {\em at least one} edge to the nodes in the other supernodes. However, in our problem formulation, we consider the exact number of edges between them.

\spara{Application-oriented graph summary.} These are graph summarization techniques for efficient query
answering and pattern mining, such as reachability, shortest path, and pattern matching queries \cite{FLWW12,TZHH11,ZMT10},
eigenvector centrality, degree, and adjacency queries \cite{LeFevreT10}, neighborhood query \cite{MP10}, keyword search \cite{WYSIY13}, distributed graph computation \cite{KTSLF11}, graph mining \cite{KE18, CLFCYH09,CH94,MP12}, information cascade and influential node discovery \cite{QLJZF14,PPKZS14,SSXSTMC16}. We demonstrate applications of our multi-relation graph summary in efficient query processing
in \S \ref{sec:app}.

\spara{Other related graph computation.} Related graph analytics problems include
sampling \cite{HL13}, sparsification \cite{BenczurK15}, clustering and community detection \cite{BBC04,NG69,WhiteS05}, graph embedding \cite{JRKKRK19},
partitioning \cite{KK95}, and dense subgraph mining \cite{GT15}. As discussed in \cite{KBB17,LYL13,JRKKRK19}, 
these problems are different from graph summarization. 

\section{Single-Relation Graph Summarization: {\sf $k$-Median} Clustering}
\label{sec:bound_single}
In this section, we provide an approximation algorithm for single-relation graph summarization problem based on {\sf $k$-Median} clustering. The {\sf $k$-Median} clustering is performed on the rows of the adjacency matrix $A_{G}$
of the input graph $G$, to create $k$ supernodes. In particular,
the goal of {\sf $k$-Median} clustering is to find a set of $k$ centers ${\bf x}=\{x_1, x_2, \ldots , x_k\}$
that minimizes the {\sf $k$-Median} cost for the node set $V \subseteq \mathbb{R}^n$.
The $k$-median cost is defined as $\sum_{v\in V}d(v,{\bf x})$, where
$d(v, {\bf x}) = \min_{x\in{\bf x}}d(v,x)$. Here, $d(v,x)$ denotes the Euclidean distance between two points (i.e., nodes) $v,x \in  \mathbb{R}^n$.
The nodes are then grouped into $k$ supernodes based on their nearest cluster center. Notice that the $k$-summary is a
graph summary with exactly $k$ supernodes. After obtaining the supernode partitioning, we include
superedges and correction list as discussed in \S~\ref{subsec:background}.
The time complexity of {\em $k$-Median} is $\bigO(m+nk\log n)$ \cite{RGB14}.

We prove that the {\sf $k$-Median} algorithm guarantees {\em 16-approximation}
to the optimal solution for the {\sf Lossless-Sum} problem with $k$ supernodes. 
In the previous study \cite{RGB14}, the {\sf $k$-Median} clustering based technique was applied to generate approximated lossy summary. We bridge the gap between the reconstruction error (Equation~\ref{eq:lp}) in lossy summary and the correction list size (Equation~\ref{eq:et}) in lossless summary, which is our problem. This enables reusing the same technique in a different problem setting (i.e.,
our problem), and it achieves a different approximation factor from that in \cite{RGB14}. Here, we only consider the
number of correction edges $|\mathcal{C}_{S}|$, since the number of superedges
$|E_s|$ can be bounded with $\bigO(k^2)$ for a $k$-summary.
Our proof is built on top of a theoretical result by Riondato et al. \cite{RGB14}, that establishes
8-approximation guarantee for a similar {\sf $k$-Median} algorithm, with respect to the quality of
a {\em lossy summarization}, known as the {\em $l_p$ reconstruction error} as below.
For a lossy summarization \cite{LeFevreT10,RGB14,BegAZK18}, only the graph summary
$G_S=(V_S,E_S)$ is created; no additional correction list is stored. The summary $G_S$ is a complete graph in this case, including all self-loops, i.e., $E_S=V_S\times V_S$.
Given a summary, the graph is approximately reconstructed by the expected adjacency matrix,
$A_{G_S}^{\uparrow}$, which is an $(n \times n)$ matrix with
$A_{G_S}^{\uparrow}(u,w)$ = $\frac{|E_{UW}|}{|U||W|}$.
Here, $U,W$ are supernodes in $V_S$ such that $u \in U$ and $w \in W$, and $E_{UW}$ is the set of edges that actually exist between $U$ and $W$ in the original graph $G$.
The quality of the summary, called the $l_p$ reconstruction error, is measured by a norm
of difference between the input adjacency matrix $A_{G}$ and the reconstructed adjacency matrix
$A_{G_S}^{\uparrow}$.

The $l_p$ reconstruction error ($RE_p$) of a summary $G_S$ for a graph $G$ is:
\begin{align}
\displaystyle RE_p(G,G_S) = \sqrt[p]{\sum_{u=1}^{|V|}\sum_{w=1}^{|V|}(|A_{G}(u,w)-A_{G_S}^{\uparrow}(u,w)|)^p}
\label{eq:lp}
\end{align}

In this paper, we use $p = 1$, that is, the $l_1$ reconstruction error. From \cite{RGB14},
we have the following theorem.

\begin{theor}\label{th:riondato}
	Let $G_{S^\#}$ be the $k$-summary induced by the $k$-Median partitioning of the rows of $A_{G}$,
	and let $G_{S^+}$ be the optimal $k$-summary for $G$ with respect to the $l_1$-reconstruction error.
	The $l_1$-reconstruction error of $G_{S^\#}$ is an 8-approximation to the best $l_1$-reconstruction error. Formally: $RE_1(G,G_{S^\#})\leq 8\cdot RE_1(G,G_{S^+})$.
\end{theor}

%Next, to prove our ultimate result that the {\sf $k$-Median} algorithm guarantees 16-approximation
%to the optimal solution for the {\sf Lossless-Sum} problem with $k$ supernodes, Lemma~\ref{th:correction} will be useful.

\begin{lem}
	Let $G_{S^+}$ be the optimal $k$-summary for $G$ with respect to the $l_1$-reconstruction error,
	and let $G_{S^*}$ be the optimal $k$-summary for $G$ with respect
	to the number of correction edges. The correction list size, $|\mathcal{C}_{S^+}|$ of $G_{S^+}$
	is a 2-approximation to the best size of correction list, $|\mathcal{C}_{S^*}|$. Formally:
$|\mathcal{C}_{S^+}|\leq 2\cdot|\mathcal{C}_{S^*}|$.
	\label{th:correction}
\end{lem}
\begin{proof}
	Let us denote by $\alpha_{UW}=\frac{|E_{UW}|}{|U||W|}$.
	From the definition of $l_1$ reconstruction error in Equation~\ref{eq:lp}, we get:

	\begin{align}
	RE_1(G,G_S) &= \sum_{u=1}^{|V|}\sum_{w=1}^{|V|}|A_{G}(u,w)-A_{G_S}^{\uparrow}(u,w)| \nonumber& \\
	\displaystyle &= 2\cdot\sum_{(U,W) \in V_S \times V_S} |U||W|\alpha_{UW}(1-\alpha_{UW})
	\label{eq:err}
	\end{align}

	The intuition behind this derivation is that there are $|U||W|$ cells corresponding to each supernode pair $(U,W)$
	in both the original adjacency matrix $A_G$ and the reconstructed adjacency matrix $A_{G_S}^{\uparrow}$. In $A_{G_S}^{\uparrow}$, all such cells are filled with $\alpha_{UW}$. In $A_G$, there are $\alpha_{UW}$ proportion of cells having value $1$, and the rest $(1-\alpha_{UW})$ proportion of cells having value $0$. The subtraction results of $|A_{G}(u,w)-A_{G_S}^{\uparrow}(u,w)|$ for the first group are all $(1-\alpha_{UW})$, and those for the second group are all $\alpha_{UW}$. Thus, we derive the second line in Equation~\ref{eq:err}.
	
	Now in the context of lossless summary, we decide whether to keep a superedge between a pair of supernodes $(U,W)$ by the edge density between them.
	If $\alpha_{UW}>0.5$, maintaining a superedge can result in less storage overhead of $\mathcal{C}_{S}^-$, than that of $\mathcal{C}_{S}^+$ without this superedge. Suppose $\mathcal{C}_{S}$ is the correction list, then its size can be calculated as below.
	\begin{align}
	|\mathcal{C}_{S}|=\frac{1}{2}\sum_{(U,W)\in V_S\times V_S}|U||W|\left\{\begin{matrix}
	(1-\alpha_{UW}),\ if \ \alpha_{UW}>0.5\\
	\alpha_{UW}\ \ \ \ \ \quad \quad ,\ otherwise
	\end{matrix}\right.
	\label{eq:et}
	\end{align}

	From Equations~\ref{eq:err} and ~\ref{eq:et}, we have:
	\begin{align}
	\frac{RE_1(G,G_S)}{|\mathcal{C}_{S}|}=\left\{\begin{matrix}
	4\alpha_{UV}\ \ \ \ ,\ if \ \alpha_{UV}>0.5\\
	4(1-\alpha_{UV}),\ otherwise
	\end{matrix}\right.
	\end{align}
	Clearly, $2\leq \frac{RE_1(G,G_S)}{|\mathcal{C}_{S}|} \leq 4$, since $\alpha_{UV}\in[0,1]$.
	In other words,
	\begin{align}
	\frac{1}{4} RE_1(G,G_S) \leq |\mathcal{C}_{S}| \leq \frac{1}{2} RE_1(G,G_S)
	\label{eq:bound}
	\end{align}
	%\begin{align}
	%|\mathcal{C}| \geq \frac{1}{4} RE_1(G,G_S)
	%\label{eq:ub}
	%\vspace{-1mm}
	%\end{align}
	
	Suppose $G_{S^+}$ be the optimal $k$-summary for $G$ with respect to the $l_1$-reconstruction error,
	and let $G_{S^*}$  be the optimal $k$-summary for $G$ with respect
	to the number of correction edges. Thus, we get:
	\begin{align}
	|\mathcal{C}_{S^+}| & \leq \frac{1}{2} RE_1(G,G_{S^+}) \quad \triangleright \quad \text{\scriptsize  by the r.h.s of Equation~\ref{eq:bound}} \nonumber & \\
	&\leq \frac{1}{2} RE_1(G,G_{S^*}) \quad \triangleright \quad \text{\scriptsize  since $G_{S^+}$ is optimal wrt $RE_1$} \nonumber  &\\
	& \leq  2\cdot|\mathcal{C}_{S^*}| \quad \triangleright \quad \text{\scriptsize  by the l.h.s of Equation~\ref{eq:bound}} &
	\end{align}
	This completes the proof.
\end{proof}

\begin{theor}
	Let $G_{S^\#}$ be the $k$-summary induced by the $k$-Median partitioning of the rows of $A_{G}$,
	and let $G_{S^*}$ be the optimal $k$-summary for $G$ with respect to the number of correction edges.
	The correction list size, $|\mathcal{C}_{{S^\#}}|$ of $G_{S^\#}$
	is a 16-approximation to the best size of correction list, $|\mathcal{C}_{{S^*}}|$. Formally: $
	|\mathcal{C}_{{S^\#}}|\leq 16\cdot |\mathcal{C}_{{S^*}}|$
	\label{th:single_sum_approx}
\end{theor}
\begin{proof}
	We denote by $G_{S^+}$ the optimal $k$-summary for $G$ with respect to the $l_1$-reconstruction error.
	Next, we derive the follows.
	\begin{align}
	|\mathcal{C}_{{S^\#}}| & \leq \frac{1}{2} RE_1(G,G_{S^\#}) \quad \triangleright \quad \text{\scriptsize  by the r.h.s of Equation~\ref{eq:bound}} \nonumber & \\
	&\leq 4\cdot RE_1(G,G_{S^+}) \quad \triangleright \text{\scriptsize  by Theorem~\ref{th:riondato}} \nonumber & \\
	&\leq 4\cdot RE_1(G,G_{S^*}) \quad \triangleright \quad \text{\scriptsize  since $G_{S^+}$ is optimal wrt $RE_1$} \nonumber  &\\
	& \leq  16\cdot|\mathcal{C}_{{S^*}}| \quad \triangleright \quad \text{\scriptsize  by the l.h.s of Equation~\ref{eq:bound}} &
	\end{align}
	Hence, the proof is completed.
\end{proof}

Note that in \S\ref{sec:k}, we discuss several empirical methods for finding a suitable $k$ for the {\sf $k$-Median} method, which helps it adapt to the general {\sf Lossless-Sum} problem. 
\section{Multi-Relation Graph Summary: Baseline Methods}
\label{sec:algo_single}
In this section, we first present several straightforward baselines for the lossless summarization of multi-relational graphs, then demonstrate how they suffer
from effectiveness issues, which
will be instrumental in developing a more accurate
and scalable solution in \S\ref{sec:algo_multi}.

Our baseline algorithms follow a two-step approach:

{\bf (1)} We explore the input graph for one relation at a time, and generate a lossless summary for each of them.

{\bf (2)} The summaries across relations are properly aggregated to obtain one uniform summary.

In the first step, our problem is same as the {\sf Lossless-Sum} problem. Given a set of summaries
$\{\mathcal{S}_1,\mathcal{S}_2,...,\mathcal{S}_q\}$, each for a specific relation, our next target
is to find a single summary, i.e., a partition into supernodes, that agrees as much as possible with the
$q$ individual summaries.

In addition to the {\sf $k$-Median} approach introduced in \S\ref{sec:bound_single}, we briefly revisit some widely-used graph summarization techniques for the {\sf Lossless-Sum} problem.
To the best of our knowledge, all existing algorithms \cite{NRS08,SGKR19} for the {\sf Lossless-Sum} problem are heuristic in nature, without any theoretical guarantee on the summary size.
In \S\ref{sec:s_agg}, we tackle the problem of summary aggregation and provide several methods to aggregate the individual summaries across relations.
Finally, in \S\ref{sec:pp} we discuss potential limitations of these two-step baseline algorithms.

\subsection{Single-relation graph summarization algorithms}
\label{sec:gs_single}

We first revisit {\sf Greedy} and {\sf Randomized} algorithms from \cite{NRS08}. Then, we discuss an advanced algorithm, {\sf SWeG}, with similar idea in recent literature \cite{SGKR19}. An example is provided to demonstrate these algorithms.

\spara{Greedy algorithm.} The {\sf Greedy} algorithm \cite{NRS08} is a heuristic, %(i.e., without any theoretical guarantee on the summary size),
bottom-up approach. It first considers every node in the input graph
as a supernode, and iteratively merges the best pair $\{u,w\}$ with the maximum reduction in summary cost. The general workflow of {\sf Greedy} algorithm is given below:
{\bf (1)} It computes the potential cost reduction for all pairs of nodes in the input graph $G$ which are 2-hops apart, and records those pairs which are positive.
{\bf (2)} The best pair of nodes $\{U,W\}$ with highest cost reduction is merged into a new node $W$.
{\bf (3)} Delete cost reduction records about $U$ or $W$ for all the nodes which is within 2-hops to $U$ or $W$, and compute their cost reduction to the new supernode $H$.
{\bf (4)} Update the cost reduction between nodes which are neighbors of $H$ ($I$ is a neighbor of $H$ if there exists any edge $\{a,b\}\in E$, $a\in H$, $b\in I$).
{\bf (5)} Repeat (2)-(4) until no positive cost reduction exists.

%Between any pair of supernodes $U$ and $W$, let $\Pi_{UW}$ be the set of all possible pairs $\{a,b\}$, such that $a\in U$ and $b\in W$. $E_{UW}\subseteq \Pi_{UW}$ is defined as the set of edges actually present in the input graph $G$, i.e., $E_{UW}=\Pi_{UW}\cap E$. Therefore, the cost of the supernode pair $(U,W)$ is calculated as:
%
%\begin{align}
%C(U,W)=min\{|\Pi_{UW}|-|E_{UW}|+1, |E_{UW}|\}
%\label{eq:cost}
%\end{align}

%The neighbor set $N(U)$ of $U$ is defined to be the set of supernodes $W$ that have such edge $\{a,b\} \in E_{UW}$.
%We define the cost of a supernode $U$ as follows.
%
%\begin{align}
%C(U)=\sum_{X\in N(U)}C(U,X)
%\end{align}

%The cost reduction due to merging supernodes $U$ and $W$ into a new supernode $H$ can be calculated as:
%
%\begin{align}
%\triangle C(U,W)=\frac{C(U)+C(W)-C(H)}{C(U)+C(W)}
%\end{align}
%
%Taking fraction instead of the absolute cost reduction in the above equation is to get rid of the bias towards nodes with higher degree.

%Based on the above notion,

Notice that {\sf Greedy} directly works with the {\sf Lossless-Sum} problem. It
can easily handle the additional input $k$ for the number of supernodes by {\bf (a)} terminating earlier when the number of supernodes becomes $k$ (even before satisfying condition (5)), or {\bf (b)} force to merge the pairs with less ``sacrifice" in summary quality (i.e., negative cost reduction) after (5), if a smaller $k$ is required.

Let $d_{av}$ be the average degree for each node, the time complexity of {\sf Greedy} is $\bigO(d_{av}^3(d_{av}+\log n+\log d_{av}))$ \cite{NRS08}.

\spara{Randomized algorithm.}
Comparing with the basic {\sf Greedy} algorithm, the {\sf Randomized} approach reduces the high computation overhead, by sacrificing the compression quality. It, in fact, has the worst performance in compression based on our experimental results in \S~\ref{sec:exp}.
%yet empirically achieving similar compression.
In each step, it randomly selects an unexplored supernode $U$,
and computes cost reduction with all its unexplored neighbors. If no positive reduction exists,
$U$ is marked explored, and the algorithm continues. Otherwise, $U$ is merged with its best neighbor $W$ (i.e., having the
highest reduction) into a new node $H$. $U$ and $W$ are removed, and $H$ is now unexplored. The algorithm stops when all nodes are explored. The complexity of the {\sf Randomized} algorithm is $\bigO(d_{av}^3)$ \cite{NRS08}.

\spara{SWeG.}
The most recent algorithm, {\sf SWeG} \cite{SGKR19}, is an advanced version of the {\sf Randomized} algorithm with further efficiency improvement.
It first divides the graph into smaller disjoint groups. Each node group contains supernodes with similar connectivity, based on the concept of the shingle of a node $u$, which is defined as:
$f(u)=\min_{w\in N_u\ or\ w=u} h(w)$. $h$ is a bijective hash function $h:V\rightarrow\{1,...,|V|\}$, and $N_v$ is the set of neighbors of node $v$ in the input graph $G$. Two nodes have the same shingle with probability equal to the Jaccard similarity of their neighbor sets \cite{BCFM00}. The shingle of a supernode $U$ is extended to be: $F(U)=\min_{u\in U}\{f(u)\}$.
Two supernode $U\neq W\in V_S$ are more likely to have the same shingle if the nodes in $U$ and those in $W$ share similar connectivity. After the nodes are partitioned into groups, a similar procedure as {\sf Randomized} is conducted within each group to merge supernodes. Moreover, this allows distributed implementation. The whole pipeline (grouping nodes, and merging nodes within each group) repeats $T$ times, each time with a different, randomly generated hash function $h$. $h$ can be easily produced by shuffling $\{1,...,|V|\}$.

The dividing step takes $\bigO(|E|)$ running time.
The cost of the merging step is same as that of the {\sf Randomized} algorithm: $\bigO(d_{av}^3)$,
but here $d_{av}$ may be smaller since it operates on a smaller graph. If we only allow sequential implementation and repeat $T$ times, it becomes $\bigO(pTd_{av}^3)$, where $p$ is the number of disjoint groups.

\begin{figure}[tb!]
	\centering
	{\includegraphics[scale=0.8,angle=90]{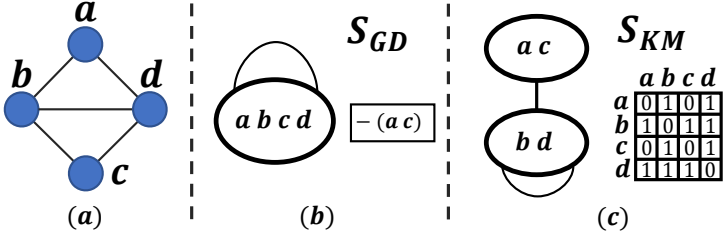}}
	\caption{Example for single-relation graph summary methods.}
	\label{fig:single_sum}
\end{figure}

\begin{exam}
Figure~\ref{fig:single_sum} provides an example for single-relation graph summarization algorithms. Given the example graph on the left, {\sf Greedy}
produces the summary $S_{GD}$. At the beginning, the cost of each node equals to the number of edges incident to this node. Merging node $a$ with $c$, or merging node $b$ with $d$ both lead to 50\% cost reduction, regardless of order. Then we achieve an intermediate result, which is the same as $S_{KM}$. The cost of the superedge between supernodes $ac$ and $bd$ are counted twice
from both side, which encourages {\sf Greedy} to continue merging. {\em This demonstrates that {\sf Greedy} tends to result in bigger-size supernodes.}
Though, this will not cause any issue in this simple example, we later demonstrate in \S\ref{sec:pp} that this may create trouble in aggregated summary finding (i.e., the
second step of baseline approaches). $S_{KM}$ is the summary for {\sf $k$-Median} when $k=2$, which can be easily obtained with the provided adjacency matrix. If $k$ is set to be $1$, {\sf $k$-Median}
will return same result as $S_{GD}$. The result of the {\sf Randomized} algorithm depends on the random order of node selection.
In this example, however, the final result of {\sf Randomized} would also be $S_{GD}$.
The costs of these two summaries are same: $S_{GD}$ has 1 superedge and 1 correction edge, while $S_{KM}$ has 2 superedges.
\end{exam}

\subsection{Summary aggregation}
\label{sec:s_agg}
The second phase of our baseline approach is aggregating summaries
across individual relations to obtain one uniform summary for the entire multi-relation graph. We formally define this problem following the well-known {\sf clustering aggregation} problem \cite{GMT07}. Both summary and clustering techniques partition the nodes (objects) into groups, and the aggregation operation further aims at finding a new uniform partition with minimum total disagreements to the current partitions.

A summary $\mathcal{S}$ corresponds to a partition of $n$ nodes.
The partition size $k$ can either be given as an input (\S~\ref{sec:bound_single}), or
can be automatically decided by the algorithm (\S~\ref{sec:gs_single}).
For any pair of nodes $(u, v)$, the indicator function $I_{\mathcal{S}}(u,v)$ returns 1 if and only if $u$ and $v$ are located in
the same supernode under summary $\mathcal{S}$ (0 otherwise). Then, we have $D_{u,v}(\mathcal{S}_i,\mathcal{S}_j)$ ($i,j\in[1,q]$)
to characterize if two summaries $\mathcal{S}_i$ and $\mathcal{S}_j$ disagree to each other on the partitioning of $u$ and $v$. Formally,
\begin{align}
D_{u,v}(\mathcal{S}_i,\mathcal{S}_j)=\left\{\begin{matrix}
1 , \,\, if\ I_{\mathcal{S}_i}(u,v)\neq I_{\mathcal{S}_j}(u,v)\\
0 ,\ \ \ \ \ \ \ \ \ \ \ \ \ \ \ \ \ \ \ otherwise
\end{matrix}\right.
\end{align}
The total disagreement between summaries $\mathcal{S}_i$ and $\mathcal{S}_j$ is:
\begin{align}
D_{V}(\mathcal{S}_i,\mathcal{S}_j)=\sum_{(u,v)\in V\times V}D_{u,v}(\mathcal{S}_i,\mathcal{S}_j)
\end{align}
In general, this metric counts the number of node pairs on which the two summaries disagree to each other.
Therefore, the {\em summary aggregation} problem is given as follows.
\begin{problem} [Sum-Agg]
	Given a set of $q$ summaries (i.e., node partitions) $\{\mathcal{S}_1,\mathcal{S}_2,...,\mathcal{S}_q\}$ on a set of nodes
	$V$, compute a new summary $\mathcal{S}$ that minimizes the total disagreements with all the given summaries, that is, it minimizes $\sum_{i=1}^{q}D_{V}(\mathcal{S},\mathcal{S}_i)$.
	\label{prob:agg}
\end{problem}

We now show that the {\sf Sum-Agg} problem can be reduced to the well-studied
correlation clustering problem \cite{BBC04}, thus standard procedures for solving correlation clustering
can be employed.
For any pair of nodes $(u,v)\in V \times V$, we define the distance between them as
$\mathcal{D}(u,v)=\frac{1}{q}\cdot |\{i:1\leq i \leq q\ and\ I_{\mathcal{S}_i}(u,v)=0\}|$,
which means the fraction of summaries that assign the pair $(u,v)$ into different supernodes.
The correlation clustering objective is to find a node partitioning $\mathcal{P}$ that minimizes
$\mathcal{D}_{\mathcal{P}}=\sum_{(u,v),\ I_\mathcal{P}(u,v)=1}\mathcal{D}(u,v)+\sum_{(u,v),\ I_\mathcal{P}(u,v)=0}(1-\mathcal{D}(u,v))$.
If the solution $\mathcal{P}$ places $u$, $v$ in the same group, it will disagree with $q\cdot\mathcal{D}(u,v)$
of the original partitionings due to individual relations.
In contrast, it will disagree with $q(1-\mathcal{D}(u,v))$ remaining partitionings if
the solution keeps $u$, $v$ separate. Thus, for any partitioning $\mathcal{P}$, we have $q\cdot \mathcal{D}_{\mathcal{P}}=\sum_{i=1}^{q}D_{V}(\mathcal{P},\mathcal{S}_i)$, which is our objective of
{\sf Sum-Agg} (i.e., Problem~\ref{prob:agg}). Due to this reduction, the following algorithms for solving
correlation clustering can be employed to solve our problem. Notice that some of them also come with provable approximation guarantees.

\spara{The BEST algorithm.} The most simple algorithm, {\sf BEST}, is to find one of the input summaries, $\mathcal{S}_i$, that minimizes the total number of disagreements to others. It can be computed in time $\bigO(q^2n)$, where $q$ is the number of input summaries (also the number of relations),
and $n$ the number of nodes in the input graph. Though simple, it yields a solution with an approximation ratio at most $2\cdot(1-1/q)$ \cite{GMT07}.

\spara{The Balls algorithm.} The {\sf Balls} algorithm \cite{CGW05} first sorts the nodes in an increasing order of the total distance
to all other nodes. Recall that for any pair of nodes $(u,v)\in V \times V$, we define the distance between them as
$\mathcal{D}(u,v)=\frac{1}{q}\cdot |\{i:1\leq i \leq q\ and\ I_{\mathcal{S}_i}(u,v)=0\}|$, that is, the fraction of summaries that
assign the pair $(u,v)$ into different supernodes. The algorithm is defined with an input parameter $\alpha$. The intuition of the algorithm
is to find a set of nodes that are close to each other, and far from other nodes. In order to find a good cluster, we take all nodes
that are close (within a ball) to a node $u$. The triangle inequality guarantees that if two nodes are close to $u$, then they are
also relatively close to each other. Once such a cluster is found, we remove it from the graph, and proceed with the rest of the
nodes. The {\sf Balls} algorithm consumes $\bigO(n^2)$ running time, while ensuring $max\{\frac{1-\alpha}{\alpha},$ $\frac{1+2\alpha}{1-2\alpha},\frac{2-2\alpha}{1-2\alpha}\}$ approximation guarantee \cite{GMT07}. The additional time
cost for pre-computing distances between all pairs of nodes is $\bigO(qn^2)$.

\begin{figure}[tb]
	\centering
	\subfigure[\small {\em Compactness}]
	{\includegraphics[scale=0.25]{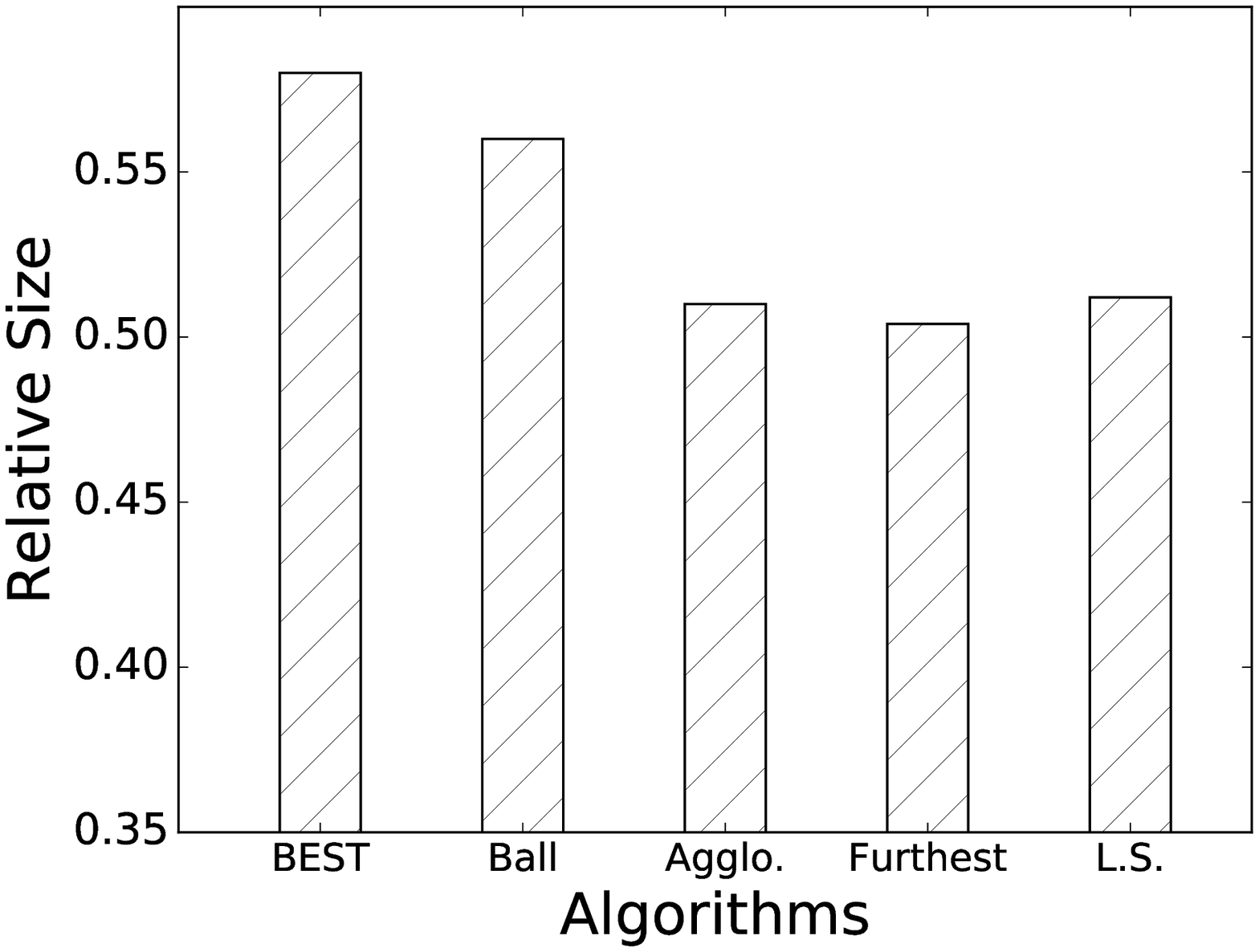}
		\label{fig:agg_c}}
	\subfigure[\small {\em Efficiency}]
	{\includegraphics[scale=0.25]{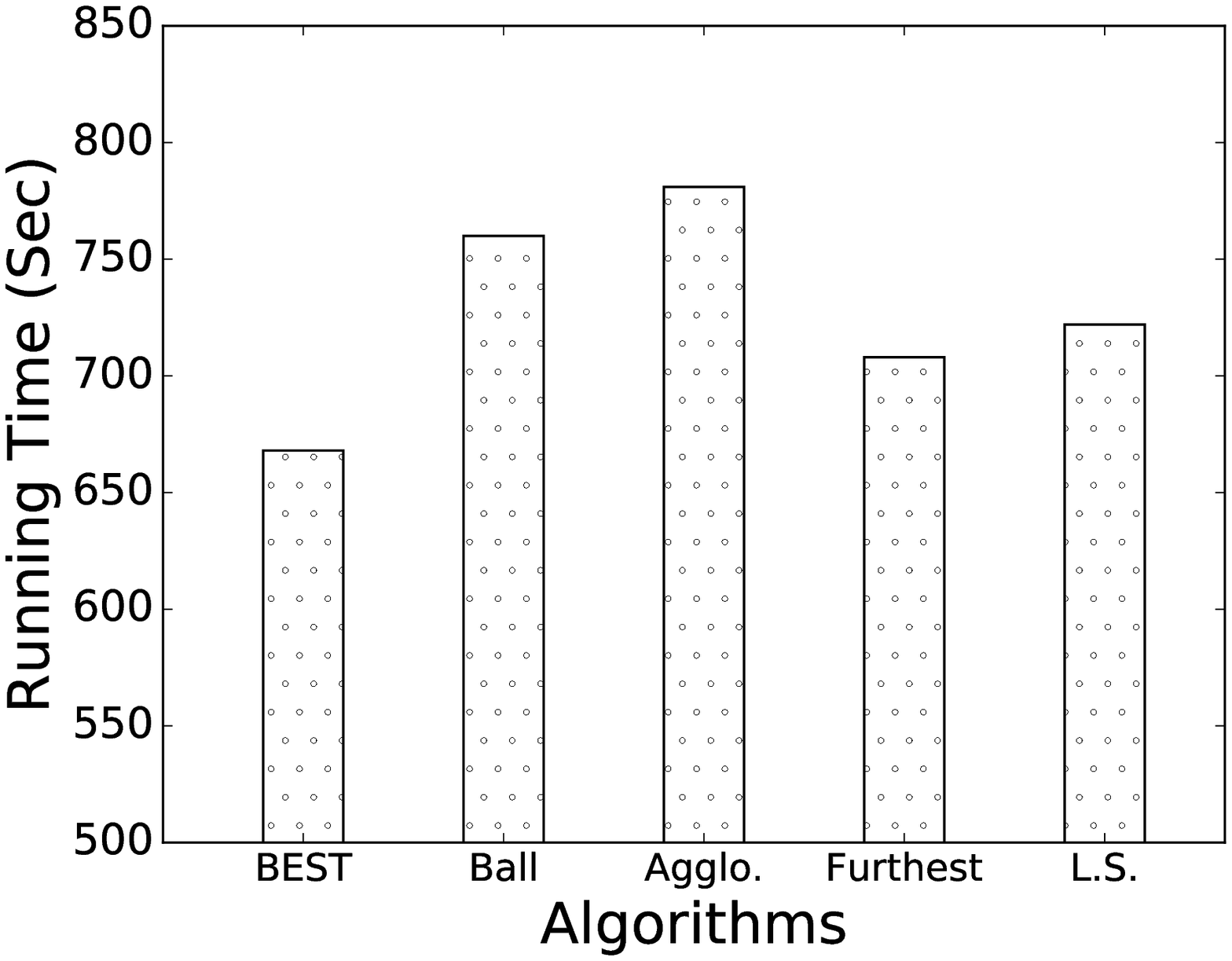}
		\label{fig:agg_t}}
	\vspace{-1mm}
	\caption{Empirical comparison among five summary aggregation algorithms over {\em DBLP\_10}.}
	\label{fig:agg}
	\vspace{-2mm}
\end{figure}

\spara{The Agglomerative algorithm.} It follows the standard bottom-up procedure. The distance $\mathcal{D}(u,v)$ between nodes $u,v$
is the same as in the {\sf Balls} algorithm. The {\sf agglomerative} algorithm first creates a singleton group for each node. Then,
in each step, it picks the pair of groups $A$ and $B$ with the smallest average distance, which is defined as the average distance
of all pair of nodes $(a,b)$, such that $a\in A$ and $b\in B$. If it is less than $1/2$, the two selected groups are merged into a single one.
The algorithm terminates when no merging is possible. The time complexity of this algorithm is $\bigO(n^2\log n)$.

\spara{The Furthest algorithm.} The {\sf Furthest} algorithm is a top-down method. At the beginning, all nodes are placed in one group.
Then, the pair of nodes which are furthest apart are found, and kept in two different groups. All remaining nodes are assigned to
the center that incurs the least cost. In the following steps, each time a new center is found to be furthest from the current centers,
and the node assignment changes accordingly. The procedure continue until the new solution induces a worse cost. Suppose that at the
end we have $k$ centers, then total running time will be $\bigO(k^2n)$.

\spara{The LocalSearch algorithm.} The {\sf LocalSearch} algorithm starts with some partition of nodes. It can be
a randomly generated one, or the result of any aforementioned algorithm. The algorithm goes through each node,
and decides to keep it still, move it to other group, or make it a singleton, by comparing various costs.
An efficient way to compute the cost of assigning a node $v$ to the cluster $C_i$ is as follows:
\begin{align}
cost(v,C_i)=M(v,C_i)+\sum_{j\neq i}(|C_j|-M(v,C_j))
\vspace{-1mm}
\end{align}
Here, $M(v,C_i)=\sum_{u\in C_i}\mathcal{D}(u,v)$. The cost of assigning a node $v$ as a
singleton is $\sum_{j}(|C_j|-M(v,C_j))$. The running time of {\sf LocalSearch} is $\bigO(Tn^2)$,
where $T$ is the number of iterations before no better move can be found.

\spara{Empirical comparison.} Figure~\ref{fig:agg} presents the experimental
comparison of the five aforementioned summary aggregation algorithms. {\em Relative size} \cite{NRS08, SGKR19} is defined as $\frac{|E_S|+|\mathcal{C}|}{|E|}$. The numerator is our objective function (Problem~\ref{prob:multi}), i.e., cost of the summary, and the denominator is constant for a given graph, i.e., graph size. Smaller relative size indicates better compactness.
All individual summaries are generated
by the {\sf Greedy} algorithm. The {\em compactness} is evaluated with the relative size of the summary to that of
the original graph. For more details on the dataset and experimental setup, we refer to \S\ref{sec:exp}.
We observe that the compactness of {\sf Agglomerative}, {\sf Furthest}, and {\sf LocalSearch} are comparable,
and they all outperform {\sf BEST} and {\sf Ball} algorithms. Considering also the running time, the {\sf Furthest}
algorithm is slightly better than others, thus it is selected as the default summary aggregation method
in our following experimental section.

\begin{figure}[tb!]
	\centering
	{\includegraphics[scale=0.85,angle=90]{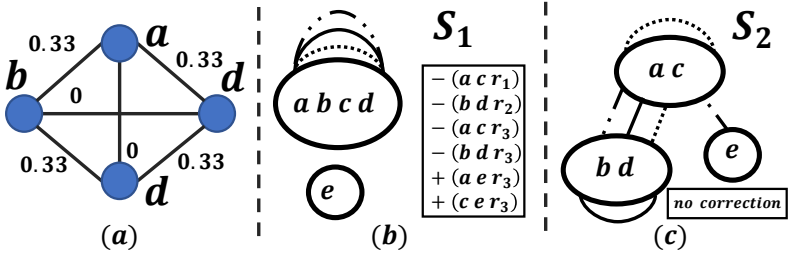}}
	\caption{(a) The graph with distances between nodes for summary aggregation computation. We omit node $E$ here since its distance to any other node is 1. (b) The aggregated summary of the individual summaries. All the summary aggregation algorithms produce the same result in this example. (c) The potentially best summary $S_2$.}
	\label{fig:summary_multi}
	\vspace{-1mm}
\end{figure}

\subsection{Limitation of baselines methods}
\label{sec:pp}
Consider again the running example multi-relation graph in Figure \ref{fig:example_m}(a) and the summaries built by {\sf Greedy} for each relation in isolation (Figure \ref{fig:s1_graphs}).

Figure~\ref{fig:summary_multi}(a) reports the distance values for the summary aggregation step.
In this setting, all the summary aggregation algorithms would return summary $S_1$, which has 3 superedges and 6 correction edges. However, there exists another summary $S_2$, which can represent the input graph with 6 superedges and without any correction edge. This example demonstrates that the two-step baselines may return a lower-quality solution.

Minimizing disagreements between the final summary and the individual summaries (obtained via summarizing on individual relations)
does not directly optimize our ultimate objective in {\sf Lossless-Sum-Multi} and {\sf $k$-Lossless-Sum-Multi} problems. Recall that the summary aggregation
operation tends to minimize disagreements between the final summary and the input set of summaries. Therefore, if the input has a majority
population of low-quality summaries, or inherently meaningless summaries for certain relations, the aggregation will be forced to agree to those useless summaries.

Notice that in Figure~\ref{fig:s1_graphs}, each of the individually optimal summaries $S_{r_1}$ and $S_{r_2}$ has two supernodes, as opposed to three supernodes in the individually optimal summary $S_{r_3}$. This is because {\sf Greedy} tends to produce larger-size supernodes. Next, when we aggregate $S_{r_1}$, $S_{r_2}$, and $S_{r_3}$ as in Figure~\ref{fig:summary_multi}(a)-(b), the resultant summary $S_1$ also has two supernodes, exactly same as the two supernodes in $S_{r_1}$ and $S_{r_2}$. Unfortunately, $S_2$ that has three supernodes (same as in $S_{r_3}$) is the optimal summary (Figure~\ref{fig:summary_multi}(c)). This demonstrates that: {\bf (1)} if the input has a majority population of low-quality summaries, the aggregation will be forced to agree to those useless summaries. {\bf (2)} Individual summaries having larger-size supernodes, as produced by Greedy, may cause trouble in the aggregation procedure.

%The aggregated summary $S_1$ compromises to $S_{r_1}$ and $S_{r_2}$ since they give 2 votes. As shown in \S~\ref{sec:single_example}, the {\sf $k$-Median} summaries with $k=3$ for $r_1$ and $r_2$ will follow the same paritioning of $S_{r_3}$ without increasing the summary cost, and results in the optimal final summary $S_2$. This demonstrates that the summaries obtained by {\sf Greedy} (usually having larger size supernodes) may cause trouble in the aggregation procedure (\S~\ref{sec:single_example}). It may also be hard to decide a good {\sf $k$-Median} individual summary for the aggregation if multiple $k$ lead to summaries with same cost.

%Similarly, compared with keeping all individual best summaries as in Figure~\ref{fig:s_graphs}, the uniform optimal summary $S_2$ provides us more insight: %althrough groups $\{ac\}$ and $\{bd\}$ are strongly related to the other in all relations, they interact with themselves in different ways. This may be %easily passed over with individual summaries, since it is normal to have only one edge absent in a strongly connected community (relation $r_1$). Moreover, %more storage cost is required for maintaining all individual summaries (\S~\ref{sec:base})
%

\section{Multi-Relation Graph Summary: Holistic Methods}
\label{sec:algo_multi}
We next present holistic algorithms that, unlike the two-step baseline approaches,
 summarize the graph in a lossless manner and considering all relations at once. In particular, we shall discuss holistic versions of {\sf $k$-Median} (\S\ref{sec:h_km}), {\sf Greedy} (\S\ref{sec:h_gr}),
and {\sf Randomized} (\S\ref{sec:h_rd}) algorithms, referred to as {\sf $k$-Median$^+$}, {\sf Greedy$^+$},
and {\sf Randomized$^+$}, respectively.

\subsection{$k$-Median$^+$}
\label{sec:h_km}
For a multi-relation graph, an adjacency matrix exists for each relation. We explore several operations for matrix aggregation,
e.g., {\sf Sum}, {\sf Or}, and {\sf Concatenate}, and formally prove that the {\sf Concatenate} operation maintains the properties for {\sf $k$-Median} based technique to return an approximated solution.

Our intuitive idea is whether it is possible to {\em aggregate the adjacency matrices due to
different relations into one aggregated matrix, and then cluster the rows of this
aggregated matrix}. Potential aggregation operations may include {\sf Sum},
{\sf Or}, and {\sf Concatenate}. Let $\{A_{G_1},A_{G_2},\ldots,A_{G_q}\}$ be a set of $q$  $ (n\times n)$ matrices, the {\sf Sum} operation produces an aggregated matrix by $A_G(i,j)=\sum_{1\leq x \leq r}A_{G_x}(i,j)$, the {\sf Or} operation produces an aggregated matrix by $A_G(i,j)=\vee_{1\leq x \leq r}A_{G_x}(i,j)$, and the {\sf Concatenate} operation produces an aggregated matrix $A_{G}=(A_{G_1}|A_{G_2}|\ldots|A_{G_q})$ by concatenating the rows. Among them, we show below that concatenation permits {\sf $k$-Median$^+$}
in achieving 16-approximation guarantee to the optimal summary size.
\begin{theor}
Let $A_{G}$ be the concatenated matrix of the adjacency matrices for individual relations, i.e.,
$A_{G}=(A_{G_1}|A_{G_2}|\ldots|$ $A_{G_q})$, where $A_{G_i}$ is the
adjacency matrix for relation $i$, $i\in(1,q)$. Let $G_{S^\#}$ be the $k$-summary induced by
the $k$-Median partitioning of the rows of $A_{G}$, and let $G_{S^*}$ be the optimal
$k$-summary for $G$ with respect to the number of correction edges.
The correction list size, $|\mathcal{C}_{{S^\#}}|$ of $G_{S^\#}$
is a 16-approximation to the best size of correction list, $|\mathcal{C}_{{S^*}}|$. Formally,
\begin{align}
	\vspace{-3mm}
|\mathcal{C}_{{S^\#}}|\leq 16\cdot |\mathcal{C}_{{S^*}}|
\end{align}
\vspace{-5mm}
\label{th:multi_sum_approx}
\end{theor}

To prove Theorem~\ref{th:multi_sum_approx}, we need to justify that Lemma~\ref{th:correction} and Theorem~\ref{th:riondato}
(mentioned in \S~\ref{sec:bound_single} in the context of {\sf $k$-Median} algorithm over single-relation graphs) still hold
with our concatenated matrix. %We only provide a high-level proof sketch below due to interest of space.

First, we verify the correctness of Lemma~\ref{th:correction} in this case. To extend our objective of counting the number of correction
edges from single-relation to multi-relation, we only need to include an additional {\sf sum} operation to Equation~\ref{eq:et}.
\begin{align}
\vspace{-1mm}
|\mathcal{C}|=\frac{1}{2}\sum_{r\in R}\;\sum_{U,W\in V_S\times V_S}|U||W|\left\{\begin{matrix}
\alpha_{UW,r}\ \ \ ,\ if \ \alpha_{UW,r}\leq 0.5\\
(1-\alpha_{UW,r}), \ otherwise
\end{matrix}\right.
\label{eq:et+}
\end{align}
Since we have a holistic summary corresponding to same set of supernodes across relations,
the edge density between supernodes $U$ and $W$ for each relation $r$, denoted by $\alpha_{UW,r}$, can be calculated in the same way.
Hence, the $l_1$ reconstruction error (Equation~\ref{eq:err}) can be modified as:
\begin{align}
RE_1(G,G_S)=2\sum_{r\in R}\sum_{(U,W)\in V_S\times V_S}|U||W|\alpha_{UW,r}(1-\alpha_{UW,r})
\end{align}
Based on the above two equations, one can prove an equivalent lemma of our earlier Lemma~\ref{th:correction}, for the multi-relation case.
That is, $|\mathcal{C}_{S^+}|\leq 2\cdot|\mathcal{C}_{S^*}|$. Here, $G_{S^+}$
is the optimal $k$-summary for $G$ with respect to the $l_1$-reconstruction error, and $|\mathcal{C}_{S^+}|$ is
the correction list size for $G_{S^+}$. Clearly, {\sf Sum} and {\sf Or} operations violate Equation~\ref{eq:et+}.

Next, we rewrite the $l_1$ reconstruction error in its original form (equivalent of Equation~\ref{eq:lp}).
\begin{align}
RE_1(G,G_S)&=\sum_{r\in R} \sum_{u=1}^{|V|}\sum_{w=1}^{|V|}\left|A_{G_r}(u,w)-A_{G_{S,r}}^{\uparrow}(u,w)\right| \nonumber
\end{align}
\begin{align}
&=\sum_{r\in R}\left|\left|A_{G_r}-A_{G_{S,r}}^{\uparrow}\right|\right|
\label{eq:multi}
\end{align}
where $A_{G_{S,r}}^{\uparrow}(u,w)$ is the edge density between the supernodes $U$ and $W$ for relation $r$, such that, $u\in U$, $w\in W$.

To prove Theorem~\ref{th:riondato}, Riondato et al. \cite{RGB14} defined an orthogonal smoothing projection $P$ for a partitioning $\mathcal{P}$ of $n$ nodes (i.e., rows)
in the adjacency matrix $A_{G}$. Since in our holistic summary, all relations would share the same partitioning $\mathcal{P}$, the same projection $P$ can be applied
to each adjacency matrix $A_{G_r}$. The $l_1$-reconstruction matrix for each relation can be computed by
$A_{G_{S,r}}^{\uparrow}(u,w)=PA_{G_r}P$. By definition, $A_{G_r}P$ is the {\sf $k$-Means} matrix. And we define the {\sf $k$-Median} matrix of relation $r$ as below:
\begin{align}
	\footnotesize
	B_{G_r}(u,w)=median(\{A(x,y)|\{x,y\}\in \Pi_{UW})
\end{align}
$\Pi_{UW}$ is the set of all possible pairs $\{x,y\}$, such that $x\in U$ and $y\in W$. Lemma 1 and Lemma 4 in \cite{RGB14} provide the inequalities bridging the $l_1$-reconstruction error $||A-PAP||$, {\sf $k$-means} cost $||A-AP||$ and the {\sf $k$-median} cost $||A-B||$.
Thus, we prove Theorem~\ref{th:riondato} for multi-relation case as follows.
\begin{align}
RE_1(G,G_{S^\#})&=\sum_{r\in R}||A_{G_r}-A_{G_{{S^\#,r}}}^{\uparrow}||  \quad \triangleright \text{\scriptsize by Equation~\ref{eq:multi}} \nonumber \\
&=\sum_{r\in R}||A_{G_r}-P_{G_{S^\#}}A_{G_r}P_{G_{S^\#}}||  \nonumber \\
&\leq 2\cdot \sum_{r\in R}||A_{G_r}-A_{G_r}P_{G_{S^\#}}||\quad \triangleright \text{\scriptsize Lemma 4, \cite{RGB14}} \nonumber \\
&\leq 4\cdot \sum_{r\in R}||A_{G_r}-B_{G_{S^\#}}||\quad \triangleright \text{\scriptsize Lemma 1, \cite{RGB14}} \nonumber  \\
&\leq 4\cdot \sum_{r\in R}||A_{G_r}-B_{G_{S^+}}||\quad \triangleright \text{\scriptsize $G_S^\#$ is best for $k$-Medain} \nonumber \\
&\leq4\cdot \sum_{r\in R}||A_{G_r}-A_{G_r}P_{G_{S^+}}||\quad \triangleright \text{\scriptsize Lemma 1, \cite{RGB14}} \nonumber \\
&\leq8\cdot \sum_{r\in R}||A_{G_r}-P_{G_{S^+}}A_{G_r}P_{G_{S^+}}||\quad \triangleright \text{\scriptsize Lemma 4, \cite{RGB14}} \nonumber \\
&=8\cdot RE_1(G,G_{S^+})
\end{align}
Since both Theorem~\ref{th:riondato} and Lemma~\ref{th:correction} hold for the multi-relation case, one can prove the correctness of 16-approximation result due to {\sf $k$-Median}
on the concatenated adjacency matrix. Therefore, Theorem~\ref{th:multi_sum_approx} follows.

\subsection{Greedy$^+$}
\label{sec:h_gr}
The {\sf Greedy} algorithm can be generalized to a holistic algorithm, {\sf Greedy$^+$} for the {\sf Lossless-Sum-Multi} problem without changing the workflow.
Between any pair of supernodes $U$ and $W$, let $\Pi_{UW}$ be the set of all possible pairs $\{a,b\}$, such that $a\in U$ and $b\in W$. $E_{UW}\subseteq \Pi_{UW}$ is defined as the set of edges actually present in the input graph $G$, i.e., $E_{UW}=\Pi_{UW}\cap E$. Obviously, $\Pi_{UW}$ is the same no matter for which relation.
Next, we define $A_{UW,r}\subseteq \Pi_{UW}$ to be the set of edges actually present in the original graph $G$ for relation $r$, i.e.,
$A_{UW,r}=\Pi_{UW}\cap E_r$. Similarly,  the cost of a supernode pair $(U,W)$ for relation $r$ can be calculated as:
\begin{align}
	C(U,W,r)=\min\{|\Pi_{UW}|-|A_{UW,r}|+1, |A_{UW,r}|\}
	\label{eq:m_cost}
\end{align}
The neighbor set $N_r(U)$ of $U$ in relation $r$ is defined to be the set of supernodes $W$ that have such edge $\{a,b\} \in A_{UW,r}$. Moreover, the cost of maintaining a supernode $U$ would add up across relations as follows:
\begin{align}
	C(U)=\sum_{r\in R}\sum_{X\in N_r(U)}C(U,X,r)
\end{align}

The cost reduction due to merging supernodes $U$ and $W$ into a new supernode $H$ can be calculated as:
\begin{align}
\triangle C(U,W)=\frac{C(U)+C(W)-C(H)}{C(U)+C(W)}
\end{align}
Taking fraction instead of the absolute cost reduction in the above equation is to get rid of the bias towards nodes with higher degree.

\begin{exam}
We use the example graph in Figure~\ref{fig:example_m}(a) for the demonstration of {\sf Greedy$^+$}. At the beginning,
the cost of each node equals to the number of edges incident to this node, for all relations. Thus, node $a$ and
$c$ will be selected for merging in the first round, resulting in $\frac{7+7-(3+3+1)}{7+7}=0.5$ cost reduction.
Similarly, the nodes $b$ and $d$ will be merged in the second round. The cost again reduces by $\frac{4+4-(3+1)}{4+4}=0.5$.
Then, it can be easily verified that no further merging can result in positive cost reduction. Thus, it returns the optimal
summary, $S_2$ in Figure~\ref{fig:summary_multi}(c), for this example.
\end{exam}

\subsection{Randomized$^+$}
\label{sec:h_rd}
{\sf Randomized$^+$} follows the same modification in cost computation as the holistic {\sf Greedy$^+$}.
The computation step does not change when comparing with the original {\sf Randomized} algorithm. For {\sf SWeG} algorithm,
it is non-trivial to extend the node set dividing step to multi-relation case. We, therefore, leave it as a future direction. 

\section{Finding Optimal Number of Supernodes}
\label{sec:k}
Our {\sf $k$-Median}-based approaches (i.e., {\sf $k$-Median} two-step baseline and holistic {\sf $k$-Median$^+$})
require $k$, a predefined number of supernodes, as an input. In practice, the user may not explicitly provide the target number of supernodes.
In this section, we suggest to use the number of supernodes returned by {\sf Greedy$^+$} algorithm as the optimal value of $k$
for {\sf $k$-Median}-based approaches, and verify its good performance via the {\sf Elbow} method \cite{Tho53} as below.

%a strategy for determining the optimal value of $k$ allows the solution of {\sf $k$-Lossless-Sum-Multi} to adapt to the more general problem, {\sf Lossless-Sum-Multi}.

\spara{The Elbow method.} The {\sf elbow} method empirically verifies the cost of the {\sf $k$-median}
clustering when varying $k$. Within each cluster, it computes the median distance between each node
to its center, and takes the sum of all median distances for all nodes, which is known as the {\em within-cluster sum-of-squares} error (WSS).
The WSS is plotted against the cluster number $k$, we select the $k$ for which WSS first starts to diminish. In the plot of WSS-versus-$k$,
this is visible as an ``elbow''. Using the elbow as a cutoff point is a common heuristic\footnote{Theoretically, {\em within-cluster sum-of-squares} error (WSS) monotonically decreases with larger cluster number $k$, thus the optimal $k$ is trivially equal to the total number of nodes in the input graph. In practice, we would like to find a clustering where the WSS no longer decreases sharply.} in optimization to choose a point where diminishing
returns are no longer worth the additional cost. In clustering, this implies that one should choose a number of clusters so that adding
another cluster will not give much better modeling of the data.

% \vspace{-0.6mm}
% \spara{The Silhouette Method.} The average {\sf silhouette} \cite{KR90} of the data is another useful criterion
% for assessing the natural number of supernodes. The silhouette of a node $v$ is defined as:
% %
% \begin{align}
% sil(v)=\frac{b(v)-a(v)}{max\{a(v),b(v)\}}
% \end{align}
% %
% where $a(v)$ is the median distance between node $v$ and all other nodes in the same supernode, and $b(v)$ is the smallest median distance of node $i$ to all nodes in any other supernode. Notice that $-1\leq sil(v)\leq 1$. We calculate the average silhouette for all nodes in a range of $k$,
% and select the $k$ with the minimum silhouette.

%\spara{The X-Median Method.} Unlike the {\sf Elbow} method, the {\sf X-Median} approach does not verify a range of $k$ one by one, and chooses the experimental best one. Rather, it works in a top-down, hierarchical clustering manner. In each step, it partitions each current cluster into two new clusters, and adopts the partition if the {\em Bayesian Information Criterion (BIC)} is optimized. %$BIC(C|X)=\log(X|C)-\frac{k}{2}\log n$.
%When no further partition can improve the $BIC$ score, the number of current $k$ is returned as the optimal.
%Similarly, all other hierarchical clustering methods {\em with a criteria of automatical termination} can be applied.

\begin{figure}[tb]
	\centering
	\vspace{-5mm}
	\subfigure[\small {\em Amazon}]
	{\includegraphics[scale=0.205]{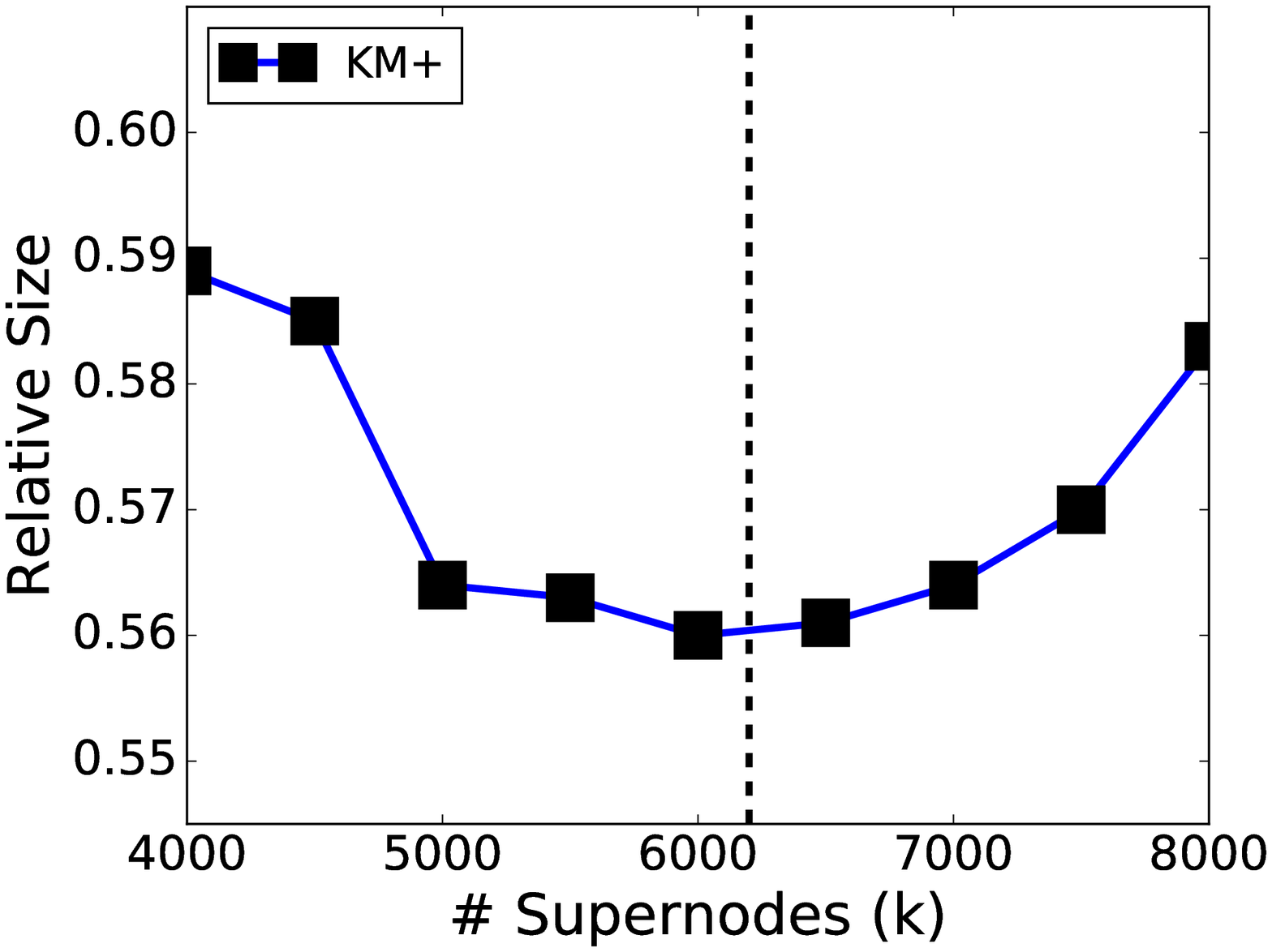}
		\label{fig:k_am}}
	\subfigure[\small {\em DBLP230}]
	{\includegraphics[scale=0.205]{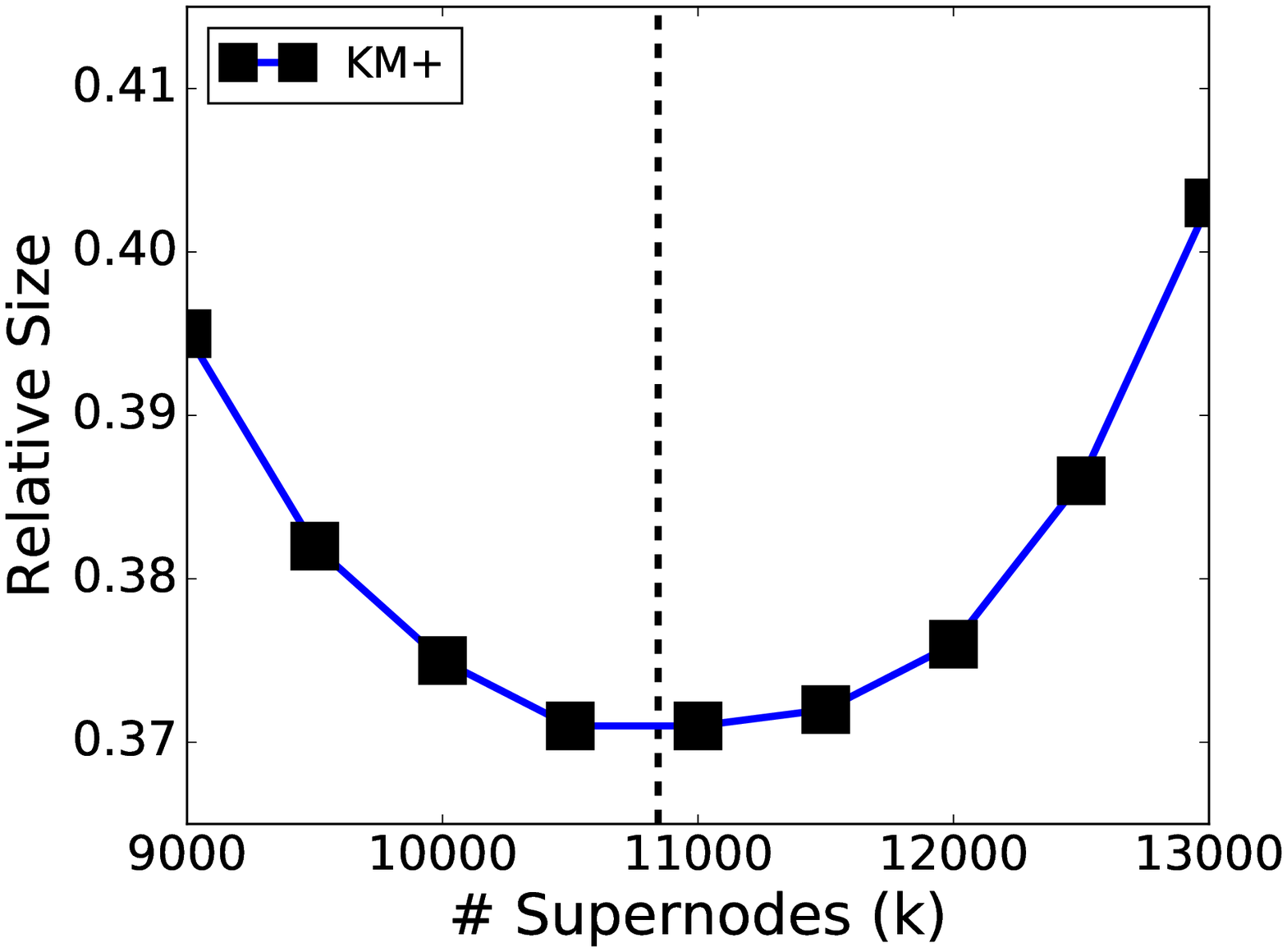}
		\label{fig:k_dblp}}
	\subfigure[\small {\em Twitter}]
	{\includegraphics[scale=0.205]{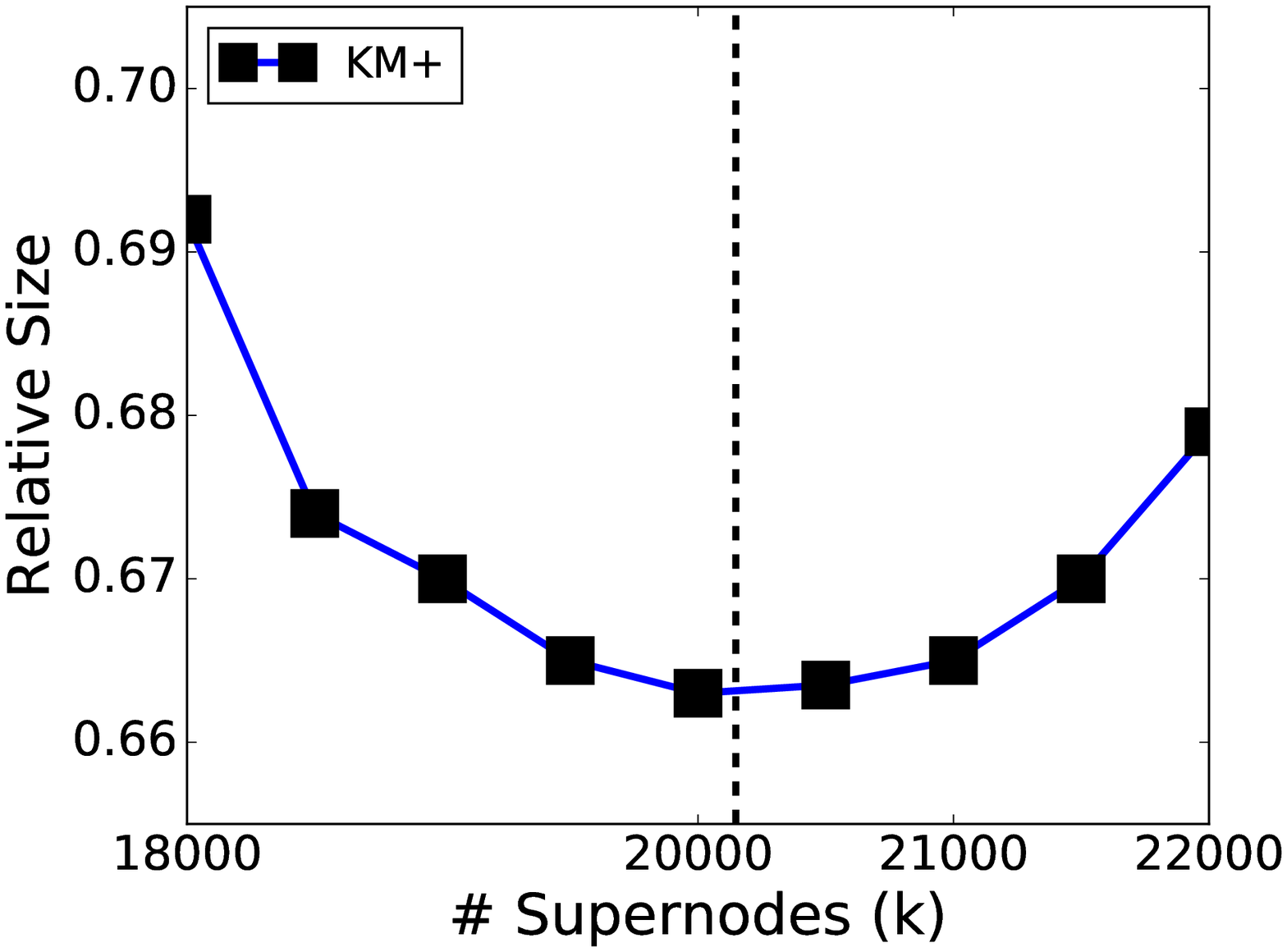}
		\label{fig:k_twitter}}
	\vspace{-3mm}
	\caption{\small Compactness analysis with \#supernodes $k$.}
	\label{fig:k_km}
	\vspace{-5mm}
\end{figure}

%\spara{Our implementation.} We adopt a combination of {\sf Greedy} and {\sf Elbow} methods.
First, we employ {\sf Greedy} %or {\sf Randomized}
to suggest a preliminary $k'$.
Then, we vary the $k$ in a range with $k'$ as the center, e.g., $[k'-2000,k'+2000]$,
and apply the {\sf Elbow} method to verify the performance of $k'$.
In practice, we can replace the clustering cost with our exact summary cost and select the optimal $k$ via Equation~\ref{eq:elbow}, where $S(k)$ denotes the {\sf $k$-Median} summary having $k$ supernodes. The exact summary cost is measured via {\em relative size} \cite{NRS08, SGKR19}, which is defined as $\frac{|E_S|+|\mathcal{C}|}{|E|}$. The numerator is our objective function (Problem~\ref{prob:multi}), i.e., cost of the summary, and the denominator is constant for a given graph, i.e., graph size. Smaller relative size means better compactness.
%The {\em Relative Size} on $y$-axis denotes the compactness, the detailed definition can be found in \S\ref{sec:exp_set}.
We show the relative size on $y$-axis and $k$ on $x$-axis, 
the curve becomes ``valley'' shape instead of ``elbow'' shape (Figure~\ref{fig:k_km}).
The vertically dashed line denotes the number of supernodes found with {\sf Greedy$^+$},
which are always located in the "valley bottom" in all our datasets. This shows that our {\sf Greedy$^+$} algorithm can suggest a good $k$ for our {\sf $k$-Median$^+$} approach.
\begin{align}
	argmin_{k\in [k'-2000,k'+2000]} \frac{|E_{S(k)}|+|C_{S(k)}|}{|E|}
	\label{eq:elbow}
\end{align}

\section{Multi-Relation Graph Summary: Hybrid Algorithm}
\label{sec:hybrid}

In this section, we discuss the shortcomings of the {\sf Greedy$^+$} and the {\sf $k$-Median$^+$} holistic algorithms, and provide a hybrid algorithm, {\sf Hybrid}, based on properly combining them, to produce the most effective summary as shown empirically in the next section.

\spara{The shortcomings of {\sf Greedy$^+$}.} The {\sf Greedy$^+$} algorithm is a bottom-up iterative approach. The subsequent rounds of {\sf Greedy$^+$} highly rely on the results of previous rounds. If the previous few rounds get trapped in some bad results, the follow-up rounds have no way to fix it.

\spara{The shortcomings of {\sf $k$-Median$^+$}.} First, the {\sf $k$-Median$^+$} algorithm requires an input number of supernodes $k$. Thus, it cannot directly solve the {\sf Lossless-Sum-Multi} problem. Second, our experimental results in \S~\ref{sec:base} present that the compactness of the {\sf $k$-Median$^+$} summaries are worse than those of {\sf Greedy$^+$} summaries in practice.
%If we zoom in the ``vally'' bottom of the compactness curve of {\sf $k$-Median$^+$} summaries, as shown in Figure~\ref{fig:k_am_plus}, it can be observed that the curve is not smooth. The compactness result rises and falls in a small range. Together with Figure~\ref{fig:k_am_size}, we can learn that {\sf $k$-Median$^+$} is confident to provide a higher quality summary with $k=7000$ than any larger $k$, and to provide a higher quality summary with $k=5000$ than any smaller $k$ in this case. But it is struggling to output one ultimately best summary with a $k\in [5000,7000]$.

Based on previous discussion, we can find that the {\sf Greedy$^+$} algorithm and the {\sf $k$-Median$^+$} algorithm are complementary: {\bf (1)} {\sf Greedy$^+$} can suggest a potentially good $k'$ for {\sf $k$-Median$^+$}; {\bf (2)} {\sf $k$-Median$^+$} is able to directly generate a summary for any $k$ around $k'$; {\bf (3)} For any {\sf $k$-Median$^+$} summary, {\sf Greedy$^+$} can further improve its quality, if possible. 

\spara{Hybrid algorithm.}
Based on aforementioned properties, we properly combine the {\sf Greedy$^+$} algorithm and the {\sf $k$-Median$^+$} algorithm as our ultimately proposed {\sf Hybrid} algorithm, as below.

{\bf (1)} As discussed in \S~\ref{sec:k}, we determine the optimal $k$ for {\sf $k$-Median$^+$}, with the help of the {\sf Greedy$^+$} method. %and the {\sf Elbow} method. %The summary cost of the {\sf $k$-Median$^+$} summaries shall be similar (e.g., within 1\%) here.
{\bf (2)} We generate a summary by {\sf $k$-Median$^+$} algorithm with the best $k$ found by {\sf Greedy$^+$} %in (1).
{\bf (3)} We conduct the {\sf Greedy$^+$} algorithm again starting from the summary generated in (2), to further improve its compactness, if possible.

Since {\sf Hybrid} applies {\sf $k$-Median$^+$}, followed by {\sf Greedy$^+$} in steps 2-3 above, it also requires $k$, a predefined number of supernodes, as an input. Therefore, in step 1, we apply the same preprocessing technique as in \S~\ref{sec:k} to determine the optimal $k$ for {\sf $k$-Median$^+$}.
However, notice that since we further apply {\sf Greedy$^+$} to improve the compactness in the third step, the optimal number of supernodes returned
by {\sf Hybrid} may eventually be reduced, in comparison with {\sf $k$-Median$^+$}.  
\section{Experimental Results}
\label{sec:exp}
%\vspace{-1mm}
%
We conduct experiments to demonstrate
the effectiveness (compactness), efficiency, and scalability of our algorithms (averaged over 10 runs).
The code is implemented in C++, and is executed on a single core, 40GB, 2.40GHz Xeon server.
\begin{table} [tb!]
	\caption{Properties of datasets.}
	\centering
	\begin{tabular} { l|c|c|c|c }
		\hline
		{\textsf{Dataset}}        & {\textsf{\#Nodes}}  & {\textsf{\#Edges}}  &  {\textsf{\#Relations}} &  {\textsf{Domain}} \\ \hline
		{\em Homo}              & 18\,222	            &  153\,923	          & 7 & genetic \\
		{\em Amazon}              & 410\,237	            &  8\,132\,507	          & 4 & co-purchasing \\
		{\em DBLP6}              & 892\,531	            &  6\,045\,859	          & 6 & co-authorship \\
		{\em DBLP230}              & 892\,531	  &  7\,207\,253	   & 230 & co-authorship \\
		{\em Twitter}              & 4\,898\,247   &  8\,053\,440  & 4 & social \\
		\hline
	\end{tabular}
	\label{tab:data}
\end{table}
\subsection{Experimental setup}
\label{sec:exp_set}

\subsubsection{Datasets} We use five real-world, multi-relation networks, whose main characteristics are listed in Table~\ref{tab:data}.

\smallskip

{\bf {\em Homo}}  (https://comunelab.fbk.eu/data.php) network describes different types of genetic interactions between genes in Homo Sapiens. Nodes are genes and edges denote their interactions. Seven different relations exist, which are: direct interaction,
physical association, suppressive genetic interaction defined by inequality, association, colocalization, additive genetic interaction defined by inequality, and synthetic genetic interaction defined by inequality.

\smallskip

{\bf {\em Amazon}} (https://snap.stanford.edu/data) is a co-purchasing temporal network with four snapshots between March and June 2003, each as a relation. Nodes are products and edges are their co-purchasing relationships.

\smallskip

{\bf {\em DBLP}} (http://dblp.uni-trier.de/xml) is a well known collaboration network. We downloaded it on Dec 31, 2020. Each node is an author and edges denote their co-authorships. We use two versions of {\em DBLP} dataset. {\em DBLP6} %\cite{EFF17}
has 6 relations for 6 representative sub-areas of computer science: data management, artificial intelligence, computer architecture, computer networks, theory of computing, and systems \& software. An edge between a node pair exists for a relation if they have published as co-authors in the top-tier venues under this sub-area. The top-tier (i.e., rank A) conferences and journals for each sub-area are given by the CCF ranking: https://www.ccf.org.cn/Academic\_Evaluation/By\_category.
{\em DBLP230} is generated based on \cite{KKC18} with the latest data. It has 230 relations for 230 keywords extracted from paper titles,
based on both their frequency and how well they can represent various sub-areas of computer science, e.g., database systems, neural networks, FGPA, etc.

\smallskip

{\bf {\em Twitter}} (https://ieee-dataport.org/open-access/usa-nov2020-election-20-mil-tweets-sentiment-and-party-name-labels-dataset) dataset is generated based on 24M US election related tweets from July 1 to November 11, 2020.  Nodes are users and edges are their re-tweet relationships. The relations stand for 4 political parties in 2020 US presidential election.

\subsubsection{Competing algorithms} Our basic two-step algorithms include {\bf (1) {\sf Greedy}} \cite{NRS08},
%It treats each node in the original graph as a supernode at the begining, and merges the best pair of supernodes with maximal positive cost reduction in each iteration, until no merging is possible.
{\bf (2) {\sf Randomized}} \cite{NRS08},
%Similar to {\sf Greedy}, it starts by keeping every node in the input graph as a supernode, and randomly pick a supernode $u$ to merge with another supernode $w$ with maximal positive cost reduction to $u$. If no positive cost reduction exists between $u$ to any other supernode, $u$ will be marked as ``explored''. The algorithm terminates when all current supernodes are explored.
{\bf (3) {\sf $k$-Median}} (proposed approximation algorithm for the {\sf Lossless-Sum} problem),
%It performs {\sf $k$-Median} clustering on the rows of the adjacency matrix $A_{G}$ of the input graph $G$, and create $k$ supernodes.
and {\bf (4) {\sf SWeG}} \cite{SGKR19}.
%It first divides the graph into smaller disjoint groups (\S~\ref{sec:gs_single}). Then a similar procedure as {\sf Randomized} is conducted within each group to merge supernodes. All methods are coupled with the {\sf Furthest} algorithm \cite{GMT07} for summary aggregation since it works best empirically (\S~\ref{sec:s_agg}), and are denoted as {\sf GD}, {\sf RD}, {\sf KM}, and {\sf SWeG}, respectively.
For fairness, we only allow sequential execution of the {\sf SWeG} algorithm. The detailed description of these methods can be found in \S\ref{sec:bound_single} and  \S\ref{sec:gs_single}.

The holistic version of {\sf Greedy}, {\sf Randomized}, and {\sf $k$-Median} are represented as {\sf GD+} ({\sf Greedy$^+$}), {\sf RD+} ({\sf Randomized$^+$}), and {\sf KM+} ({\sf $k$-Median$^+$}), respectively. Our final algorithm, {\sf Hybrid}, for lossless, multi-relation graph summarization is represented as {\sf HY}.
In addition, we also consider a method, denoted as {\sf ALL}, that stores individually optimal summaries (following the {\sf $k$-Median} algorithm) for all relations.

{The optimal number of supernodes (reported in Table~\ref{tab:k})
are automatically decided by {\sf GD+} and {\sf RD+}. For {\sf $k$-Median}, {\sf KM+}, and {\sf HY},
we decide the optimal number of supernodes empirically by {\sf Greedy} methods, as discussed in \S\ref{sec:k}.}

%\pagebreak

\subsubsection{Evaluation metrics used} We adopt the following evaluation metrics:

\begin{itemize}
\item {\bf Relative size.} We measure the compactness of the obtained summaries by {\em relative size} \cite{NRS08, SGKR19}, which is defined as $\frac{|E_S|+|\mathcal{C_S}|}{|E|}$. Recall that $E$ denotes the set of edges in the input graph,
$E_S$ the set of superedges between supernodes, and $C_S$ the set of correction edges.
The numerator is our objective function (Problem~\ref{prob:multi}), i.e., cost of the summary, and the denominator is constant for a given graph, i.e., graph size.
Smaller relative size means better compactness.

\item {\bf Running time.} The total running time for generating a graph summary is reported. For two-step methods, it contains both the time of producing summaries for all individual relations and the time of aggregating them.

\item {\bf Storage cost.} We report the exact storage cost for original graphs and the corresponding summaries. The mappings from the node set $V$ to the supernode set $V_S$ are included in summaries.
\end{itemize}

\subsection{Performance analyses}
\label{sec:base}
\begin{figure*}[h!]
	\centering
	\subfigure[\small {\em Homo}]
	{\includegraphics[scale=0.28]{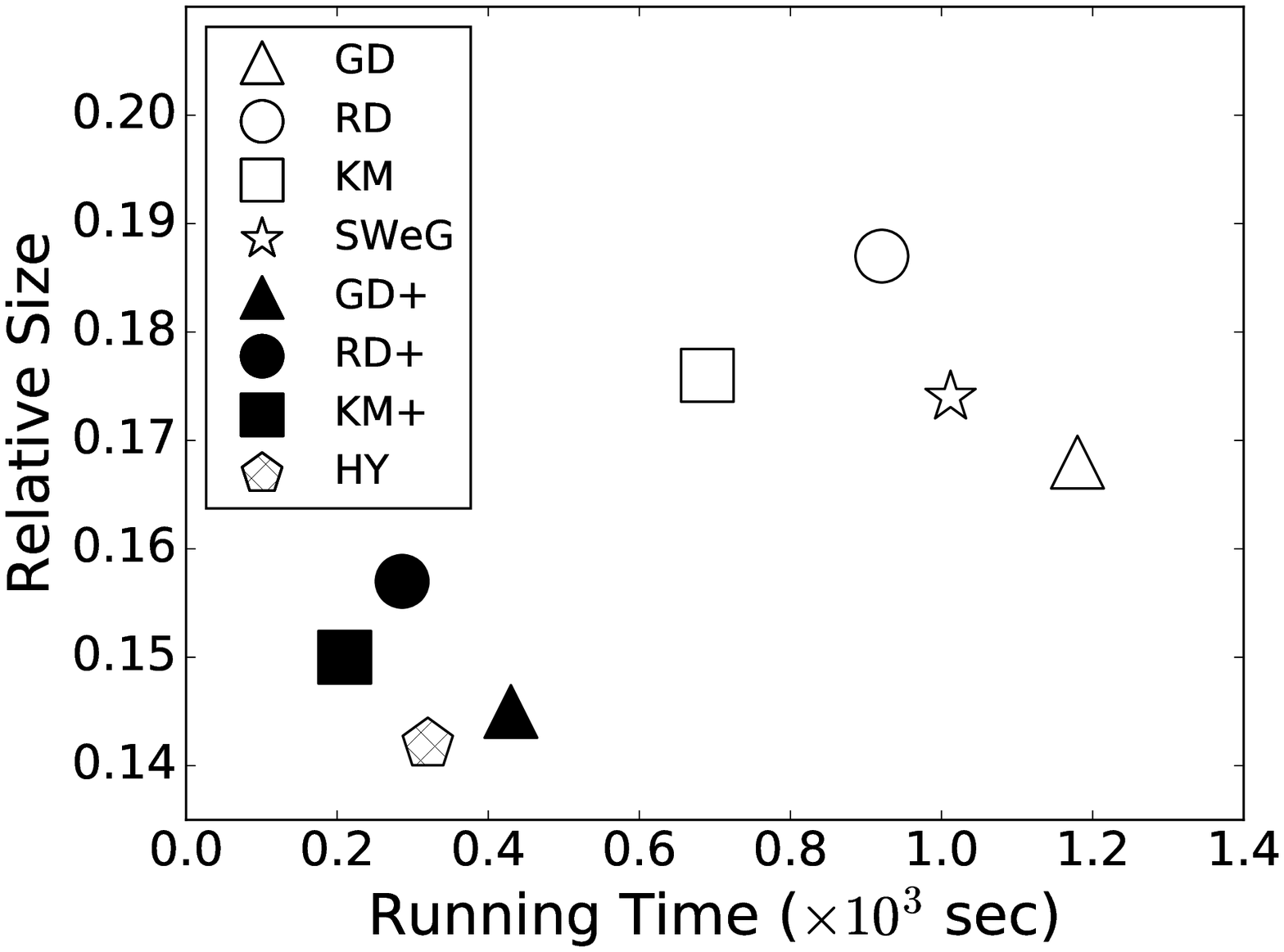}
		\label{fig:homo_t}}
	\subfigure[\small {\em Amazon}]
	{\includegraphics[scale=0.28]{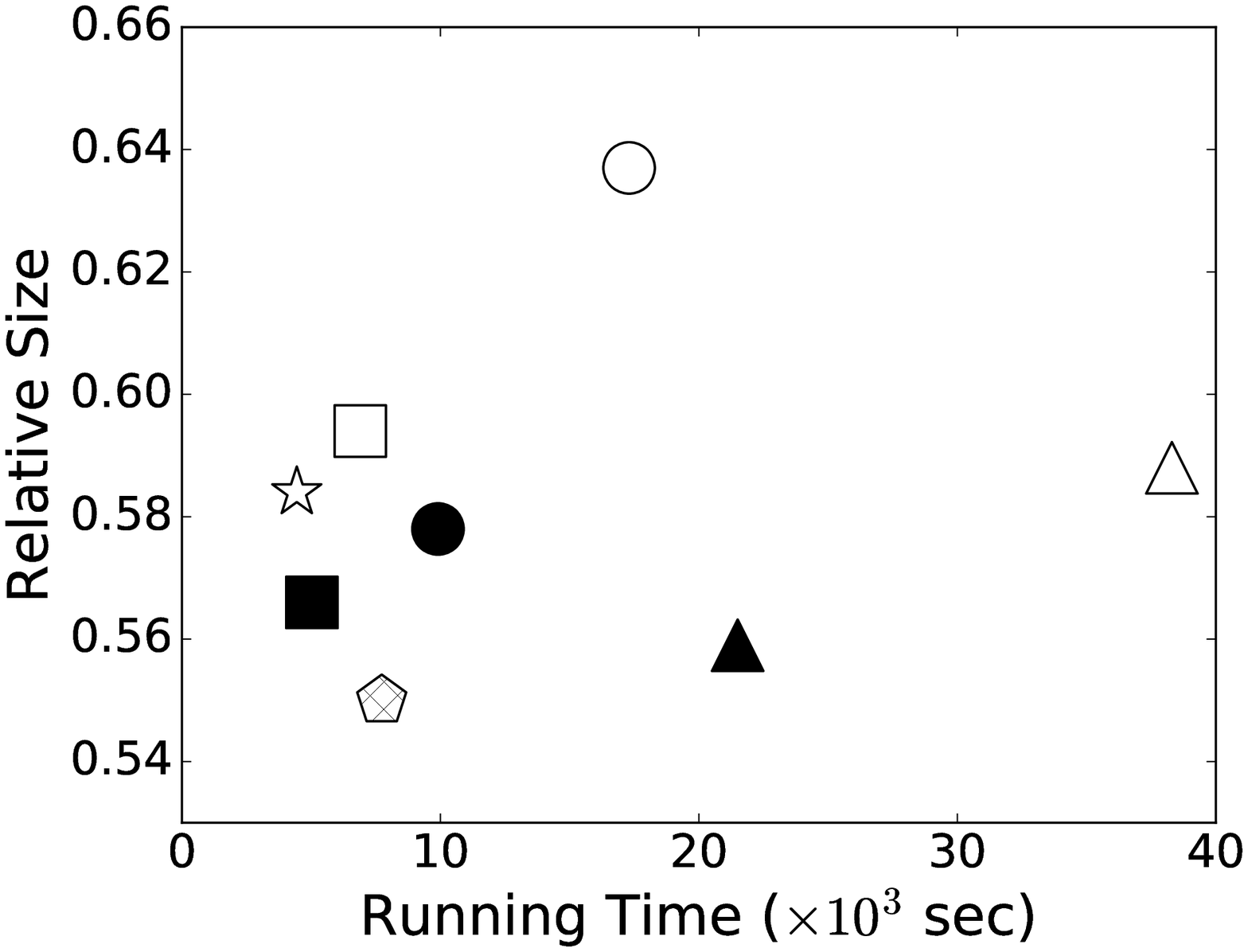}
		\label{fig:amazon_t}}
	\subfigure[\small {\em DBLP6}]
	{\includegraphics[scale=0.28]{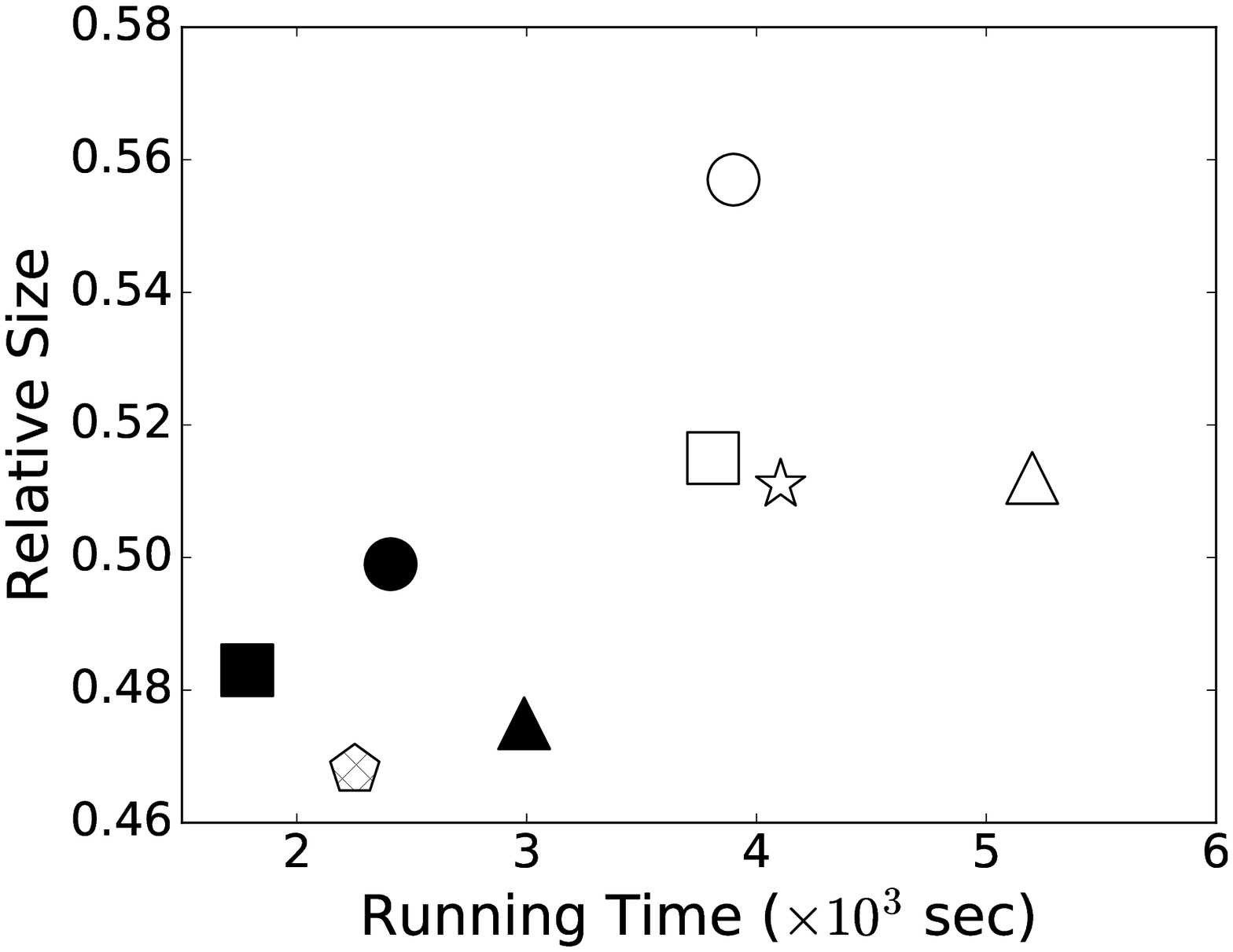}
		\label{fig:dblp10_t}}
	\subfigure[\small {\em DBLP230}]
	{\includegraphics[scale=0.28]{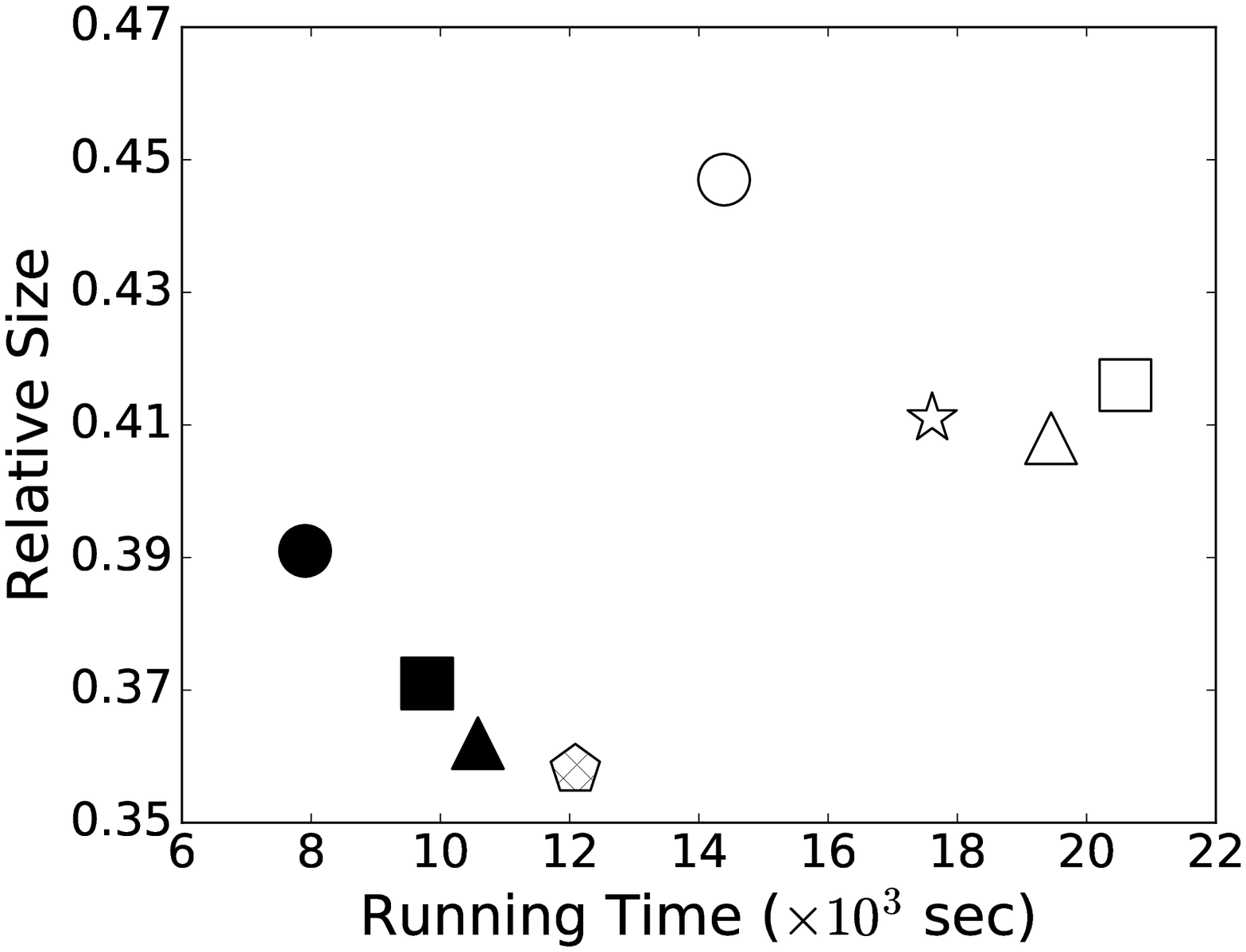}
		\label{fig:dblp230_t}}
	\subfigure[\small {\em Twitter}]
	{\includegraphics[scale=0.28]{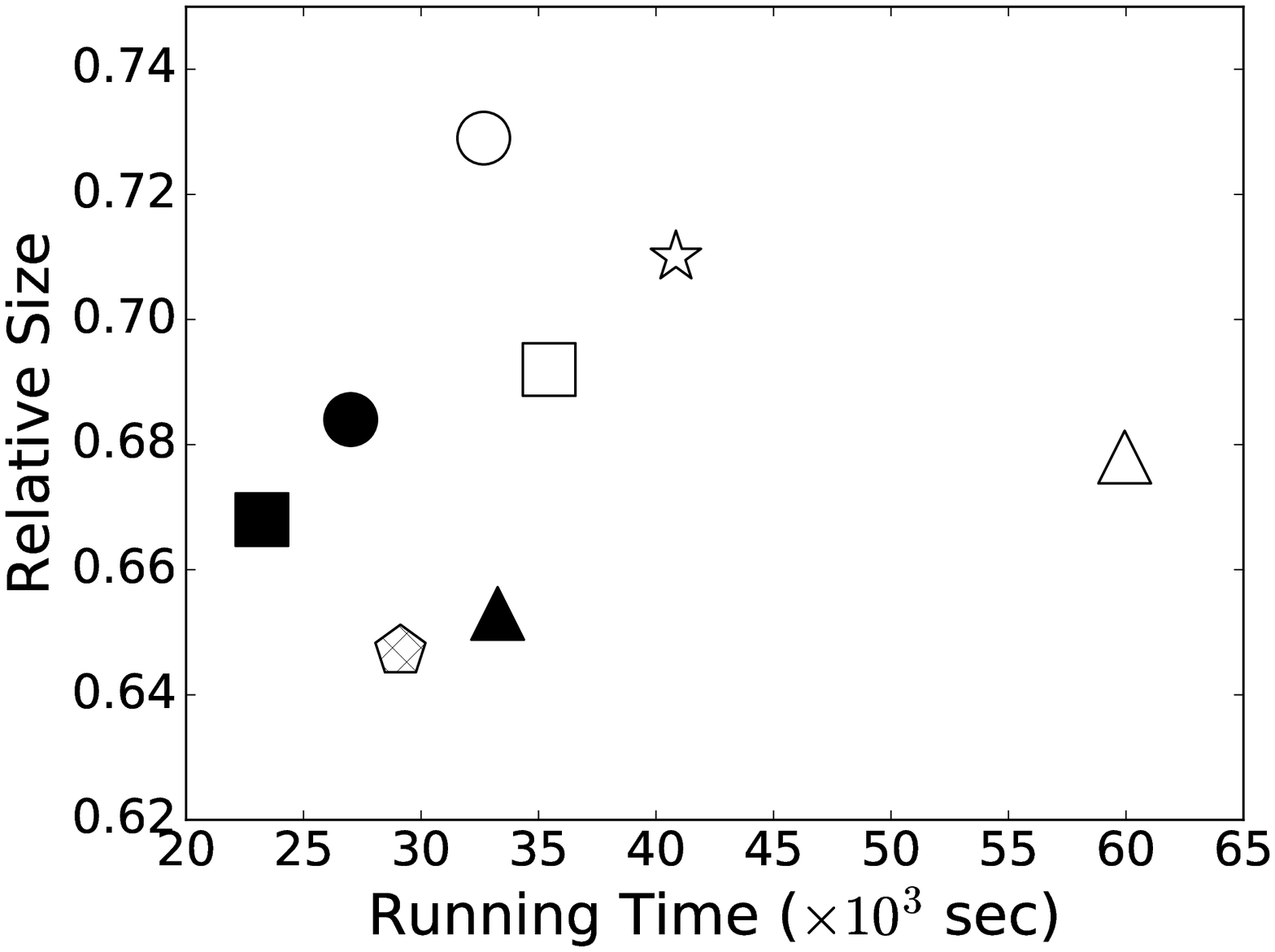}
		\label{fig:twitter_t}}
	\subfigure[\small {\em Breakup} ({\sf HY})]
	{\includegraphics[scale=0.28]{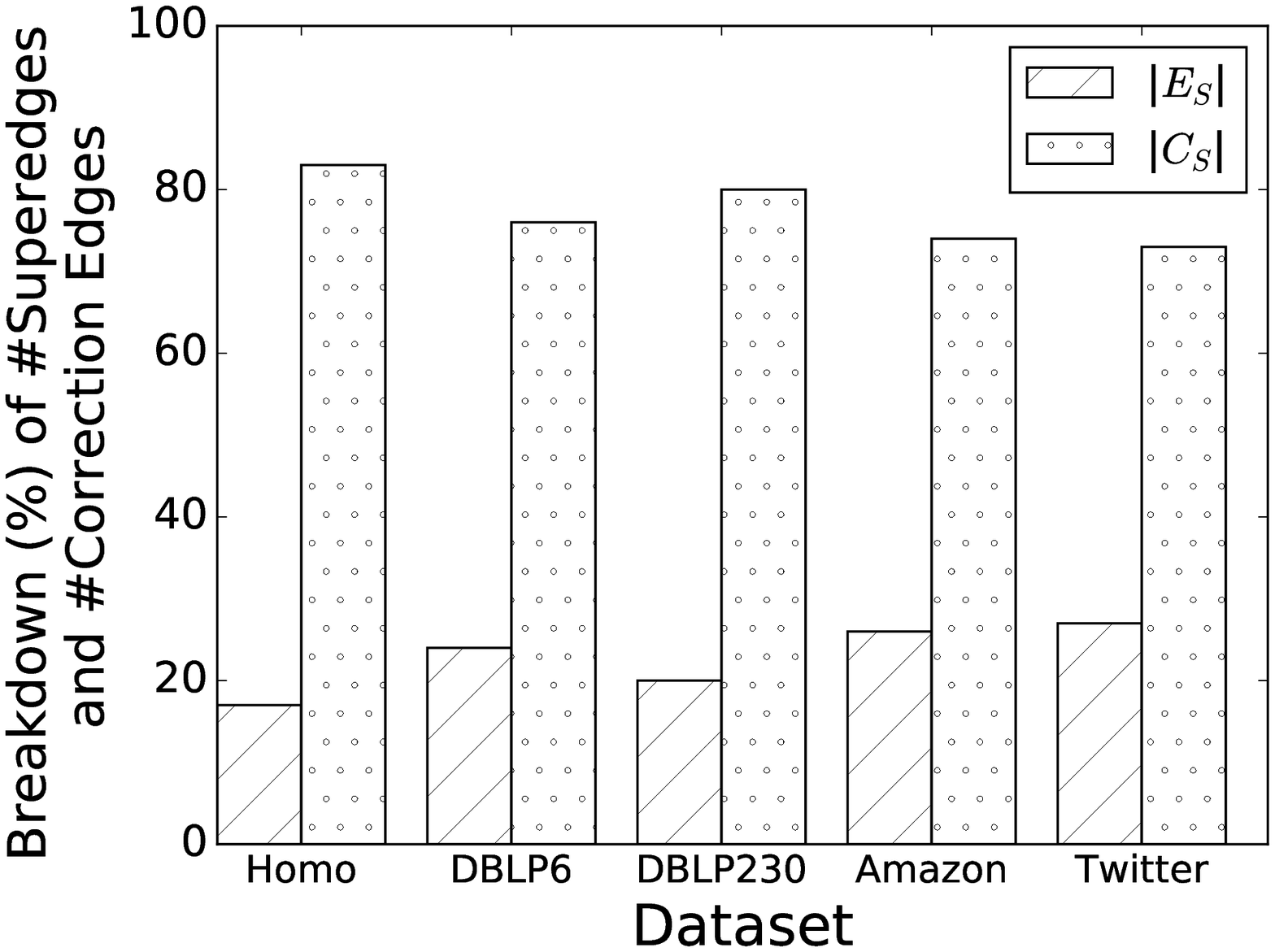}
		\label{fig:breakup_cost}}
	\vspace{-2mm}
	\caption{Trade-off between relative size and running time. {Sub-figure (f) presents breakup on \#superedges and \#correction-edges for summaries}.}
	\label{fig:trade_off}
	\vspace{-2mm}
\end{figure*}

In Figure~\ref{fig:trade_off}, the $y$-axis presents the relative size. Meanwhile, the $x$-axis reports
the running time for summary construction. We want the summary to be as compact as possible, and the construction to be as fast as possible. Thus, {\em better solution would be closer to the origin point}
in these plots.

The solid markers in Figure~\ref{fig:trade_off} stand for our proposed holistic algorithms, while the hollow markers represent the two-step methods.
We observe that the solid markers are closer to the origin point on all datasets, which confirms the superiority of our proposed holistic algorithms, based on both summary compactness and its construction efficiency. Our %ultimately
{\sf Hybrid} algorithm is shown with the shadow marker. In general, the shadow markers are below all others, demonstrating that {\em the {\sf Hybrid} algorithm produces the most compact summaries}.

{For {\sf $k$-Median}-based methods ({\sf $k$-Median}, {\sf KM+}, and {\sf HY}), $k$ must be given as an input,
whereas the optimal number of supernodes are automatically decided by {\sf Greedy}, {\sf Randomized}, {\sf SWeG}, {\sf GD+}, and {\sf RD+}.
For fairness of comparison, we find the optimal $k$ for {\sf $k$-Median}, {\sf KM+}, and {\sf HY} with an additional preprocessing step (\S\ref{sec:k}),
and report our results for {\sf $k$-Median}, {\sf KM+}, and {\sf HY} in Figure~\ref{fig:trade_off} with this optimal $k$ as input.
This additional preprocessing time for determining the optimal $k$ is provided in Table~\ref{tab:pre}.
The preprocessing time for {\sf $k$-Median} is higher than that of {\sf KM+}, since we need to identify an optimal $k$ for each relation in {\sf $k$-Median}, while we only require a single optimal $k$ for {\sf KM+}. {\sf HY} consumes exactly the same preprocessing time as {\sf KM+}, since its
preprocessing step is same as that of {\sf KM+}. However, in the third step of {\sf HY} (\S \ref{sec:hybrid}),
it further applies {\sf GD+} to improve the compactness, thus the final optimal number of supernodes may reduce for {\sf HY},
in comparison with {\sf KM+} (as reported in Table~\ref{tab:k}).}

% \begin{table}[tb!]
% 	\small
% 	\centering
% 	\caption{\small \#supernodes, and (\#superedges, \#correction-edges) in obtained summaries via our holistic methods.}
% 	\vspace{-2mm}
% 	\begin{tabular} { c||c|c|c }
% 		\hline
% 		{\sf Graph}        & {\sf GD$^+$}   & {\sf KM$^+$} & {\sf HY}\\ \hline
% 		{\em Homo}       & 440, (4K,18K)  & 452, (4K,19K) &   444, (4K,18K) \\
% 		{\em Amazon}      & 6K, (956K,3.6M)     & 6K, (960K,3.6M)   &  6K, (957K,3.6M)\\
% 		{\em DBLP6}    & 12K, (399K,2.6M)     & 12K, (398K,2.8M)   &  12K, (398K,2.6M) \\
% 		{\em DBLP230}    & 11K, (412K,2.3M)     &  11K, (419K,2.4M) & 11K, (413K,2.2M)\\
% 		{\em Twitter}     & 20K, (1.3M,4.2M)    & 20K, (1.3M,4.2M)  &  20K, (1.3M,4.1M)\\
% 		\hline
% 	\end{tabular}
% 	\label{tab:k}
% 	\vspace{-2mm}
% \end{table}

% \begin{table}[tb!]
% 	\small
% 	\centering
% 	\caption{\small Additional preprocessing time ($\times 10^3$ sec) of determining optimal $k$ for {\sf $k$-Median}-based methods.}
% 	\vspace{-2mm}
% 	\begin{tabular} { c||c|c|c }
% 		\hline
% 		{\sf Graph}  & {\sf KM}   & {\sf KM$^+$} & {\sf HY}\\ \hline
% 		{\em Homo}   & 6.6  & 2.7  & 2.7  \\
% 		{\em Amazon}   &  92    &  56  & 56  \\
% 		{\em DBLP6}    &   32   & 19  & 19  \\
% 		{\em DBLP230}    &   262   & 137  & 137\\
% 		{\em Twitter}     &   469  & 281  & 281 \\
% 		\hline
% 	\end{tabular}
% 	\label{tab:pre}
% 	\vspace{-2mm}
% \end{table}

\begin{table}[h!]
	\centering
		\caption{{\#supernodes, and (\#superedges, \#correction-edges) in obtained summaries via our holistic methods}.		\label{tab:k}}
		\begin{tabular} { c||c|c|c }
			\hline
			{\sf Graph}      & {\sf GD+}           & {\sf KM+}          & {\sf HY}\\ \hline
			{\em Homo}       & 440, (4.1K,18.2K)       & 440, (4.2K,18.7K)      &  428, (4.1K,18.1K) \\
			{\em Amazon}     & 6.1K, (956K,3.6M)     & 6.1K, (959K,3.6M)    &  6.0K, (957K,3.6M)\\
			{\em DBLP6}      & 12.2K, (399K,2.6M)    & 12.2K, (398K,2.8M)   &  12.1K, (398K,2.6M) \\
			{\em DBLP230}    & 11.3K, (412K,2.3M)    &  11.3K, (418K,2.4M)  & 11.3K, (413K,2.2M)\\
			{\em Twitter}    & 20.2K, (1.3M,4.2M)    & 20.2K, (1.3M,4.2M)   &  20.1K, (1.3M,4.1M)\\
			\hline
		\end{tabular}

\medskip \medskip

		\caption{Additional preprocessing time ($\times 10^3$ sec) of determining optimal $k$ for {\sf $k$-Median}-based methods.		\label{tab:pre}}
		\begin{tabular} { c||c|c|c }
			\hline
			{\sf Graph}  & {\sf KM}   & {\sf KM+} & {\sf HY}\\ \hline
			{\em Homo}   & 1.0  & 0.4  & 0.4  \\
			{\em Amazon}   &  30    &  21  & 21  \\
			{\em DBLP6}    &   4.8   & 2.9  & 2.9  \\
			{\em DBLP230}    &   18   & 10  & 10\\
			{\em Twitter}     &   51  & 32  & 32 \\
			\hline
		\end{tabular}
	\vspace{-2mm}
\end{table}

\spara{Compactness.} For the same kind of algorithm, {\em the proposed holistic version can return up to 5\% more compact summary than the corresponding two-step algorithm},
e.g., the {\sf KM+} summary always has at least 3\% smaller relative size than the {\sf KM} summary, over all datasets. Our ultimate {\sf HY} method can result in about 1\% more compact summaries than the best holistic method on all datasets.% except {\em DBLP230}.

Within each algorithm group, we find that {\bf (1)} {\sf Greedy} returns the most compact summaries, while {\sf Randomized} produces the worst results; {\bf (2)} the compactness of
the summaries by {\sf SWeG} and {\sf $k$-Median} are in the middle range. Usually, they are comparable to the summaries by {\sf Greedy}. These two observations hold both within the two-step
methods and within the holistic algorithm groups.

%Table~\ref{tab:size} presents the actual storage cost of the original graph files and the summary returned by our proposed holistic algorithms, whose ratio generally follows the relative size.

{The objective function of our summary has two components, the number of superedges $|E_S|$ and the number of correction edges $|\mathcal{C_S}|$.
Figure~\ref{fig:breakup_cost} demonstrates that {\em the cost of correction list dominates the total summary cost}. It is always about 80\% of the total cost.}

\spara{Efficiency.} For efficiency, we have the following observations. {\bf (1)} The holistic algorithms (with solid markers) tend to consume less running time than the two-step methods.
Recalling the complexity analysis listed in \S~\ref{sec:gs_single}, the two-step methods repeat the summary generation for every relation. This multiplies the time complexity of single summary
computation by $q$, where $q$ is the total number of relations. For example, the total time cost for producing $q$ single-relation summaries will be $\bigO((m+nk\log n)\cdot q)$ for {\sf $k$-Median}.
An additional time for summary aggregation is also required. In contrast, for {\sf KM+}, the time complexity is only $\bigO(m'+nk\log nq)$. The first term $m'$ is the total number of edges
across all relations, which is similar to $mq$. However, the second term $\log nq$ is much smaller than $q\log n$. This explains why our proposed holistic algorithms are faster than the two-step ones.
%
%For the {\em Greedy}, the {\em Randomized}, and the {\em SWeG} algorithms, the average degree of each node empirically increase by about 1 (The degree in multi-relation graph is defined by the number of neighbors of a node in any relation. If a neighbor is incident to this node with more than one relations, it only counts once). Thus the complexity term $\bigO(d_{av}^3)$ do not increase as much as multiplying $q$.
%
{\bf (2)} Usually, {\sf KM+} is the fastest among the proposed holistic methods, and {\sf GD+} is the slowest.
% However, {\sf $k$-Median$^+$} takes more time than {\sf Greedy$^+$} in {\em DBLP\_230}. This dataset has the large number of nodes, and the graph is sparse. Recalling the complexity analysis, {\sf $k$-Median$^+$} is more sensitive to the number of nodes, while {\sf Greedy$^+$} is more sensitive to the average degree of each node. On the other hand, the efficiency results on a much denser graph, e.g., {\em Amazon}, show that {\sf Greedy$^+$} has higher running time.
{\bf (3)} The running time of {\sf HY} is always higher than that of {\sf KM+}, since it requires to run {\sf KM+} at first, then applies {\sf GD+} to further improve the compactness.

\begin{table}[tb!]
	\centering
	\caption{ Actual storage cost (MB) for graphs and summaries. The actual storage cost for summaries includes the supernode mapping(s). We also report summary storage percentage w.r.t. original graph storage.}
	\begin{tabular} { c|c|c|c|c|c }
		\hline
		{\sf Graph} & {\sf Original}  & {\sf GD+} & {\sf KM+} & {\sf HY} & {\sf ALL}  \\ \hline
		{\em Homo} & 1.7              &  0.31 (18\%)    & 0.32 (19\%) & {\bf 0.30} (18\%) & 0.58 (34\%)\\
		{\em Amazon} & 116.9          &  68.8 (59\%)    &  69.3 (59\%)    & 68.4 (58\%) & {\bf 68.2} (58\%)\\
		{\em DBLP6} & 74.3         &  36.2 (49\%)   & 37.2 (50\%)  & {\bf 35.1} (47\%) & 37.2 (50\%)\\
		{\em DBLP230} & 82.5  &  30.3 (37\%)  & 31.1 (38\%) & {\bf 30.0} (37\%) & 137.8 (167\%)\\
		{\em Twitter} & 237.5      &   163.9 (69\%) &  165.0 (69\%) & {\bf 160.7} (68\%) & 182.9 (77\%) \\
		\hline
	\end{tabular}
	\label{tab:size1}
\end{table}

\begin{table}[tb!]
	\centering
	\caption{Actual storage cost (MB) for graphs and summaries with further compression: (super)edges between the same set of (super)nodes over multiple relations are stored as $<node\_i, node\_j,relation\_x,relation\_y,...,$ $relation\_z>$. The actual storage cost for summaries includes the supernode mapping(s).}
	\begin{tabular} { c|c|c|c|c|c }
		\hline
		{\sf Graph} & {\sf Original}  & {\sf GD+} & {\sf KM+} & {\sf HY} & {\sf ALL}  \\ \hline
		{\em Homo} & 1.3   &  {\bf 0.17} (13\%) & 0.18 (14\%)&{\bf 0.17} (13\%) & 0.54 (42\%) \\
		{\em Amazon} & 94.9           &   50.8 (54\%)& 51.1 (54\%) & {\bf 48.9} (52\%) &  60.6 (64\%) \\
		{\em DBLP6} & 63.1         &   27.8 (44\%) & 28.2 (45\%) & {\bf 26.7} (42\%) & 29.7 (47\%)\\
		{\em DBLP230} & 70.6        &   24.8 (35\%)& 25.8 (36\%)  &  {\bf 24.1} (34\%) & 126.0 (178\%)\\
		{\em Twitter} & 211.5         &   137.6 (65\%)& 139.8 (66\%) &{\bf 132.9} (63\%) & 151.4 (72\%)\\
		\hline
	\end{tabular}
	\label{tab:size2}
\end{table}

\begin{table}[tb!]
	\centering
	\caption{Breakup of the actual storage cost (MB) for summaries with further compression (shown in Table~\ref{tab:size2}). $E_S$ denotes superedges, $C$ denotes correction edges, and $M$ denotes the node mapping.}
	\begin{tabular} { c||c|c|c||c|c|c }
		\hline
		\multirow{2}{*}{\sf Graph}  & \multicolumn{3}{c||}{\sf HY}  & \multicolumn{3}{c}{\sf ALL}  \\ \cline{2-7}
		& {$|E_S|$} & {$|C|$} & {$|M|$} & {$|E_S|$} & {$|C|$} & {$|M|$}
		\\ \hline
		{\em \small Homo} &0.02 & 0.10 & 0.05 & 0.05 &0.15 &0.34\\
		{\em \small Amazon} &10.5 & 37.0 & 1.4 & 15.2 & 36.6 & 8.8\\
		{\em \small DBLP6} & 5.2 & 19.9 & 1.5 & 5.4 & 20.0 & 4.3\\
		{\em \small DBLP230}& 4.7 & 17.8 & 1.5 & 6.6 & 22.9 & 96.1\\
		{\em \small Twitter} &29.5 & 128.5 & 1.7 & 25.2 & 114.8 & 11.4\\
		\hline
	\end{tabular}
	\label{tab:size4}
\end{table}

%\begin{table}[tb!]
%	\centering
%	\small
%	\caption{\small Breakup of the actual storage cost for summaries with original format (MB).}
%	\vspace{1mm}
%	\begin{tabular} { c||c|c|c||c|c|c }
%		\hline
%		\multirow{2}{*}{\sf Dataset}  & \multicolumn{3}{c||}{\em GD+}  & \multicolumn{3}{c}{\em All}  \\ \cline{2-7}
%		& $|E|$ & $|C|$ & $M$ & $|E|$ & $|C|$ & $M$
%		\\ \hline
%		{\em Homo} & 0.05 & 0.21 & 0.05 & 0.05 & 0.19 & 0.34\\
%		{\em DBLP\_10} & 1.9 & 5.8 & 0.6 & 2.0 & 4.9 & 4.4 \\
%		{\em DBLP\_230} & 5.2 & 21.0 & 0.8 & 7.9 & 16.2 & 85.2 \\
%		{\em Amazon} &18.3 & 48.6 & 1.4 & 15.2 &44.6 & 8.8\\
%		{\em Twitter} &75.0 & 341.5 & 3.6 & 73.3 & 330.5 & 742.5\\
%		\hline
%	\end{tabular}
%	\label{tab:size3}
%	\vspace{-4mm}
%\end{table}

%

\subsection{Exact storage cost}
\label{sec:storage}

We empirically study the exact storage cost of the original graphs and the summaries in the memory.
Here, we further compare our uniform summary for all relations, against maintaining all the single-relation optimal summaries, denoted as {\sf ALL}. For the exact storage, in addition to our objective of superedges and correction list, we need to store the mapping $M$ from the original node set $V$ to the supernode set $V_S$. In the single-relation graph summary, and in our uniform summary for all relations, this mapping $M$ simply has the same size as the cardinality of the original node set $V$. However, there exist $q$ mappings for the {\sf ALL} method, since the supernode partitioning can be different across relations. As shown in Table~\ref{tab:size1}, in practice, the {\sf ALL} summaries require more storage overhead than our holistic {\sf GD+} or {\sf HY} summary. When the relation number is large, e.g., on {\em DBLP230}, the {\sf ALL} summaries have even larger storage cost than the original graphs. Table~\ref{tab:size4} provides more insight through decomposing the storage cost.

In the storage format so far (reported in Table~\ref{tab:size1}), each entry is represented as $<node\_i, node\_j, relation>$, following our definition in \S~\ref{subsec:background}. However, the storage cost can further reduce when there exist (super)edges between the same set of (super)nodes over multiple relations. In such scenarios, a simple way to further reduce the exact storage cost is to keep them as $<node\_i, node\_j, relation\_x, relation\_y, ..., relation\_z>$. As shown in Table~\ref{tab:size2}, our proposed summaries, both {\sf GD+} and {\sf HY}, benefit more from this type of storage format, compared to the {\sf ALL} summary. %Notice that the superedges part of the {\sf All} method cannot be compressed, since the superedges are established between different set of supernodes in various relations. %For the correction list, the storage cost reduction via the compression for our uniform summary {\sf GD+} is also slightly better than that of {\sf All}.

\subsection{Scalability analysis}
\label{sec:sca}
We analyze the scalability of our methods on the larger datasets, {\em Twitter} and {\em DBLP}.

\begin{figure}[tb!]
	\centering
	\subfigure[\small {\em Varying graph size (Twitter)}]
	{\includegraphics[scale=0.28]{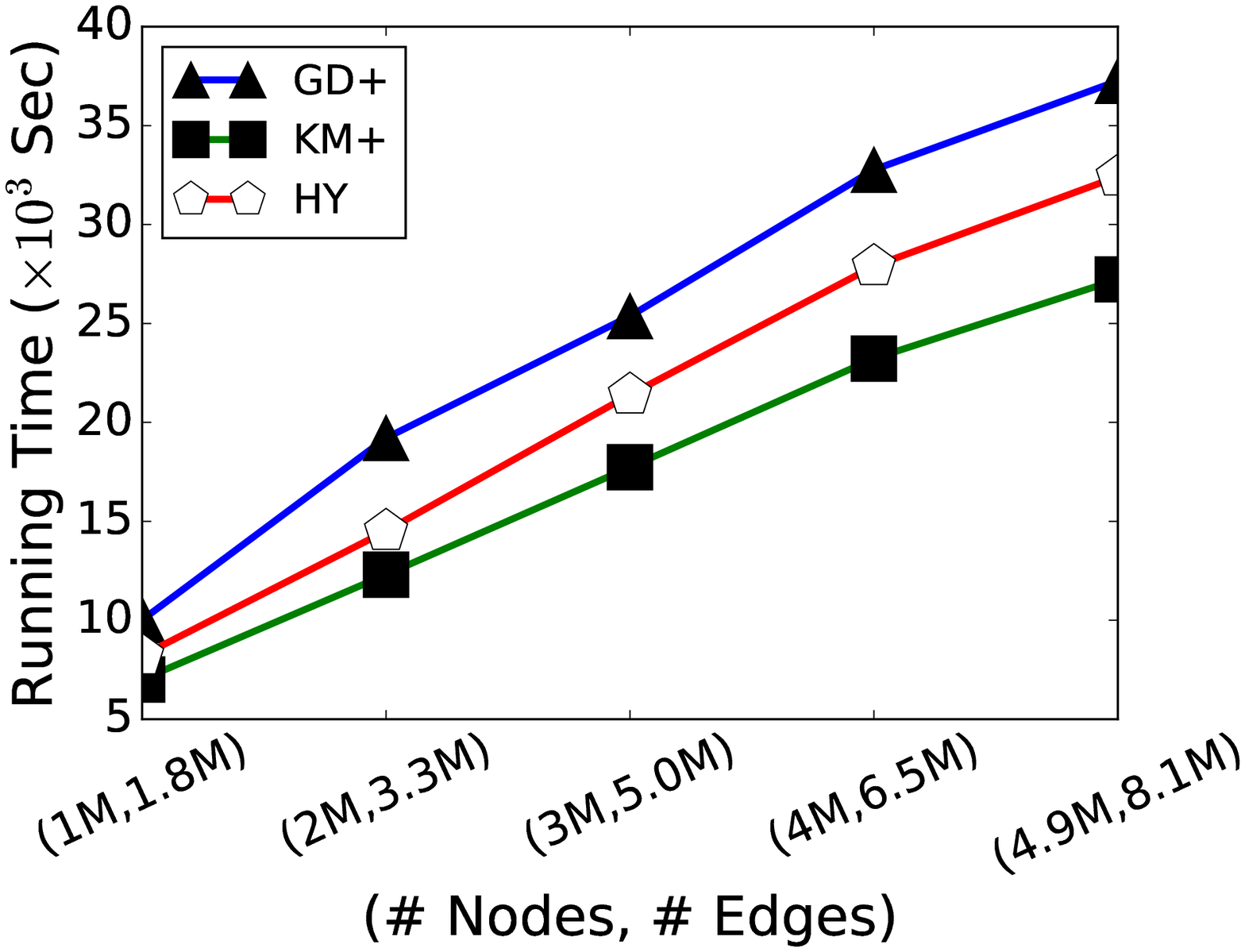}
		\label{fig:sca1}}
	\subfigure[\small {\em Varying number of relations (DBLP\_230)}]
	{\includegraphics[scale=0.28]{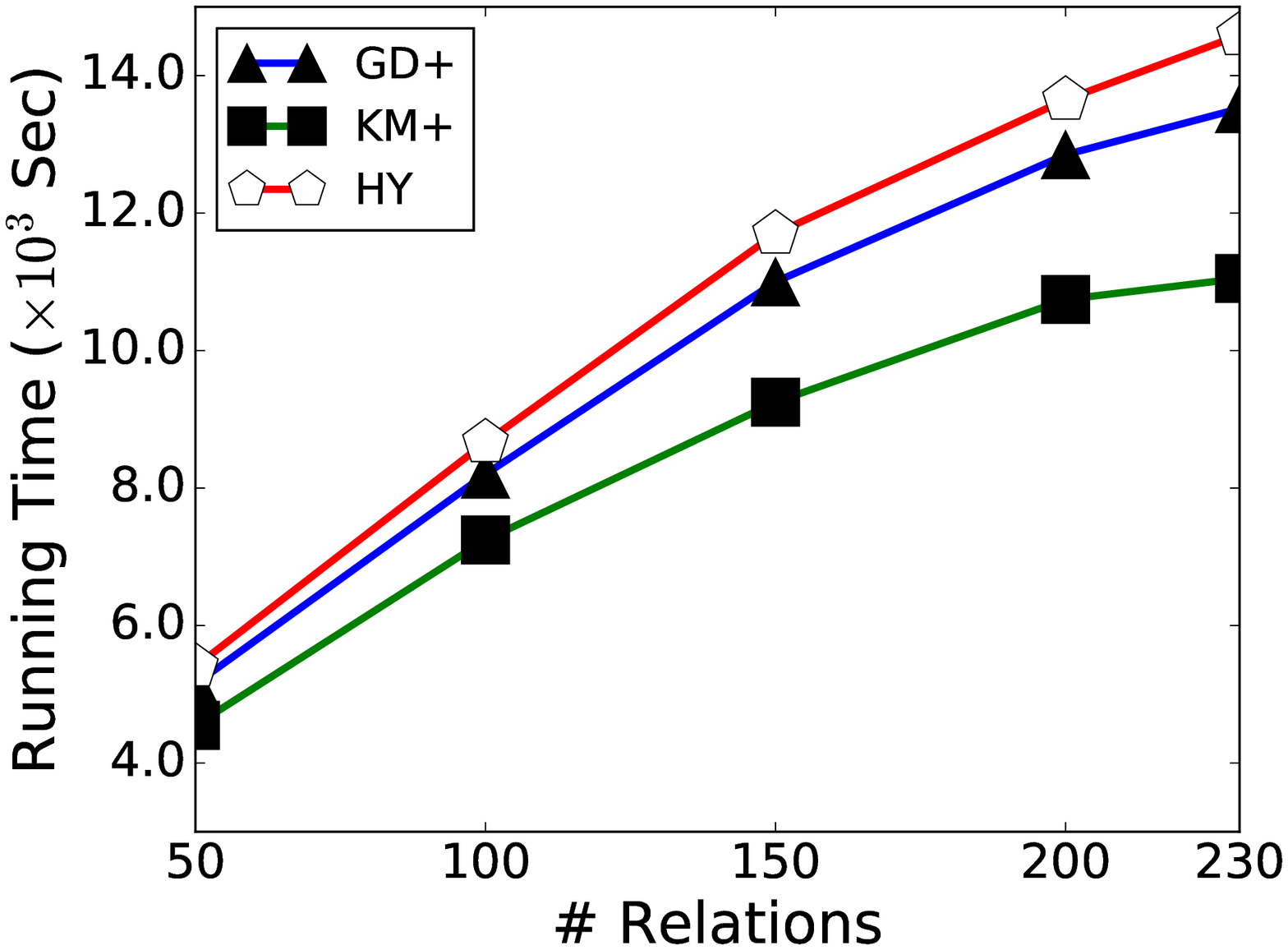}
		\label{fig:sca2}}
	\vspace{-3mm}
	\caption{\small Scalability analysis.}
	\label{fig:sca}
	\vspace{-2mm}
\end{figure}

\spara{Graph size.} The {\em Twitter} dataset has about 4.9 million nodes, we select 1M, 2M, 3M, 4M, and all 4.9M nodes uniformly at random to generate five graphs considering all relations, and apply our algorithms on them. Figure~\ref{fig:sca1} demonstrates that {\em all our proposed holistic and hybrid algorithms scale linearly in graph size (i.e., number of nodes).}

\spara{Number of relations.} {\em DBLP\_230} dataset has 230 relations. We randomly choose 50, 100, 150, 200, and all 230 relations, and apply the algorithms on the full graphs of selected relations. Figure~\ref{fig:sca2} shows that {\em the running times of all our algorithms increase linearly with more relations}. In the time complexity of {\sf KM+}: $\bigO(m'+nk\log nq)$,
$m'$ scales about linearly in the number ($q$) of relations, while the second term $nk\log nq$ keeps nearly the same with increasing $q$.
%For the {\em Greedy$^+$} and the {\em Randomized$^+$} algorithms, the average degree increase more when the number of relations is small.
%Overall, the experiments confirm good scalability of our holistic and hybrid methods.
%
% \begin{figure}[tb]
% 	\centering
% 	\vspace{-3mm}
% 	\subfigure[\small {\em Relative Size}]
% 	{\includegraphics[scale=0.18]{images/exp/k/Homo}
% 		\label{fig:k_homo_size}}
% 	\subfigure[\small {\em Running Time}]
% 	{\includegraphics[scale=0.18]{images/exp/k/Homo_time}
% 		\label{fig:k_homo_time}}
% 	\vspace{-4.5mm}
% 	\caption{\small Sensitivity analysis with \#supernodes $k$, {\em Homo}}
% 	\label{fig:k_homo}
% 	\vspace{-2mm}
% \end{figure}
%

%
% \vspace{-1mm}
% \subsection{Varying Number of Supernodes}
% \label{sec:exp_k}
% \vspace{-0.5mm}
% %
% We further investigate the sensitivity with respect to the number of supernodes, $k$, for the {\sf k-Median$^+$} algorithm.
% In Figures~\ref{fig:k_am_size}, we find that the relative size plots are valley-shaped. Too small or too large $k$
% both result in less compact summaries. The vertically dashed lines denote the number of supernodes found with the {\sf Greedy$^+$} algorithm, which are
% located in the optimal ranges of $k$ for {\sf $k$-Median$^+$}. This demonstrates that starting from the supernode number $k'$ of {\sf Greedy$^+$}, we can vary the $k$ in a small range centered at $k'$ to find the optimal one for {\sf $k$-Median$^+$}. The running time is about linear in the number of supernodes, $k$. Due to the space limit, we omit the results on other datasets since they are in general similar to that on {\em Amazon} dataset.

\subsection{Summary and recommendation}
\label{sec:rec}

\begin{table}
	\centering
	\caption{Summary and recommendation.}
	\scriptsize
	\vspace{-4mm}
	\begin{tabular} {r|c|c|c|c}
		%\hline \hline
	%	\multicolumn{4}{c}{Two-step Methods} \\ \hline \hline
		Method & Summary Compactness &  Summary Construction Time &  Optimal $k$ Finding (Preprocessing) Time & Approx. Guarantee \\
		\hline
		GD  & $\bigstar \bigstar \bigstar$ & $\bigstar $    & not required & \XSolidBrush \\
		RD & $ \bigstar$ & $\bigstar \bigstar \bigstar$  & not required & \XSolidBrush \\
		KM & $ \bigstar \bigstar$ & $\bigstar \bigstar \bigstar$  &  $\bigstar$  & \CheckmarkBold \\
		SWeG & $\bigstar \bigstar$ & $\bigstar \bigstar \bigstar$   & not required & \XSolidBrush  \\
		\hline% \hline
%		\multicolumn{4}{c}{Holistic Methods} \\ \hline \hline
		GD+ & $\bigstar \bigstar \bigstar \bigstar$ & $\bigstar  \bigstar$  & not required & \XSolidBrush  \\
		RD+  & $\bigstar \bigstar$ & $\bigstar \bigstar \bigstar \bigstar$  & not required & \XSolidBrush  \\
		KM+  & $\bigstar \bigstar \bigstar$ & $\bigstar \bigstar \bigstar \bigstar \bigstar$  &  $\bigstar \bigstar \bigstar$ & \CheckmarkBold \\
		\hline %\hline
	%	\multicolumn{4}{c}{Hybrid} \\ \hline \hline
		HY & $\bigstar \bigstar \bigstar \bigstar \bigstar$ & $\bigstar  \bigstar \bigstar$  & $\bigstar \bigstar \bigstar$ & \CheckmarkBold \\ \hline %\hline
	\end{tabular}
	\label{tab:rec}
	\vspace{-1mm}
\end{table}

Table~\ref{tab:rec} summarizes the recommendation level of each method according to different performance metrics. The scale is from 1 to 5 stars, and larger star number stands for higher ranking. Clearly, there is no single winner. The holistic algorithms tend to produce more compact summary and consume less running time than the corresponding two-step version. {\sf {\em k}-Median}-based approaches ({\sf KM}, {\sf KM+}, {\sf HY}) have approximation guarantees on the summary compactness. According to empirical evaluations, {\sf GD+} returns the most compact summary among the holistic methods. However, {\sf HY} applies {\sf GD+} to further improve the practical quality of the approximated solution returned by {\sf KM+}, and is empirically
shown to produce the best quality summaries (\S~\ref{sec:base}). For efficiency, {\sf KM+} is the fastest one.
{\sf RD+} is always faster than {\sf GD+} , and {\sf HY} is always slower than {\sf KM+}. The {\sf $k$-Median}-based methods ({\sf KM}, {\sf KM+}, and {\sf HY}) require an additional preprocessing step to identify the optimal $k$, and {\sf KM} consumes the most amount of preprocessing time.

Based on application requirements, a user can decide to adopt a specific algorithm as per our summary in Table~\ref{tab:rec}.
In general case, considering various trade-offs, we recommend {\sf Hybrid} method ({\sf HY}) for multi-relation graph summarization.
It has good performance in summary compactness from both theoretical and practical perspectives, and its efficiency lies in the middle range.

\section{Applications and case studies}
\label{sec:app}

\subsection{Efficient query processing on graph summaries}
Since our graph summary is lossless, we can always answer a graph query using the summary as precisely as in the original graph. Thus, we focus on the efficiency analyses.
We present the comparison for the {\em Neighborhood Query} \cite{LeFevreT10, RGB14}: Given a node $v$ and a graph $G=(V,E,R)$, find the set of nodes $N_v=\{u|(v,u,r)\in E\}$, and return the distribution of relations in $N_v$.
Notice that the {\em Degree Query} and the {\em Eigenvector-Centrality Query} in \cite{LeFevreT10, RGB14} can be answered
based on the results of neighborhood query; we do not consider them in the current study.

%(2) {\sf Labelled Triangle-Counting Query}: Given three relations $(r_1, r_2, r_3)$ and a graph $G=(V,E,R)$, count the number of triangles (cycles of size 3) satisfying the relation constraint. We consider the naive exact counting by checking all triples of nodes.

\begin{figure}[tb!]
	\centering
	\subfigure[\small {\em DBLP\_230}]
	{\includegraphics[scale=0.25]{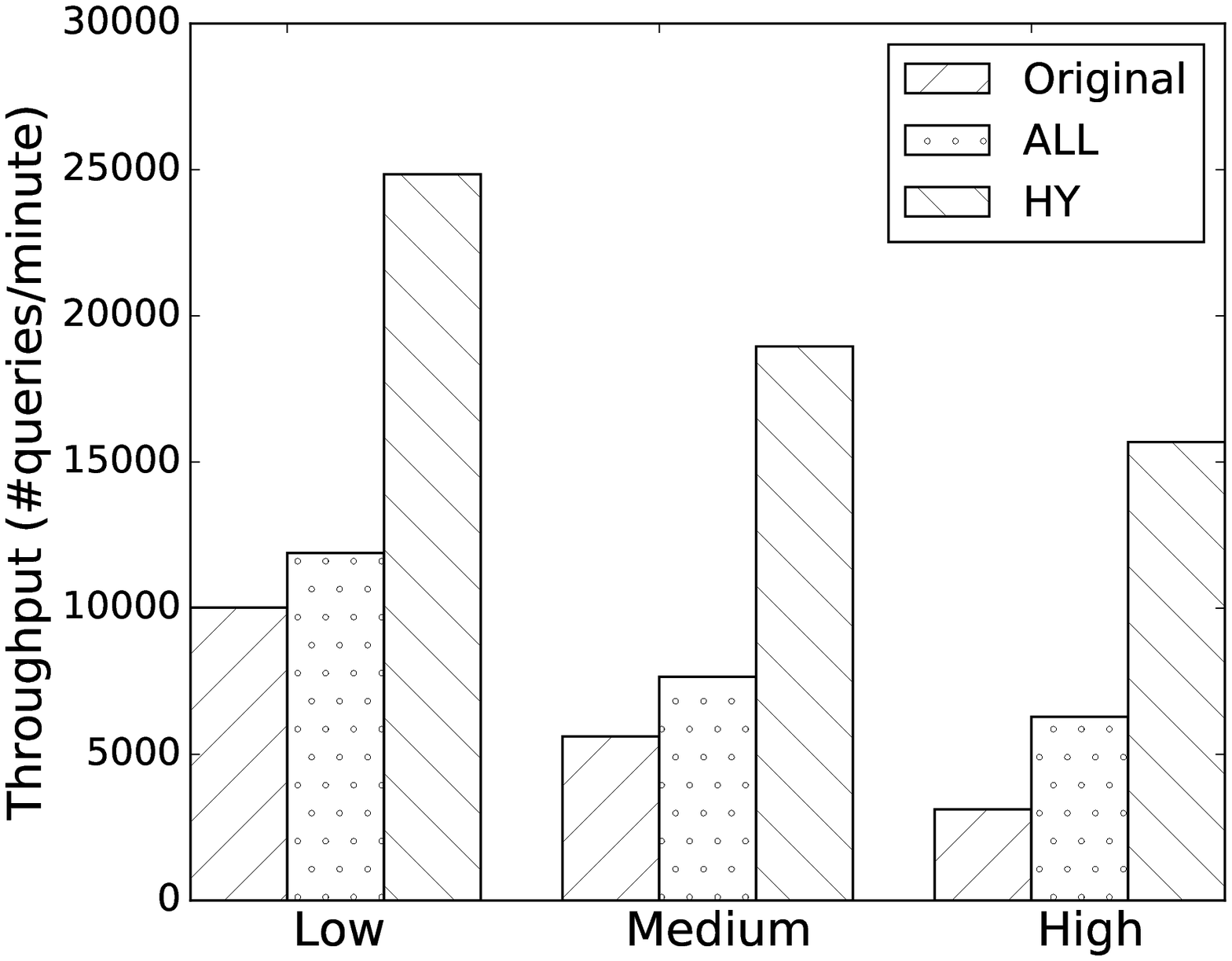}
		\label{fig:nei1}}
	\subfigure[\small {\em Twitter}]
	{\includegraphics[scale=0.25]{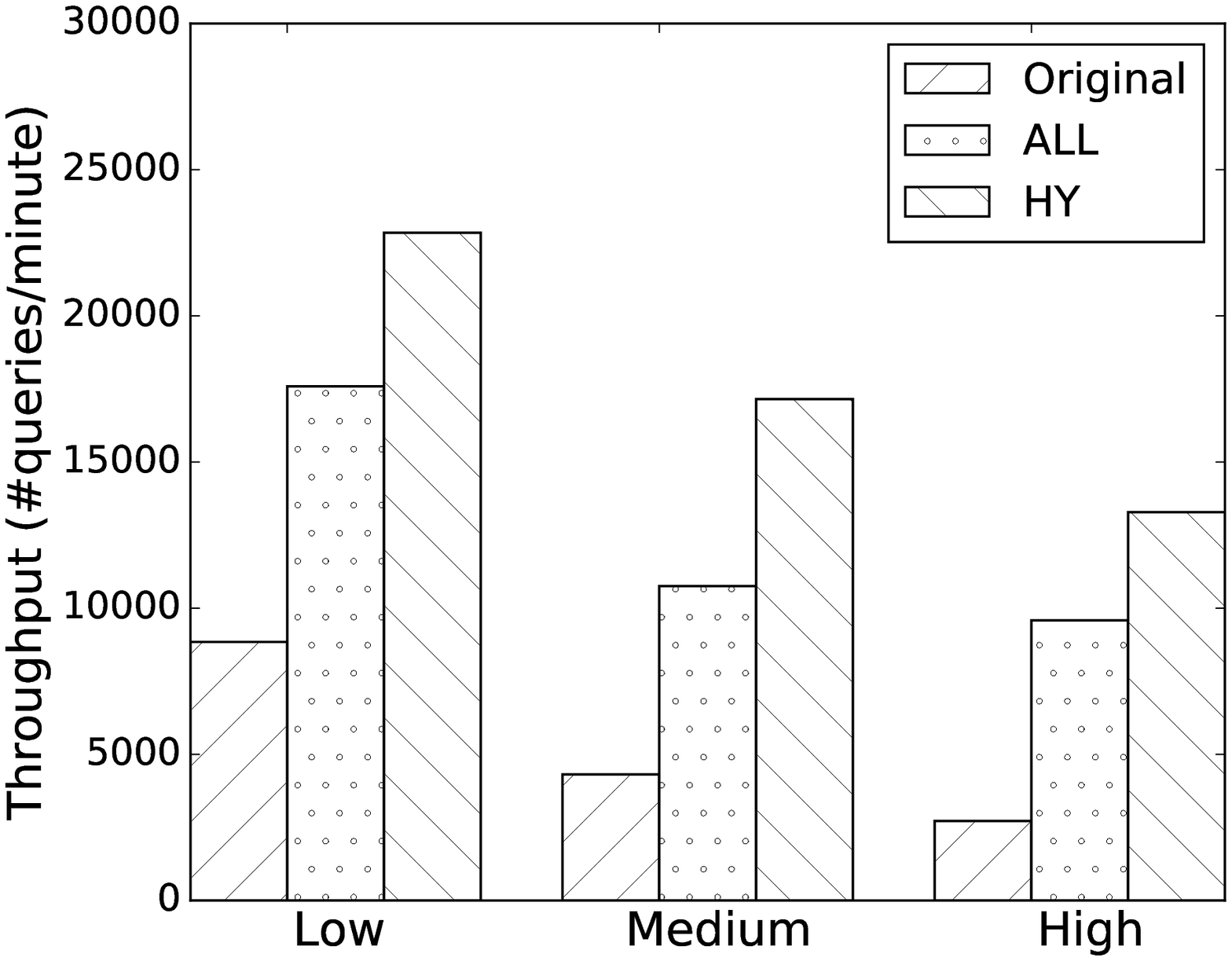}
		\label{fig:nei2}}
	\vspace{-3mm}
	\caption{\small Throughput comparison for neighborhood query answering on original graph and on different graph summaries.}
	\label{fig:nei}
	\vspace{-3mm}
\end{figure}

Figure~\ref{fig:nei} presents the throughput comparison for neighborhood query on the original graph and on the graph summaries ({\sf ALL} and {\sf HY}),  using two larger datasets. Query nodes are divided into three groups: low ($\leq 5$), medium ($(5,20]$), and high ($>20$), based on their out-degrees.
For an hour, we continue to answer neighborhood queries for query nodes selected uniformly at random, and then report the average throughput (per minute). The neighborhood query can be processed more efficiently on graph summary since we  explore the superedges and corrections instead of exact edges linked to the query node, and our objective ensures that the former tends to have smaller size than the latter. We have the following observations: {\bf (1)} the throughput on {\sf HY} summary is about 2.5$\times$ of that on the original graph for low-degree query nodes. It increases to be around 5$\times$ for high-degree nodes, since their edges are more likely to be wrapped within superedges. The benefit of using {\sf ALL} summary to answer
neighborhood query is 1.5$\times$ to 3$\times$ in throughput against using the original graph; {\bf (2)} the efficiency improvement is more significant when the graph summary is more compact, e.g., in {\em DBLP\_230}. The intuition is that the edges of the query nodes are more likely to be represented via superedges, rather than corrections; and {\bf (3)} when the number of relations is high, e.g., in {\em DBLP\_230}, the efficiency improvement for answering the query with {\sf HY} summary is more significant than with {\sf ALL} summary. This is because we need to repeatedly identify the supernode containing the query node in each relation for {\sf ALL} summary, and the degree of the query node may be low in each relation, which makes the individual summaries within
{\sf ALL} less beneficial for query answering.

\begin{figure}[tb!]
	\centering
	{\includegraphics[scale=0.47,angle=270]{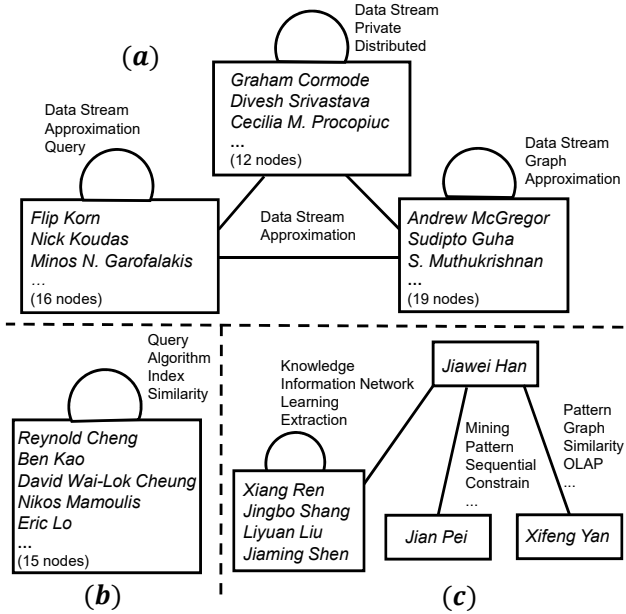}}
	\caption{Case study on {\em DBLP}.}
	\label{fig:case}
\end{figure}

\subsection{Visualization case study on DBLP}
We visualize the summary of {\em DBLP230} dataset, and present some interesting case studies in Figure~\ref{fig:case}.
In the first case shown in Figure~\ref{fig:case}(a), we find that all researchers in each supernode actively collaborate with others on the topics such as ``data stream'' and ``approximation''. They have both self-loops and links to each other with such topics. However, a few different topics also appear within each group. For example, {\em Graham Cormode} also has quite a few ``private''-related (i.e., privacy) joint works with {\em Divesh Srivastava}. Another interesting finding is that both the left-side and the top supernodes contain researchers from database community, since they have more publications in {\em SIGMOD} and {\em VLDB}, while the right-side supernode consists of theory community researchers, who publishes more in {\em STOC}, {\em SODA}, etc.

The second case in Figure~\ref{fig:case}(b) presents a group of people with close collaboration. One can verify that they are all from same geographical location, i.e., {\em Hong Kong} in this example. This is a frequent pattern in {\em DBLP}.

Figure~\ref{fig:case}(c) first shows that very senior researchers tend to be kept as a single node, since they work on quite diverse topics and with many researchers. Here, {\em Jiawei Han} and his ex-students, {\em Jian Pei} and {\em Xifeng Yan}, are all kept as a singleton supernode, and {\em Jiawei Han} has different collaboration topics with them. His recently graduated student, Xiang Ren (in 2018), and some other current students are grouped together, since they collaborate frequently with each other on topics related to knowledge extraction. Such correlations across different relations and node set could
not be immediately inferred if one keeps individual summaries for all relations separately.

\begin{figure}[tb!]
	\centering
	%\vspace{-4mm}
	{\includegraphics[scale=0.75,angle=90]{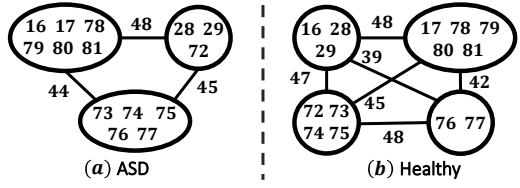}}
	\caption{Case study on {\em Austism}. The partitioning of the same set of 14 nodes are quite different for ASD brain networks and for healthy brain networks. The numbers on superedges denote the number of relations having such superedges.}
	\label{fig:brain}
\end{figure}

\subsection{Visualization and classification case studies on brain networks}
\label{sec:brain}

For our second case study, we use the {\em Austism} dataset \cite{CBCCE13} containing 96 brain networks. 48 of them are collected from ASD (Autism spectrum disorder) patients, and the other 48 brain networks are of healthy people. In the original dataset, there were 49 networks of ASD patients and 52 networks of healthy people. To maintain a balanced number of networks in each group, and for the ease of our comparison, we only keep the first 48 networks in each group here. All networks share the same set of 116 nodes. The average number of edges in each network is 1336.8 (1336.2 for ASD patients, and 1337.5 for healthy people). If we treat the 48 ASD brain networks as a multi-relation graph, {\sf HY} returns a summary with 0.514 relative size (0.516 by {\sf GD+}, 0.517 by {\sf KM+}). Meanwhile, the relative size of {\sf HY} multi-relation summary for the 48 healthy people is 0.526 (0.526 by {\sf GD+}, 0.529 by {\sf KM+}). However, if we randomly choose 24 networks from each group, and generate a multi-relation summary for them, the relative size will be 0.559 by {\sf HY} (0.559 by {\sf GD+}, 0.560 by {\sf KM+}, both are the average value over 20 attempts). Clearly, it is harder to summarize the mixed group, which implies that {\em the networks of ASD patients have different structure compared to those of healthy people}, and inspires us to {\em utilize the multi-relation summaries for ASD patient detection}.

\spara{Classification.} We randomly choose two brain networks, one for ASD patient and one for healthy people. A multi-relation summary $S_1$ is generated for the rest 47 ASD brain networks, and another multi-relation summary $S_2$ for the rest 47 healthy ones. Then, we alternatively apply $S_1$ and $S_2$ as summaries for the two selected networks, and suggest the label ``ASD'' to a network if $S_1$ results in lower cost than $S_2$. We repeat the procedure 20 times. Within the 20 true ASD patients, 19 of them are correctly detected. For the 20 healthy people, 3 of them are wrongly labelled as ``ASD''. It can be easily calculated that our classification precision is 0.86, recall is 0.95, and F1-score is 0.90. This demonstrates the effectiveness of our multi-relation summaries in this classification task.

\spara{Visualization.} Finally, we visualize some subgraphs of our multi-relation summaries for the two groups of brain networks. As shown in Figure~\ref{fig:brain}, the partitionings of the same set of 14 nodes are quite different for ASD brain networks and for healthy brain networks. Although these supernodes are well-connected to each other, they can not be further merged since they have different superedges.

\vspace{-2mm}
\section{Conclusions}
\label{sec:conclusion}
\vspace{-1mm}
In this paper, we first revisited the classic single-relation graph summarization problem, and provided the first polynomial-time approximation algorithm based on the {\sf $k$-Median} clustering. Then we introduced and investigated the novel problem of multi-relation graph summarization. %We studied the complexity of the problem, and clearly distinguished it with correlation clustering and graph partitioning.
To solve the problem, we first studied the baseline two-step approaches: first generate a summary for each relation, and then properly aggregate them.
%When revisiting the single relation graph summarization algorithms, we proved that the {\sf $k$-Median} clustering on the rows of adjacency matrix  provides a bounded solution to our objective.
We further demonstrated and discussed the limitations of these baselines, and proposed holistic solutions to overcome them. Among them, the holistic {\sf $k$-Median$^+$} is able to maintain the approximation guarantee over multi-relation graphs. Finally, we developed {\sf Hybrid} algorithm as our ultimate solution, by utilizing both the strengths of {\sf $k$-Median$^+$} and {\sf Greedy$^+$}. Our experimental results and case studies validated the effectiveness and efficiency of our algorithms.

\begin{acks}
Arijit Khan is supported by MOE Tier1 and Tier2 grants RG117/19 and MOE2019T2-2-042.
\end{acks}

\bibliographystyle{ACM-Reference-Format}
\bibliography{ref}


\begin{thebibliography}{87}

%%% the Web address.

\ifx \showCODEN    \undefined \def \showCODEN     #1{\unskip}     \fi
\ifx \showDOI      \undefined \def \showDOI       #1{#1}\fi
\ifx \showISBNx    \undefined \def \showISBNx     #1{\unskip}     \fi
\ifx \showISBNxiii \undefined \def \showISBNxiii  #1{\unskip}     \fi
\ifx \showISSN     \undefined \def \showISSN      #1{\unskip}     \fi
\ifx \showLCCN     \undefined \def \showLCCN      #1{\unskip}     \fi
\ifx \shownote     \undefined \def \shownote      #1{#1}          \fi
\ifx \showarticletitle \undefined \def \showarticletitle #1{#1}   \fi
\ifx \showURL      \undefined \def \showURL       {\relax}        \fi
% invisible to TeX
\providecommand\bibfield[2]{#2}
\providecommand\bibinfo[2]{#2}
\providecommand\natexlab[1]{#1}
\providecommand\showeprint[2][]{arXiv:#2}

\bibitem[\protect\citeauthoryear{Ahn, Guha, and McGregor}{Ahn
  et~al\mbox{.}}{2012a}]%
        {AhnGM12}
\bibfield{author}{\bibinfo{person}{K.~J. Ahn}, \bibinfo{person}{S. Guha}, {and}
  \bibinfo{person}{A. McGregor}.} \bibinfo{year}{2012}\natexlab{a}.
\newblock \showarticletitle{{Analyzing Graph Structure via Linear
  Measurements}}. In \bibinfo{booktitle}{\emph{SODA}}.
\newblock


\bibitem[\protect\citeauthoryear{Ahn, Guha, and McGregor}{Ahn
  et~al\mbox{.}}{2012b}]%
        {AGM12}
\bibfield{author}{\bibinfo{person}{K.~J. Ahn}, \bibinfo{person}{S. Guha}, {and}
  \bibinfo{person}{A. McGregor}.} \bibinfo{year}{2012}\natexlab{b}.
\newblock \showarticletitle{{Graph Sketches: Sparsification, Spanners, and
  Subgraphs}}. In \bibinfo{booktitle}{\emph{PODS}}.
\newblock


\bibitem[\protect\citeauthoryear{Bandyopadhyay, Mehta, Kuo, Sung, Chuang,
  Jaehnig, Bodenmiller, Licon, Copeland, Shales, Fiedler, Dutkowski,
  Gu\'{e}nol\'{e}, van Attikum, Shokat, Kolodner, Huh, Aebersold, Keogh,
  Krogan, and Ideker}{Bandyopadhyay et~al\mbox{.}}{2010}]%
        {Bandyopadhyay10}
\bibfield{author}{\bibinfo{person}{S. Bandyopadhyay}, \bibinfo{person}{M.
  Mehta}, \bibinfo{person}{D. Kuo}, \bibinfo{person}{M.-K. Sung},
  \bibinfo{person}{R. Chuang}, \bibinfo{person}{E.~J. Jaehnig},
  \bibinfo{person}{B. Bodenmiller}, \bibinfo{person}{K. Licon},
  \bibinfo{person}{W. Copeland}, \bibinfo{person}{M. Shales},
  \bibinfo{person}{D. Fiedler}, \bibinfo{person}{J. Dutkowski},
  \bibinfo{person}{A. Gu\'{e}nol\'{e}}, \bibinfo{person}{H. van Attikum},
  \bibinfo{person}{K.~M. Shokat}, \bibinfo{person}{R.~D. Kolodner},
  \bibinfo{person}{W.-K. Huh}, \bibinfo{person}{R. Aebersold},
  \bibinfo{person}{M.-C. Keogh}, \bibinfo{person}{N.~J. Krogan}, {and}
  \bibinfo{person}{T. Ideker}.} \bibinfo{year}{2010}\natexlab{}.
\newblock \showarticletitle{{Rewiring of Genetic Networks in Response to DNA
  Damage}}.
\newblock \bibinfo{journal}{\emph{Science}} \bibinfo{volume}{330},
  \bibinfo{number}{6009} (\bibinfo{year}{2010}).
\newblock


\bibitem[\protect\citeauthoryear{Bansal, Blum, and S.}{Bansal
  et~al\mbox{.}}{2004}]%
        {BBC04}
\bibfield{author}{\bibinfo{person}{N. Bansal}, \bibinfo{person}{A. Blum}, {and}
  \bibinfo{person}{Chawla S.}} \bibinfo{year}{2004}\natexlab{}.
\newblock \showarticletitle{{Correlation Clutering}}.
\newblock \bibinfo{journal}{\emph{Machine Learning}}  \bibinfo{volume}{56}
  (\bibinfo{year}{2004}), \bibinfo{pages}{89--113}.
\newblock


\bibitem[\protect\citeauthoryear{Beg, Ahmad, Zaman, and Khan}{Beg
  et~al\mbox{.}}{2018}]%
        {BegAZK18}
\bibfield{author}{\bibinfo{person}{M.~A. Beg}, \bibinfo{person}{M. Ahmad},
  \bibinfo{person}{A. Zaman}, {and} \bibinfo{person}{I. Khan}.}
  \bibinfo{year}{2018}\natexlab{}.
\newblock \showarticletitle{{Scalable Approximation Algorithm for Graph
  Summarization}}. In \bibinfo{booktitle}{\emph{PAKDD}}.
\newblock


\bibitem[\protect\citeauthoryear{Bencz{\'{u}}r and Karger}{Bencz{\'{u}}r and
  Karger}{2015}]%
        {BenczurK15}
\bibfield{author}{\bibinfo{person}{A.~A. Bencz{\'{u}}r} {and}
  \bibinfo{person}{D.~R. Karger}.} \bibinfo{year}{2015}\natexlab{}.
\newblock \showarticletitle{{Randomized Approximation Schemes for Cuts and
  Flows in Capacitated Graphs}}.
\newblock \bibinfo{journal}{\emph{{SIAM} J. Comput.}} \bibinfo{volume}{44},
  \bibinfo{number}{2} (\bibinfo{year}{2015}), \bibinfo{pages}{290--319}.
\newblock


\bibitem[\protect\citeauthoryear{Besta and Hoefler}{Besta and Hoefler}{2018}]%
        {BH18}
\bibfield{author}{\bibinfo{person}{M. Besta} {and} \bibinfo{person}{T.
  Hoefler}.} \bibinfo{year}{2018}\natexlab{}.
\newblock \showarticletitle{{Survey and Taxonomy of Lossless Graph Compression
  and Space-Efficient Graph Representations}}.
\newblock \bibinfo{journal}{\emph{CoRR}}  \bibinfo{volume}{abs/1806.01799}
  (\bibinfo{year}{2018}).
\newblock


\bibitem[\protect\citeauthoryear{Boden, G\"{u}nnemann, Hoffmann, and
  Seidl}{Boden et~al\mbox{.}}{2012}]%
        {BGHS12}
\bibfield{author}{\bibinfo{person}{B. Boden}, \bibinfo{person}{S.
  G\"{u}nnemann}, \bibinfo{person}{H. Hoffmann}, {and} \bibinfo{person}{T.
  Seidl}.} \bibinfo{year}{2012}\natexlab{}.
\newblock \showarticletitle{{Mining Coherent Subgraphs in Multi-layer Graphs
  with Edge Labels}}. In \bibinfo{booktitle}{\emph{KDD}}.
\newblock


\bibitem[\protect\citeauthoryear{Boldi, Rosa, Santini, and Vigna}{Boldi
  et~al\mbox{.}}{2011}]%
        {BRSV11}
\bibfield{author}{\bibinfo{person}{P. Boldi}, \bibinfo{person}{M. Rosa},
  \bibinfo{person}{M. Santini}, {and} \bibinfo{person}{S. Vigna}.}
  \bibinfo{year}{2011}\natexlab{}.
\newblock \showarticletitle{{Layered Label Propagation: A Multiresolution
  Coordinate-Free Ordering for Compressing Social Networks}}. In
  \bibinfo{booktitle}{\emph{WWW}}.
\newblock


\bibitem[\protect\citeauthoryear{Boldi and Vigna}{Boldi and Vigna}{2004}]%
        {BV04}
\bibfield{author}{\bibinfo{person}{P. Boldi} {and} \bibinfo{person}{S. Vigna}.}
  \bibinfo{year}{2004}\natexlab{}.
\newblock \showarticletitle{{The Webgraph Framework I: Compression
  Techniques}}. In \bibinfo{booktitle}{\emph{WWW}}.
\newblock


\bibitem[\protect\citeauthoryear{Brisaboa, Ladra, and Navarro}{Brisaboa
  et~al\mbox{.}}{2009}]%
        {BLN09}
\bibfield{author}{\bibinfo{person}{N.~R. Brisaboa}, \bibinfo{person}{S. Ladra},
  {and} \bibinfo{person}{G. Navarro}.} \bibinfo{year}{2009}\natexlab{}.
\newblock \showarticletitle{{k2-Trees for Compact Web Graph Representation}}.
  In \bibinfo{booktitle}{\emph{SPIRE}}.
\newblock


\bibitem[\protect\citeauthoryear{Broder, Charikar, Frieze, and
  Mitzenmacher}{Broder et~al\mbox{.}}{2000}]%
        {BCFM00}
\bibfield{author}{\bibinfo{person}{A.~Z. Broder}, \bibinfo{person}{M.
  Charikar}, \bibinfo{person}{A.~M. Frieze}, {and} \bibinfo{person}{M.
  Mitzenmacher}.} \bibinfo{year}{2000}\natexlab{}.
\newblock \showarticletitle{{Min-wise Independent Permutations}}.
\newblock \bibinfo{journal}{\emph{J. Comput. System Sci.}}
  \bibinfo{volume}{60} (\bibinfo{year}{2000}), \bibinfo{pages}{630--659}.
\newblock


\bibitem[\protect\citeauthoryear{Buehrer and Chellapilla}{Buehrer and
  Chellapilla}{2008}]%
        {BC08}
\bibfield{author}{\bibinfo{person}{G. Buehrer} {and} \bibinfo{person}{K.
  Chellapilla}.} \bibinfo{year}{2008}\natexlab{}.
\newblock \showarticletitle{{A Scalable Pattern Mining Approach to Web Graph
  Compression with Communities}}. In \bibinfo{booktitle}{\emph{WSDM}}.
\newblock


\bibitem[\protect\citeauthoryear{Cardillo, G{\'{o}}mez{-}Garde{\~{n}}es, Zanin,
  Romance, Papo, del Pozo, and Boccaletti}{Cardillo et~al\mbox{.}}{2012}]%
        {CGZRPPB12}
\bibfield{author}{\bibinfo{person}{A. Cardillo}, \bibinfo{person}{J.
  G{\'{o}}mez{-}Garde{\~{n}}es}, \bibinfo{person}{M. Zanin},
  \bibinfo{person}{M. Romance}, \bibinfo{person}{D. Papo}, \bibinfo{person}{F.
  del Pozo}, {and} \bibinfo{person}{S. Boccaletti}.}
  \bibinfo{year}{2012}\natexlab{}.
\newblock \showarticletitle{{Emergence of Network Features from Multiplexity}}.
\newblock \bibinfo{journal}{\emph{CoRR}}  \bibinfo{volume}{abs/1212.2153}
  (\bibinfo{year}{2012}).
\newblock


\bibitem[\protect\citeauthoryear{Charikar, Guruswami, and Wirth}{Charikar
  et~al\mbox{.}}{2005}]%
        {CGW05}
\bibfield{author}{\bibinfo{person}{M. Charikar}, \bibinfo{person}{V.
  Guruswami}, {and} \bibinfo{person}{A. Wirth}.}
  \bibinfo{year}{2005}\natexlab{}.
\newblock \showarticletitle{{Clustering with Qualitative Information}}.
\newblock \bibinfo{journal}{\emph{J. Comput. Syst. Sci.}} \bibinfo{volume}{71},
  \bibinfo{number}{3} (\bibinfo{year}{2005}), \bibinfo{pages}{360--383}.
\newblock


\bibitem[\protect\citeauthoryear{Chen, Lin, Fredrikson, Christodorescu, Yan,
  and Han}{Chen et~al\mbox{.}}{2009}]%
        {CLFCYH09}
\bibfield{author}{\bibinfo{person}{C. Chen}, \bibinfo{person}{C.~X. Lin},
  \bibinfo{person}{M. Fredrikson}, \bibinfo{person}{M. Christodorescu},
  \bibinfo{person}{X. Yan}, {and} \bibinfo{person}{J. Han}.}
  \bibinfo{year}{2009}\natexlab{}.
\newblock \showarticletitle{{Mining Graph Patterns Efficiently via Randomized
  Summaries}}.
\newblock \bibinfo{journal}{\emph{{PVLDB}}} \bibinfo{volume}{2},
  \bibinfo{number}{1} (\bibinfo{year}{2009}), \bibinfo{pages}{742--753}.
\newblock


\bibitem[\protect\citeauthoryear{Chen, Lin, Fredrikson, Christodorescu, Yan,
  and Han}{Chen et~al\mbox{.}}{2010}]%
        {LFCY010}
\bibfield{author}{\bibinfo{person}{C. Chen}, \bibinfo{person}{C.~X. Lin},
  \bibinfo{person}{M. Fredrikson}, \bibinfo{person}{M. Christodorescu},
  \bibinfo{person}{X. Yan}, {and} \bibinfo{person}{J. Han}.}
  \bibinfo{year}{2010}\natexlab{}.
\newblock \showarticletitle{{Mining Large Information Networks by Graph
  Summarization}}.
\newblock In \bibinfo{booktitle}{\emph{Link Mining: Models, Algorithms, and
  Applications}}. \bibinfo{pages}{475--501}.
\newblock


\bibitem[\protect\citeauthoryear{Chen, Yan, Zhu, Han, and Yu}{Chen
  et~al\mbox{.}}{2008}]%
        {CYZHY08}
\bibfield{author}{\bibinfo{person}{C. Chen}, \bibinfo{person}{X. Yan},
  \bibinfo{person}{F. Zhu}, \bibinfo{person}{J. Han}, {and}
  \bibinfo{person}{P.~S. Yu}.} \bibinfo{year}{2008}\natexlab{}.
\newblock \showarticletitle{{Graph OLAP: Towards Online Analytical Processing
  on Graphs}}. In \bibinfo{booktitle}{\emph{ICDM}}.
\newblock


\bibitem[\protect\citeauthoryear{Chierichetti, Kumar, Lattanzi, Mitzenmacher,
  Panconesi, and Raghavan}{Chierichetti et~al\mbox{.}}{2009}]%
        {CKLMPR09}
\bibfield{author}{\bibinfo{person}{F. Chierichetti}, \bibinfo{person}{R.
  Kumar}, \bibinfo{person}{S. Lattanzi}, \bibinfo{person}{M. Mitzenmacher},
  \bibinfo{person}{A. Panconesi}, {and} \bibinfo{person}{P. Raghavan}.}
  \bibinfo{year}{2009}\natexlab{}.
\newblock \showarticletitle{{On Compressing Social Networks}}. In
  \bibinfo{booktitle}{\emph{KDD}}.
\newblock


\bibitem[\protect\citeauthoryear{Choi and Szpankowski}{Choi and
  Szpankowski}{2012}]%
        {CS12}
\bibfield{author}{\bibinfo{person}{Y. Choi} {and} \bibinfo{person}{W.
  Szpankowski}.} \bibinfo{year}{2012}\natexlab{}.
\newblock \showarticletitle{{Compression of Graphical Structures: Fundamental
  Limits, Algorithms, and Experiments}}.
\newblock \bibinfo{journal}{\emph{{IEEE} Trans. Information Theory}}
  \bibinfo{volume}{58}, \bibinfo{number}{2} (\bibinfo{year}{2012}),
  \bibinfo{pages}{620--638}.
\newblock


\bibitem[\protect\citeauthoryear{Cook and Holder}{Cook and Holder}{1994}]%
        {CH94}
\bibfield{author}{\bibinfo{person}{D.~J. Cook} {and} \bibinfo{person}{L.~B.
  Holder}.} \bibinfo{year}{1994}\natexlab{}.
\newblock \showarticletitle{{Substructure Discovery Using Minimum Description
  Length and Background Knowledge}}.
\newblock \bibinfo{journal}{\emph{J. Artif. Intell. Res.}}  \bibinfo{volume}{1}
  (\bibinfo{year}{1994}), \bibinfo{pages}{231--255}.
\newblock


\bibitem[\protect\citeauthoryear{Cormode and Muthukrishnan}{Cormode and
  Muthukrishnan}{2005}]%
        {CormodeM05}
\bibfield{author}{\bibinfo{person}{G. Cormode} {and} \bibinfo{person}{S.
  Muthukrishnan}.} \bibinfo{year}{2005}\natexlab{}.
\newblock \showarticletitle{{Space efficient mining of multigraph streams}}. In
  \bibinfo{booktitle}{\emph{PODS}}.
\newblock


\bibitem[\protect\citeauthoryear{Coscia, Rossetti, Pennacchioli, Ceccarelli,
  and Giannotti}{Coscia et~al\mbox{.}}{2013}]%
        {CosciaRPCG13}
\bibfield{author}{\bibinfo{person}{M. Coscia}, \bibinfo{person}{G. Rossetti},
  \bibinfo{person}{D. Pennacchioli}, \bibinfo{person}{D. Ceccarelli}, {and}
  \bibinfo{person}{F. Giannotti}.} \bibinfo{year}{2013}\natexlab{}.
\newblock \showarticletitle{{"You Know because I Know": A Multidimensional
  Network Approach to Human Resources Problem}}. In
  \bibinfo{booktitle}{\emph{ASONAM}}.
\newblock


\bibitem[\protect\citeauthoryear{Craddock, Benhajali, Chu, Chouinard, Evans,
  Jakab, Khundrakpam, Lewis, Li, Miham, Yan, and Bellec}{Craddock
  et~al\mbox{.}}{2013}]%
        {CBCCE13}
\bibfield{author}{\bibinfo{person}{C. Craddock}, \bibinfo{person}{Y.
  Benhajali}, \bibinfo{person}{C. Chu}, \bibinfo{person}{F. Chouinard},
  \bibinfo{person}{A. Evans}, \bibinfo{person}{A. Jakab},
  \bibinfo{person}{B.~S. Khundrakpam}, \bibinfo{person}{J.~D. Lewis},
  \bibinfo{person}{Q. Li}, \bibinfo{person}{M. Miham}, \bibinfo{person}{C.
  Yan}, {and} \bibinfo{person}{P. Bellec}.} \bibinfo{year}{2013}\natexlab{}.
\newblock \showarticletitle{{The Neuro Bureau Preprocessing Initiative: Open
  Sharing of Preprocessed Neuroimaging Data and Derivatives}}.
\newblock \bibinfo{journal}{\emph{Frontiers in Neuroinformatics}}
  \bibinfo{volume}{41} (\bibinfo{year}{2013}).
\newblock


\bibitem[\protect\citeauthoryear{Dickison, Magnani, and Rossi}{Dickison
  et~al\mbox{.}}{2016}]%
        {dickison_magnani_rossi_2016}
\bibfield{author}{\bibinfo{person}{M.~E. Dickison}, \bibinfo{person}{M.
  Magnani}, {and} \bibinfo{person}{L. Rossi}.} \bibinfo{year}{2016}\natexlab{}.
\newblock \bibinfo{booktitle}{\emph{{Multilayer Social Networks}}}.
\newblock \bibinfo{publisher}{Cambridge University Press}.
\newblock


\bibitem[\protect\citeauthoryear{Fan, Li, Wang, and Wu}{Fan
  et~al\mbox{.}}{2012}]%
        {FLWW12}
\bibfield{author}{\bibinfo{person}{W. Fan}, \bibinfo{person}{J. Li},
  \bibinfo{person}{X. Wang}, {and} \bibinfo{person}{Y. Wu}.}
  \bibinfo{year}{2012}\natexlab{}.
\newblock \showarticletitle{{Query Preserving Graph Compression}}. In
  \bibinfo{booktitle}{\emph{SIGMOD}}.
\newblock


\bibitem[\protect\citeauthoryear{Feigenbaum, Kannan, McGregor, Suri, and
  Zhang}{Feigenbaum et~al\mbox{.}}{2008}]%
        {FeigenbaumKMSZ08}
\bibfield{author}{\bibinfo{person}{J. Feigenbaum}, \bibinfo{person}{S. Kannan},
  \bibinfo{person}{A. McGregor}, \bibinfo{person}{S. Suri}, {and}
  \bibinfo{person}{J. Zhang}.} \bibinfo{year}{2008}\natexlab{}.
\newblock \showarticletitle{{Graph Distances in the Data-Stream Model}}.
\newblock \bibinfo{journal}{\emph{{SIAM} J. Comput.}} \bibinfo{volume}{38},
  \bibinfo{number}{5} (\bibinfo{year}{2008}), \bibinfo{pages}{1709--1727}.
\newblock


\bibitem[\protect\citeauthoryear{Galimberti, Bonchi, and Gullo}{Galimberti
  et~al\mbox{.}}{2017}]%
        {GBG17}
\bibfield{author}{\bibinfo{person}{E. Galimberti}, \bibinfo{person}{F. Bonchi},
  {and} \bibinfo{person}{F. Gullo}.} \bibinfo{year}{2017}\natexlab{}.
\newblock \showarticletitle{{Core Decomposition and Densest Subgraph in
  Multilayer Networks}}. In \bibinfo{booktitle}{\emph{CIKM}}.
\newblock


\bibitem[\protect\citeauthoryear{Galimberti, Bonchi, Gullo, and
  Lanciano}{Galimberti et~al\mbox{.}}{2020}]%
        {GalimbertiBGL20}
\bibfield{author}{\bibinfo{person}{Edoardo Galimberti},
  \bibinfo{person}{Francesco Bonchi}, \bibinfo{person}{Francesco Gullo}, {and}
  \bibinfo{person}{Tommaso Lanciano}.} \bibinfo{year}{2020}\natexlab{}.
\newblock \showarticletitle{Core Decomposition in Multilayer Networks: Theory,
  Algorithms, and Applications}.
\newblock \bibinfo{journal}{\emph{{ACM} Trans. Knowl. Discov. Data}}
  \bibinfo{volume}{14}, \bibinfo{number}{1} (\bibinfo{year}{2020}),
  \bibinfo{pages}{11:1--11:40}.
\newblock


\bibitem[\protect\citeauthoryear{Gionis, Mannila, and Tsaparas}{Gionis
  et~al\mbox{.}}{2007}]%
        {GMT07}
\bibfield{author}{\bibinfo{person}{A. Gionis}, \bibinfo{person}{H. Mannila},
  {and} \bibinfo{person}{P. Tsaparas}.} \bibinfo{year}{2007}\natexlab{}.
\newblock \showarticletitle{{Clustering Aggregation}}.
\newblock \bibinfo{journal}{\emph{TKDD}}  \bibinfo{volume}{1}
  (\bibinfo{year}{2007}).
\newblock


\bibitem[\protect\citeauthoryear{Gionis and Tsourakakis}{Gionis and
  Tsourakakis}{2015}]%
        {GT15}
\bibfield{author}{\bibinfo{person}{A. Gionis} {and} \bibinfo{person}{C.~E.
  Tsourakakis}.} \bibinfo{year}{2015}\natexlab{}.
\newblock \showarticletitle{{Dense Subgraph Discovery: KDD 2015 Tutorial}}. In
  \bibinfo{booktitle}{\emph{KDD}}.
\newblock


\bibitem[\protect\citeauthoryear{Gou, Zou, Zhao, and Yang}{Gou
  et~al\mbox{.}}{2019}]%
        {Gou0Z019}
\bibfield{author}{\bibinfo{person}{X. Gou}, \bibinfo{person}{L. Zou},
  \bibinfo{person}{C. Zhao}, {and} \bibinfo{person}{T. Yang}.}
  \bibinfo{year}{2019}\natexlab{}.
\newblock \showarticletitle{{Fast and Accurate Graph Stream Summarization}}. In
  \bibinfo{booktitle}{\emph{ICDE}}.
\newblock


\bibitem[\protect\citeauthoryear{Hassanlou, Shoaran, and Thomo}{Hassanlou
  et~al\mbox{.}}{2013}]%
        {HassanlouST13}
\bibfield{author}{\bibinfo{person}{N. Hassanlou}, \bibinfo{person}{M. Shoaran},
  {and} \bibinfo{person}{A. Thomo}.} \bibinfo{year}{2013}\natexlab{}.
\newblock \showarticletitle{Probabilistic Graph Summarization}. In
  \bibinfo{booktitle}{\emph{WAIM}} \emph{(\bibinfo{series}{Lecture Notes in
  Computer Science}, Vol.~\bibinfo{volume}{7923})}.
  \bibinfo{publisher}{Springer}.
\newblock


\bibitem[\protect\citeauthoryear{Hu and Lau}{Hu and Lau}{2013}]%
        {HL13}
\bibfield{author}{\bibinfo{person}{P. Hu} {and} \bibinfo{person}{W.~C. Lau}.}
  \bibinfo{year}{2013}\natexlab{}.
\newblock \showarticletitle{{A Survey and Taxonomy of Graph Sampling}}.
\newblock \bibinfo{journal}{\emph{CoRR}}  \bibinfo{volume}{abs/1308.5865}
  (\bibinfo{year}{2013}).
\newblock


\bibitem[\protect\citeauthoryear{Jayaram, Khan, Li, Yan, and Elmasri}{Jayaram
  et~al\mbox{.}}{2015}]%
        {JKLYE15}
\bibfield{author}{\bibinfo{person}{N. Jayaram}, \bibinfo{person}{A. Khan},
  \bibinfo{person}{C. Li}, \bibinfo{person}{X. Yan}, {and} \bibinfo{person}{R.
  Elmasri}.} \bibinfo{year}{2015}\natexlab{}.
\newblock \showarticletitle{{Querying Knowledge Graphs by Example Entity
  Tuples}}.
\newblock \bibinfo{journal}{\emph{{IEEE} Trans. Knowl. Data Eng.}}
  \bibinfo{volume}{27}, \bibinfo{number}{10} (\bibinfo{year}{2015}),
  \bibinfo{pages}{2797--2811}.
\newblock


\bibitem[\protect\citeauthoryear{Jin, Gao, He, Jin, and Li}{Jin
  et~al\mbox{.}}{2020}]%
        {JGHJL20}
\bibfield{author}{\bibinfo{person}{B. Jin}, \bibinfo{person}{C. Gao},
  \bibinfo{person}{X. He}, \bibinfo{person}{D. Jin}, {and} \bibinfo{person}{Y.
  Li}.} \bibinfo{year}{2020}\natexlab{}.
\newblock \showarticletitle{{Multi-behavior Recommendation with Graph
  Convolutional Networks}}. In \bibinfo{booktitle}{\emph{SIGIR}}.
\newblock


\bibitem[\protect\citeauthoryear{Jin, Rossi, Koh, Kim, Rao, and Koutra}{Jin
  et~al\mbox{.}}{2019}]%
        {JRKKRK19}
\bibfield{author}{\bibinfo{person}{D. Jin}, \bibinfo{person}{R.~A. Rossi},
  \bibinfo{person}{E. Koh}, \bibinfo{person}{S. Kim}, \bibinfo{person}{A. Rao},
  {and} \bibinfo{person}{D. Koutra}.} \bibinfo{year}{2019}\natexlab{}.
\newblock \showarticletitle{{Latent Network Summarization: Bridging Network
  Embedding and Summarization}}. In \bibinfo{booktitle}{\emph{KDD}}.
\newblock


\bibitem[\protect\citeauthoryear{k.~Lee, Jo, Ko, Lim, and Shin}{k.~Lee
  et~al\mbox{.}}{2020}]%
        {LJKLS20}
\bibfield{author}{\bibinfo{person}{k. Lee}, \bibinfo{person}{H. Jo},
  \bibinfo{person}{J. Ko}, \bibinfo{person}{S. Lim}, {and} \bibinfo{person}{K.
  Shin}.} \bibinfo{year}{2020}\natexlab{}.
\newblock \showarticletitle{{SSumM: Sparse Summarization of Massive Graphs}}.
  In \bibinfo{booktitle}{\emph{KDD}}.
\newblock


\bibitem[\protect\citeauthoryear{Kang and Faloutsos}{Kang and
  Faloutsos}{2011}]%
        {KF11}
\bibfield{author}{\bibinfo{person}{U. Kang} {and} \bibinfo{person}{C.
  Faloutsos}.} \bibinfo{year}{2011}\natexlab{}.
\newblock \showarticletitle{{Beyond 'Caveman Communities': Hubs and Spokes for
  Graph Compression and Mining}}. In \bibinfo{booktitle}{\emph{ICDM}}.
\newblock


\bibitem[\protect\citeauthoryear{Kang, Tong, Sun, Lin, and Faloutsos}{Kang
  et~al\mbox{.}}{2011}]%
        {KTSLF11}
\bibfield{author}{\bibinfo{person}{U. Kang}, \bibinfo{person}{H. Tong},
  \bibinfo{person}{J. Sun}, \bibinfo{person}{C.{-}Y. Lin}, {and}
  \bibinfo{person}{C. Faloutsos}.} \bibinfo{year}{2011}\natexlab{}.
\newblock \showarticletitle{{GBASE: A Scalable and General Graph Management
  System}}. In \bibinfo{booktitle}{\emph{KDD}}.
\newblock


\bibitem[\protect\citeauthoryear{Karypis and Kumar}{Karypis and Kumar}{1995}]%
        {KK95}
\bibfield{author}{\bibinfo{person}{G. Karypis} {and} \bibinfo{person}{V.
  Kumar}.} \bibinfo{year}{1995}\natexlab{}.
\newblock \showarticletitle{{Analysis of Multilevel Graph Partitioning}}. In
  \bibinfo{booktitle}{\emph{Supercomputing}}.
\newblock


\bibitem[\protect\citeauthoryear{Ke, Khan, and Cong}{Ke et~al\mbox{.}}{2018}]%
        {KKC18}
\bibfield{author}{\bibinfo{person}{X. Ke}, \bibinfo{person}{A. Khan}, {and}
  \bibinfo{person}{G. Cong}.} \bibinfo{year}{2018}\natexlab{}.
\newblock \showarticletitle{{Finding Seeds and Relevant Tags Jointly: For
  Targeted Influence Maximization in Social Networks}}. In
  \bibinfo{booktitle}{\emph{SIGMOD}}.
\newblock


\bibitem[\protect\citeauthoryear{Khan and Aggarwal}{Khan and Aggarwal}{2017}]%
        {KhanA17}
\bibfield{author}{\bibinfo{person}{A. Khan} {and} \bibinfo{person}{C.~C.
  Aggarwal}.} \bibinfo{year}{2017}\natexlab{}.
\newblock \showarticletitle{{Toward Query-Friendly Compression of Rapid Graph
  Streams}}.
\newblock \bibinfo{journal}{\emph{Social Netw. Analys. Mining}}
  \bibinfo{volume}{7}, \bibinfo{number}{1} (\bibinfo{year}{2017}),
  \bibinfo{pages}{23:1--23:19}.
\newblock


\bibitem[\protect\citeauthoryear{Khan, Bhowmick, and Bonchi}{Khan
  et~al\mbox{.}}{2017}]%
        {KBB17}
\bibfield{author}{\bibinfo{person}{A. Khan}, \bibinfo{person}{S.~S. Bhowmick},
  {and} \bibinfo{person}{F. Bonchi}.} \bibinfo{year}{2017}\natexlab{}.
\newblock \showarticletitle{Summarizing Static and Dynamic Big Graphs}.
\newblock \bibinfo{journal}{\emph{{PVLDB}}} \bibinfo{volume}{10},
  \bibinfo{number}{12} (\bibinfo{year}{2017}), \bibinfo{pages}{1981--1984}.
\newblock


\bibitem[\protect\citeauthoryear{Khan, Nawaz, and Lee}{Khan
  et~al\mbox{.}}{2014}]%
        {KNL14}
\bibfield{author}{\bibinfo{person}{K.~U. Khan}, \bibinfo{person}{W. Nawaz},
  {and} \bibinfo{person}{Y.-K. Lee}.} \bibinfo{year}{2014}\natexlab{}.
\newblock \showarticletitle{{Set-Based Unified Approach for Attributed Graph
  Summarization}}. In \bibinfo{booktitle}{\emph{BDCLOUD}}.
\newblock


\bibitem[\protect\citeauthoryear{Koutra, Kang, Vreeken, and Faloutsos}{Koutra
  et~al\mbox{.}}{2014}]%
        {KKVF14}
\bibfield{author}{\bibinfo{person}{D. Koutra}, \bibinfo{person}{U. Kang},
  \bibinfo{person}{J. Vreeken}, {and} \bibinfo{person}{C. Faloutsos}.}
  \bibinfo{year}{2014}\natexlab{}.
\newblock \showarticletitle{{VOG: Summarizing and Understanding Large Graphs}}.
  In \bibinfo{booktitle}{\emph{SDM}}.
\newblock


\bibitem[\protect\citeauthoryear{Koutra, Kang, Vreeken, and Faloutsos}{Koutra
  et~al\mbox{.}}{2015}]%
        {KKVF15}
\bibfield{author}{\bibinfo{person}{D. Koutra}, \bibinfo{person}{U. Kang},
  \bibinfo{person}{J. Vreeken}, {and} \bibinfo{person}{C. Faloutsos}.}
  \bibinfo{year}{2015}\natexlab{}.
\newblock \showarticletitle{{Summarizing and Understanding Large Graphs}}.
\newblock \bibinfo{journal}{\emph{Statistical Analysis and Data Mining}}
  \bibinfo{volume}{8}, \bibinfo{number}{3} (\bibinfo{year}{2015}),
  \bibinfo{pages}{183--202}.
\newblock


\bibitem[\protect\citeauthoryear{Koutra, Vreeken, and Bonchi}{Koutra
  et~al\mbox{.}}{2018}]%
        {KoutraVB18}
\bibfield{author}{\bibinfo{person}{D. Koutra}, \bibinfo{person}{J. Vreeken},
  {and} \bibinfo{person}{F. Bonchi}.} \bibinfo{year}{2018}\natexlab{}.
\newblock \showarticletitle{{Summarizing Graphs at Multiple Scales: New
  Trends}}. In \bibinfo{booktitle}{\emph{ICDM}}.
\newblock


\bibitem[\protect\citeauthoryear{Kumar and Efstathopoulos}{Kumar and
  Efstathopoulos}{2018}]%
        {KE18}
\bibfield{author}{\bibinfo{person}{K.~A. Kumar} {and} \bibinfo{person}{P.
  Efstathopoulos}.} \bibinfo{year}{2018}\natexlab{}.
\newblock \showarticletitle{{Utility-Driven Graph Summarization}}.
\newblock \bibinfo{journal}{\emph{{PVLDB}}} \bibinfo{volume}{12},
  \bibinfo{number}{4} (\bibinfo{year}{2018}), \bibinfo{pages}{335--347}.
\newblock


\bibitem[\protect\citeauthoryear{LeFevre and Terzi}{LeFevre and Terzi}{2010}]%
        {LeFevreT10}
\bibfield{author}{\bibinfo{person}{K. LeFevre} {and} \bibinfo{person}{E.
  Terzi}.} \bibinfo{year}{2010}\natexlab{}.
\newblock \showarticletitle{{GraSS: Graph Structure Summarization}}. In
  \bibinfo{booktitle}{\emph{SDM}}.
\newblock


\bibitem[\protect\citeauthoryear{Lin, Yeh, and Li}{Lin et~al\mbox{.}}{2013}]%
        {LYL13}
\bibfield{author}{\bibinfo{person}{S.-D. Lin}, \bibinfo{person}{M.-Y. Yeh},
  {and} \bibinfo{person}{C.-T. Li}.} \bibinfo{year}{2013}\natexlab{}.
\newblock \showarticletitle{{Sampling and Summarization for Social Networks}}.
  In \bibinfo{booktitle}{\emph{SDM}}.
\newblock


\bibitem[\protect\citeauthoryear{Liu, Tian, He, Lee, and McPherson}{Liu
  et~al\mbox{.}}{2014}]%
        {LTHLM14}
\bibfield{author}{\bibinfo{person}{X. Liu}, \bibinfo{person}{Y. Tian},
  \bibinfo{person}{Q. He}, \bibinfo{person}{W.-C. Lee}, {and}
  \bibinfo{person}{J. McPherson}.} \bibinfo{year}{2014}\natexlab{}.
\newblock \showarticletitle{{Distributed Graph Summarization}}. In
  \bibinfo{booktitle}{\emph{CIKM}}.
\newblock


\bibitem[\protect\citeauthoryear{Liu, Safavi, Dighe, and Koutra}{Liu
  et~al\mbox{.}}{2018}]%
        {LiuSDK18}
\bibfield{author}{\bibinfo{person}{Y. Liu}, \bibinfo{person}{T. Safavi},
  \bibinfo{person}{A. Dighe}, {and} \bibinfo{person}{D. Koutra}.}
  \bibinfo{year}{2018}\natexlab{}.
\newblock \showarticletitle{{Graph Summarization Methods and Applications: A
  Survey}}.
\newblock \bibinfo{journal}{\emph{{ACM} Comput. Surv.}} \bibinfo{volume}{51},
  \bibinfo{number}{3} (\bibinfo{year}{2018}), \bibinfo{pages}{62:1--62:34}.
\newblock


\bibitem[\protect\citeauthoryear{Mao, Lu, Zhang, and Zhang}{Mao
  et~al\mbox{.}}{2017}]%
        {MaoLZZ17}
\bibfield{author}{\bibinfo{person}{M. Mao}, \bibinfo{person}{J. Lu},
  \bibinfo{person}{G. Zhang}, {and} \bibinfo{person}{J. Zhang}.}
  \bibinfo{year}{2017}\natexlab{}.
\newblock \showarticletitle{{Multirelational Social Recommendations via
  Multigraph Ranking}}.
\newblock \bibinfo{journal}{\emph{{IEEE} Trans. Cybernetics}}
  \bibinfo{volume}{47}, \bibinfo{number}{12} (\bibinfo{year}{2017}),
  \bibinfo{pages}{4049--4061}.
\newblock


\bibitem[\protect\citeauthoryear{Maserrat and Pei}{Maserrat and Pei}{2010}]%
        {MP10}
\bibfield{author}{\bibinfo{person}{H. Maserrat} {and} \bibinfo{person}{J.
  Pei}.} \bibinfo{year}{2010}\natexlab{}.
\newblock \showarticletitle{{Neighbor Query Friendly Compression of Social
  Networks}}. In \bibinfo{booktitle}{\emph{KDD}}.
\newblock


\bibitem[\protect\citeauthoryear{Maserrat and Pei}{Maserrat and Pei}{2012}]%
        {MP12}
\bibfield{author}{\bibinfo{person}{H. Maserrat} {and} \bibinfo{person}{J.
  Pei}.} \bibinfo{year}{2012}\natexlab{}.
\newblock \showarticletitle{{Community Preserving Lossy Compression of Social
  Networks}}. In \bibinfo{booktitle}{\emph{ICDM}}.
\newblock


\bibitem[\protect\citeauthoryear{Navlakha, Rastogi, and Shrivastava}{Navlakha
  et~al\mbox{.}}{2008}]%
        {NRS08}
\bibfield{author}{\bibinfo{person}{S. Navlakha}, \bibinfo{person}{R. Rastogi},
  {and} \bibinfo{person}{N. Shrivastava}.} \bibinfo{year}{2008}\natexlab{}.
\newblock \showarticletitle{{Graph Summarization with Bounded Error}}. In
  \bibinfo{booktitle}{\emph{SIGMOD}}.
\newblock


\bibitem[\protect\citeauthoryear{Newman and Girvan}{Newman and Girvan}{2004}]%
        {NG69}
\bibfield{author}{\bibinfo{person}{M.~E.~J. Newman} {and} \bibinfo{person}{M.
  Girvan}.} \bibinfo{year}{2004}\natexlab{}.
\newblock \showarticletitle{{Finding and Evaluating Community Structure in
  Networks}}.
\newblock \bibinfo{journal}{\emph{Phys. Rev. E}}  \bibinfo{volume}{69}
  (\bibinfo{year}{2004}), \bibinfo{pages}{026113}.
\newblock
Issue 2.


\bibitem[\protect\citeauthoryear{Phuoc, Quoc, Quoc, Nhat, and Hauswirth}{Phuoc
  et~al\mbox{.}}{2016}]%
        {PhuocQQNH16}
\bibfield{author}{\bibinfo{person}{D.~L. Phuoc}, \bibinfo{person}{H.~N.~M.
  Quoc}, \bibinfo{person}{H.~N. Quoc}, \bibinfo{person}{T.~T. Nhat}, {and}
  \bibinfo{person}{M. Hauswirth}.} \bibinfo{year}{2016}\natexlab{}.
\newblock \showarticletitle{{The Graph of Things: A Step towards the Live
  Knowledge Graph of Connected Things}}.
\newblock \bibinfo{journal}{\emph{J. Web Semant.}}  \bibinfo{volume}{37-38}
  (\bibinfo{year}{2016}), \bibinfo{pages}{25--35}.
\newblock


\bibitem[\protect\citeauthoryear{Purohit, Prakash, Kang, Zhang, and
  Subrahmanian}{Purohit et~al\mbox{.}}{2014}]%
        {PPKZS14}
\bibfield{author}{\bibinfo{person}{M. Purohit}, \bibinfo{person}{B.~A.
  Prakash}, \bibinfo{person}{C. Kang}, \bibinfo{person}{Y. Zhang}, {and}
  \bibinfo{person}{V.~S. Subrahmanian}.} \bibinfo{year}{2014}\natexlab{}.
\newblock \showarticletitle{{Fast Influence-based Coarsening for Large
  Networks}}. In \bibinfo{booktitle}{\emph{KDD}}.
\newblock


\bibitem[\protect\citeauthoryear{Qu, Liu, Jensen, Zhu, and Faloutsos}{Qu
  et~al\mbox{.}}{2014}]%
        {QLJZF14}
\bibfield{author}{\bibinfo{person}{Q. Qu}, \bibinfo{person}{S. Liu},
  \bibinfo{person}{C.~S. Jensen}, \bibinfo{person}{F. Zhu}, {and}
  \bibinfo{person}{C. Faloutsos}.} \bibinfo{year}{2014}\natexlab{}.
\newblock \showarticletitle{{Interestingness-Driven Diffusion Process
  Summarization in Dynamic Networks}}. In \bibinfo{booktitle}{\emph{ECML
  PKDD}}.
\newblock


\bibitem[\protect\citeauthoryear{Raghavan and Garcia{-}Molina}{Raghavan and
  Garcia{-}Molina}{2003}]%
        {RG03}
\bibfield{author}{\bibinfo{person}{S. Raghavan} {and} \bibinfo{person}{H.
  Garcia{-}Molina}.} \bibinfo{year}{2003}\natexlab{}.
\newblock \showarticletitle{{Representing Web Graphs}}. In
  \bibinfo{booktitle}{\emph{ICDE}}.
\newblock


\bibitem[\protect\citeauthoryear{Riondato, Garc\'{i}a-Soriano, and
  Bonchi}{Riondato et~al\mbox{.}}{2014}]%
        {RGB14}
\bibfield{author}{\bibinfo{person}{M. Riondato}, \bibinfo{person}{D.
  Garc\'{i}a-Soriano}, {and} \bibinfo{person}{F. Bonchi}.}
  \bibinfo{year}{2014}\natexlab{}.
\newblock \showarticletitle{{Graph Summarization with Quality Guarantees}}. In
  \bibinfo{booktitle}{\emph{ICDM}}.
\newblock


\bibitem[\protect\citeauthoryear{Riondato, Garc{\'{\i}}a{-}Soriano, and
  Bonchi}{Riondato et~al\mbox{.}}{2017}]%
        {RGB17}
\bibfield{author}{\bibinfo{person}{Matteo Riondato}, \bibinfo{person}{David
  Garc{\'{\i}}a{-}Soriano}, {and} \bibinfo{person}{Francesco Bonchi}.}
  \bibinfo{year}{2017}\natexlab{}.
\newblock \showarticletitle{Graph summarization with quality guarantees}.
\newblock \bibinfo{journal}{\emph{Data Min. Knowl. Discov.}}
  \bibinfo{volume}{31}, \bibinfo{number}{2} (\bibinfo{year}{2017}),
  \bibinfo{pages}{314--349}.
\newblock


\bibitem[\protect\citeauthoryear{Rissanen}{Rissanen}{1978}]%
        {JR78}
\bibfield{author}{\bibinfo{person}{J. Rissanen}.}
  \bibinfo{year}{1978}\natexlab{}.
\newblock \showarticletitle{{Modelling by the Shortest Data Description}}.
\newblock \bibinfo{journal}{\emph{Automatica}}  \bibinfo{volume}{14}
  (\bibinfo{year}{1978}).
\newblock


\bibitem[\protect\citeauthoryear{Rossi and Zhou}{Rossi and Zhou}{2018}]%
        {RZ18}
\bibfield{author}{\bibinfo{person}{R.~A. Rossi} {and} \bibinfo{person}{R.
  Zhou}.} \bibinfo{year}{2018}\natexlab{}.
\newblock \showarticletitle{{GraphZIP: A Clique-based Sparse Graph Compression
  Method}}.
\newblock \bibinfo{journal}{\emph{J. Big Data}}  \bibinfo{volume}{5}
  (\bibinfo{year}{2018}), \bibinfo{pages}{10}.
\newblock


\bibitem[\protect\citeauthoryear{Seah, Bhowmick, and Jr}{Seah
  et~al\mbox{.}}{2014}]%
        {SBD14}
\bibfield{author}{\bibinfo{person}{B.-S. Seah}, \bibinfo{person}{S.~S.
  Bhowmick}, {and} \bibinfo{person}{C.~F.~Dewey Jr}.}
  \bibinfo{year}{2014}\natexlab{}.
\newblock \showarticletitle{{DiffNet: Automatic Differential Functional
  Summarization of dE-MAP Networks}}.
\newblock \bibinfo{journal}{\emph{Methods}} \bibinfo{volume}{63},
  \bibinfo{number}{3} (\bibinfo{year}{2014}).
\newblock


\bibitem[\protect\citeauthoryear{Seah, Bhowmick, Jr, and Yu}{Seah
  et~al\mbox{.}}{2012}]%
        {SBDY12}
\bibfield{author}{\bibinfo{person}{B.-S. Seah}, \bibinfo{person}{S.~S.
  Bhowmick}, \bibinfo{person}{C.~F.~Dewey Jr}, {and} \bibinfo{person}{H. Yu}.}
  \bibinfo{year}{2012}\natexlab{}.
\newblock \showarticletitle{{FUSE: A Profit Maximization Approach for
  Functional Summarization of Biological Networks}}.
\newblock \bibinfo{journal}{\emph{BMC Bioinformatics}}  \bibinfo{volume}{13}
  (\bibinfo{year}{2012}).
\newblock


\bibitem[\protect\citeauthoryear{Shah, Koutra, Zou, Gallagher, and
  Faloutsos}{Shah et~al\mbox{.}}{2015}]%
        {SKZGF15}
\bibfield{author}{\bibinfo{person}{N. Shah}, \bibinfo{person}{D. Koutra},
  \bibinfo{person}{T. Zou}, \bibinfo{person}{B. Gallagher}, {and}
  \bibinfo{person}{C. Faloutsos}.} \bibinfo{year}{2015}\natexlab{}.
\newblock \showarticletitle{{TimeCrunch: Interpretable Dynamic Graph
  Summarization}}. In \bibinfo{booktitle}{\emph{KDD}}.
\newblock


\bibitem[\protect\citeauthoryear{Shi, Sun, Xuan, Su, Tong, Ma, and Chen}{Shi
  et~al\mbox{.}}{2016}]%
        {SSXSTMC16}
\bibfield{author}{\bibinfo{person}{L. Shi}, \bibinfo{person}{S. Sun},
  \bibinfo{person}{Y. Xuan}, \bibinfo{person}{Y. Su}, \bibinfo{person}{H.
  Tong}, \bibinfo{person}{S. Ma}, {and} \bibinfo{person}{Y. Chen}.}
  \bibinfo{year}{2016}\natexlab{}.
\newblock \showarticletitle{{TOPIC: Toward Perfect Influence Graph
  Summarization}}. In \bibinfo{booktitle}{\emph{ICDE}}.
\newblock


\bibitem[\protect\citeauthoryear{Shin, Ghoting, Kim, and Raghavan}{Shin
  et~al\mbox{.}}{2019}]%
        {SGKR19}
\bibfield{author}{\bibinfo{person}{K. Shin}, \bibinfo{person}{A. Ghoting},
  \bibinfo{person}{M. Kim}, {and} \bibinfo{person}{H. Raghavan}.}
  \bibinfo{year}{2019}\natexlab{}.
\newblock \showarticletitle{{SWeG: Lossless and Lossy Summariation of Web-Scale
  Graphs}}. In \bibinfo{booktitle}{\emph{WWW}}.
\newblock


\bibitem[\protect\citeauthoryear{Stella, Andreazzi, Selakovic, Goudarzi, and
  Antonioni}{Stella et~al\mbox{.}}{2017}]%
        {SASGA17}
\bibfield{author}{\bibinfo{person}{M. Stella}, \bibinfo{person}{C.~S.
  Andreazzi}, \bibinfo{person}{S. Selakovic}, \bibinfo{person}{A. Goudarzi},
  {and} \bibinfo{person}{A. Antonioni}.} \bibinfo{year}{2017}\natexlab{}.
\newblock \showarticletitle{{Parasite Spreading in Spatial Ecological Multiplex
  Networks}}.
\newblock \bibinfo{journal}{\emph{J. Complex Networks}} \bibinfo{volume}{5},
  \bibinfo{number}{3} (\bibinfo{year}{2017}), \bibinfo{pages}{486--511}.
\newblock


\bibitem[\protect\citeauthoryear{Tang, Chen, and Mitra}{Tang
  et~al\mbox{.}}{2016}]%
        {TCM16}
\bibfield{author}{\bibinfo{person}{N. Tang}, \bibinfo{person}{Q. Chen}, {and}
  \bibinfo{person}{P. Mitra}.} \bibinfo{year}{2016}\natexlab{}.
\newblock \showarticletitle{{Graph Stream Summarization: From Big Bang to Big
  Crunch}}. In \bibinfo{booktitle}{\emph{SIGMOD}}.
\newblock


\bibitem[\protect\citeauthoryear{Thorndike}{Thorndike}{1953}]%
        {Tho53}
\bibfield{author}{\bibinfo{person}{R.~L. Thorndike}.}
  \bibinfo{year}{1953}\natexlab{}.
\newblock \showarticletitle{{Who Belongs in the Family?}}
\newblock \bibinfo{journal}{\emph{Psychometrika}}  \bibinfo{volume}{18}
  (\bibinfo{year}{1953}), \bibinfo{pages}{267--276}.
\newblock


\bibitem[\protect\citeauthoryear{Tian, Hankins, and Patel}{Tian
  et~al\mbox{.}}{2008}]%
        {THP08}
\bibfield{author}{\bibinfo{person}{Y. Tian}, \bibinfo{person}{R.~A. Hankins},
  {and} \bibinfo{person}{J.~M. Patel}.} \bibinfo{year}{2008}\natexlab{}.
\newblock \showarticletitle{{Efficient Aggregation for Graph Summarization}}.
  In \bibinfo{booktitle}{\emph{SIGMOD}}.
\newblock


\bibitem[\protect\citeauthoryear{Tian and Patel}{Tian and Patel}{2010}]%
        {TianP10}
\bibfield{author}{\bibinfo{person}{Y. Tian} {and} \bibinfo{person}{J.~M.
  Patel}.} \bibinfo{year}{2010}\natexlab{}.
\newblock \showarticletitle{{Interactive Graph Summarization}}.
\newblock In \bibinfo{booktitle}{\emph{Link Mining: Models, Algorithms, and
  Applications}}. \bibinfo{pages}{389--409}.
\newblock


\bibitem[\protect\citeauthoryear{Toivonen, Zhou, Hartikainen, and
  Hinkka}{Toivonen et~al\mbox{.}}{2011}]%
        {TZHH11}
\bibfield{author}{\bibinfo{person}{H. Toivonen}, \bibinfo{person}{F. Zhou},
  \bibinfo{person}{A. Hartikainen}, {and} \bibinfo{person}{A. Hinkka}.}
  \bibinfo{year}{2011}\natexlab{}.
\newblock \showarticletitle{{Compression of Weighted Graphs}}. In
  \bibinfo{booktitle}{\emph{KDD}}.
\newblock


\bibitem[\protect\citeauthoryear{Tsalouchidou, Morales, Bonchi, and
  Baeza{-}Yates}{Tsalouchidou et~al\mbox{.}}{2016}]%
        {TMBB16}
\bibfield{author}{\bibinfo{person}{I. Tsalouchidou},
  \bibinfo{person}{G.~De~Francisci Morales}, \bibinfo{person}{F. Bonchi}, {and}
  \bibinfo{person}{R. Baeza{-}Yates}.} \bibinfo{year}{2016}\natexlab{}.
\newblock \showarticletitle{{Scalable dynamic graph summarization}}. In
  \bibinfo{booktitle}{\emph{BigData}}.
\newblock


\bibitem[\protect\citeauthoryear{White and Smyth}{White and Smyth}{2005}]%
        {WhiteS05}
\bibfield{author}{\bibinfo{person}{S. White} {and} \bibinfo{person}{P. Smyth}.}
  \bibinfo{year}{2005}\natexlab{}.
\newblock \showarticletitle{{A Spectral Clustering Approach To Finding
  Communities in Graph}}. In \bibinfo{booktitle}{\emph{SDM}}.
\newblock


\bibitem[\protect\citeauthoryear{Wu, Yang, Srivatsa, Iyengar, and Yan}{Wu
  et~al\mbox{.}}{2013}]%
        {WYSIY13}
\bibfield{author}{\bibinfo{person}{Y. Wu}, \bibinfo{person}{S. Yang},
  \bibinfo{person}{M. Srivatsa}, \bibinfo{person}{A. Iyengar}, {and}
  \bibinfo{person}{X. Yan}.} \bibinfo{year}{2013}\natexlab{}.
\newblock \showarticletitle{{Summarizing Answer Graphs Induced by Keyword
  Queries}}.
\newblock \bibinfo{journal}{\emph{{PVLDB}}} \bibinfo{volume}{6},
  \bibinfo{number}{14} (\bibinfo{year}{2013}), \bibinfo{pages}{1774--1785}.
\newblock


\bibitem[\protect\citeauthoryear{Xia, Huang, Xu, Dai, Zhang, Yang, Pei, and
  Bo}{Xia et~al\mbox{.}}{2021}]%
        {XHXDZYPB21}
\bibfield{author}{\bibinfo{person}{L. Xia}, \bibinfo{person}{C. Huang},
  \bibinfo{person}{Y. Xu}, \bibinfo{person}{P. Dai}, \bibinfo{person}{X.
  Zhang}, \bibinfo{person}{H. Yang}, \bibinfo{person}{J. Pei}, {and}
  \bibinfo{person}{L. Bo}.} \bibinfo{year}{2021}\natexlab{}.
\newblock \showarticletitle{{Knowledge-enhanced Hierarchical Graph Transformer
  Network for Multi-behavior Recommendation}}. In
  \bibinfo{booktitle}{\emph{AAAI}}.
\newblock


\bibitem[\protect\citeauthoryear{Yan, Zhou, and Han}{Yan et~al\mbox{.}}{2005}]%
        {YanZH05}
\bibfield{author}{\bibinfo{person}{X. Yan}, \bibinfo{person}{X.~J. Zhou}, {and}
  \bibinfo{person}{J. Han}.} \bibinfo{year}{2005}\natexlab{}.
\newblock \showarticletitle{{Mining Closed Relational Graphs with Connectivity
  Constraints}}. In \bibinfo{booktitle}{\emph{KDD}}.
\newblock


\bibitem[\protect\citeauthoryear{{Yang}, {You}, and {Wan}}{{Yang}
  et~al\mbox{.}}{2021}]%
        {YYW21}
\bibfield{author}{\bibinfo{person}{J. {Yang}}, \bibinfo{person}{J. {You}},
  {and} \bibinfo{person}{X. {Wan}}.} \bibinfo{year}{2021}\natexlab{}.
\newblock \showarticletitle{Graph Embedding via Graph Summarization}.
\newblock \bibinfo{journal}{\emph{IEEE Access}}  \bibinfo{volume}{9}
  (\bibinfo{year}{2021}), \bibinfo{pages}{45163--45174}.
\newblock


\bibitem[\protect\citeauthoryear{Zhang, Bhowmick, Nguyen, Choi, and Zhu}{Zhang
  et~al\mbox{.}}{2015}]%
        {ZhangBNCZ15}
\bibfield{author}{\bibinfo{person}{J. Zhang}, \bibinfo{person}{S.~S. Bhowmick},
  \bibinfo{person}{H.~H. Nguyen}, \bibinfo{person}{B. Choi}, {and}
  \bibinfo{person}{F. Zhu}.} \bibinfo{year}{2015}\natexlab{}.
\newblock \showarticletitle{{DaVinci: Data-driven Visual Interface Construction
  for Subgraph Search in Graph Databases}}. In
  \bibinfo{booktitle}{\emph{ICDE}}.
\newblock


\bibitem[\protect\citeauthoryear{Zhang and {\"{O}}zsu}{Zhang and
  {\"{O}}zsu}{2019}]%
        {ZhangO19}
\bibfield{author}{\bibinfo{person}{X. Zhang} {and} \bibinfo{person}{T.
  {\"{O}}zsu}.} \bibinfo{year}{2019}\natexlab{}.
\newblock \showarticletitle{{Correlation Constraint Shortest Path over Large
  Multi-Relation Graphs}}.
\newblock \bibinfo{journal}{\emph{{PVLDB}}} \bibinfo{volume}{12},
  \bibinfo{number}{5} (\bibinfo{year}{2019}), \bibinfo{pages}{488--501}.
\newblock


\bibitem[\protect\citeauthoryear{Zhao, Aggarwal, and Wang}{Zhao
  et~al\mbox{.}}{2011}]%
        {ZhaoAW11}
\bibfield{author}{\bibinfo{person}{P. Zhao}, \bibinfo{person}{C.~C. Aggarwal},
  {and} \bibinfo{person}{M. Wang}.} \bibinfo{year}{2011}\natexlab{}.
\newblock \showarticletitle{{gSketch: On Query Estimation in Graph Streams}}.
\newblock \bibinfo{journal}{\emph{{PVLDB}}} \bibinfo{volume}{5},
  \bibinfo{number}{3} (\bibinfo{year}{2011}), \bibinfo{pages}{193--204}.
\newblock


\bibitem[\protect\citeauthoryear{Zhou, Mahler, and Toivonen}{Zhou
  et~al\mbox{.}}{2010}]%
        {ZMT10}
\bibfield{author}{\bibinfo{person}{F. Zhou}, \bibinfo{person}{S. Mahler}, {and}
  \bibinfo{person}{H. Toivonen}.} \bibinfo{year}{2010}\natexlab{}.
\newblock \showarticletitle{{Network Simplification with Minimal Loss of
  Connectivity}}. In \bibinfo{booktitle}{\emph{ICDM}}.
\newblock


\end{thebibliography}


\end{document}